\documentclass[twocolumn]{aastex631}

\usepackage{CJK}
\usepackage{lmodern}
\usepackage{slantsc}
\usepackage{gensymb}
\usepackage{amsmath}
\usepackage{bm}

\newcommand{\hi}{\textrm{H\textsc{i}}}
\newcommand{\althi}{\textsl{H\textsc{i}}}
\newcommand{\secref}[1]{\hyperref[#1]{Section~\ref*{#1}}}
\newcommand{\appref}[1]{\hyperref[#1]{Appendix~\ref*{#1}}}

\defcitealias{2025MNRAS.537.3632M}{MK25}

\begin{document}
\begin{CJK*}{UTF8}{gbsn}
\title{Emission-line Stacking of 21\,cm Intensity Maps with MeerKLASS: Inference Pipeline and Application to the \textit{L}-band Deep-field Data}

\correspondingauthor{Zhaoting Chen}
\email{zhaoting.chen@roe.ac.uk}

\author[0000-0002-4965-8239]{Zhaoting Chen (陈兆庭)}
\affiliation{Institute for Astronomy, The University of Edinburgh, Royal Observatory, Edinburgh EH9 3HJ, UK}

\author[0000-0001-6594-107X]{Steven Cunnington}
\affiliation{Institute of Cosmology \& Gravitation, University of Portsmouth, Dennis Sciama Building,
Portsmouth, PO1 3FX, UK}
\affiliation{Jodrell Bank Centre for Astrophysics, Department of Physics \& Astronomy, The University of Manchester, Manchester M13 9PL, UK}

\author[0000-0001-9110-5550]{Alkistis Pourtsidou}
\affiliation{Institute for Astronomy, The University of Edinburgh, Royal Observatory, Edinburgh EH9 3HJ, UK}
\affiliation{Higgs Centre for Theoretical Physics, School of Physics and Astronomy, Edinburgh EH9 3FD, UK}

\author[0000-0003-3334-3037]{Laura Wolz}
\affiliation{Jodrell Bank Centre for Astrophysics, Department of Physics \& Astronomy, The University of Manchester, Manchester M13 9PL, UK}

\author[0000-0003-0148-3254]{Marta Spinelli}
\affiliation{Observatoire de la C\^ote d'Azur, Laboratoire Lagrange, Bd de l'Observatoire, CS 34229, 06304 Nice cedex 4, France}
\affiliation{Department of Physics and Astronomy, University of the Western Cape, Robert Sobukwe Road, Cape Town 7535, South Africa}

\author[0000-0002-0961-4653]{Jos\'e Luis Bernal}
\affiliation{Instituto de F\'isica de Cantabria (IFCA), CSIC-Univ. de Cantabria, Avda. de los Castros s/n, E-39005 Santander, Spain}

\author[0009-0007-8964-5807]{Matilde Barberi-Squarotti}
\affiliation{Universit\`a degli Studi di Milano, Via Celoria 16, 20133 Milan, Italy}
\affiliation{INAF -- Istituto Nazionale di Astrofisica, Osservatorio Astrofisico di Brera-Merate, Via Brera 28, 20121, Milan, Italy}
\affiliation{INFN -- Istituto Nazionale di Fisica Nucleare, Sezione di Milano, Via Celoria 16, 20133 Milan, Italy}

\author[0000-0003-3399-3574]{Stefano Camera}
\affiliation{Dipartimento di Fisica, Universit\`a degli Studi di Torino, Via P.\ Giuria 1, 10125 Torino, Italy}
\affiliation{INFN -- Istituto Nazionale di Fisica Nucleare, Sezione di Torino, Via P.\ Giuria 1, 10125 Torino, Italy}
\affiliation{INAF -- Istituto Nazionale di Astrofisica, Osservatorio Astrofisico di Torino, Strada Osservatorio 20, 10025 Pino Torinese, Italy}
\affiliation{Department of Physics \& Astronomy, University of the Western Cape, 7535 Cape Town, South Africa}

\author[0000-0001-5287-0065]{Isabella P. Carucci}
\affiliation{INAF - Osservatorio Astronomico di Trieste, Via G.B. Tiepolo 11, 34131 Trieste, Italy}
\affiliation{IFPU - Institute for Fundamental Physics of the Universe, Via Beirut 2, 34151 Trieste, Italy}

\author[0000-0003-0549-1614]{Jos\'e Fonseca}
\affiliation{Instituto de Astrofisica e Ci\^{e}ncias do Espa\c{c}o, Universidade do Porto CAUP, Rua das Estrelas, PT4150-762 Porto, Portugal}
\affiliation{Department of Physics and Astronomy, University of the Western Cape, Robert Sobukwe Road, Cape Town 7535, South Africa}

\author[0000-0002-6780-1406]{Keith Grainge}
\affiliation{Jodrell Bank Centre for Astrophysics, Department of Physics \& Astronomy, The University of Manchester, Manchester M13 9PL, UK}

\author[0000-0003-2021-7357]{Melis O. Irfan}
\affiliation{Institute of Astronomy, University of Cambridge, Cambridge, CB3 0HA}

\author[0000-0003-3892-3073]{Mario G. Santos}
\affiliation{Department of Physics and Astronomy, University of the Western Cape, Robert Sobukwe Road, Cape Town 7535, South Africa}

\author[0000-0002-5598-2668]{Jingying Wang (王婧颖)}
\affiliation{Shanghai Astronomical Observatory, Chinese Academy of Sciences, 80 Nandan Road, Shanghai, 200030, China}
\affiliation{Department of Physics and Astronomy, University of the Western Cape, Robert Sobukwe Road, Cape Town 7535, South Africa}

\collaboration{20}{(MeerKLASS Collaboration)}


\begin{abstract}
We present a novel analysis of observational systematics through the emission-line stacking of the MeerKLASS \textit{L}-band deep-field intensity maps, following the detection in \cite{2025MNRAS.537.3632M}. 
A stacking signal is obtained by stacking the 21\,cm intensity map cubelets around the galaxy positions from the GAMA survey at $0.39\lesssim z \lesssim0.46$.
An extensive simulation framework is built to study the viability of the stacking detection, the covariance estimation, and the model inference, which are then applied to the data.
The statistical significance of the detection is $8.66\sigma$ when averaged into an angular map, and $7.45\sigma$ when averaged into a spectrum. 
The stacked spectrum exhibits an oscillating component of systematics, and we provide evidence that these systematics are a convolutional effect on the map data. 
The oscillation frequency matches the diffraction from the secondary reflector into the primary beam of the MeerKAT telescope.
Bayesian inference can be used to constrain the systematics and the average \hi\ emission of the galaxies. 
The fitting of the parameters gives a constraint on the systematics frequency $\nu_{\rm sys}\,[{\rm MHz}] = 17.90^{+6.53}_{-4.27}$. 
{The posterior of the systematics amplitude reaches the wide prior and gives $A_{\rm sys}=0.50^{+0.33}_{-0.33}$.}
A tentative measurement of the average \hi\ mass of the sources is achieved at $\log_{10}[\langle M_{\hi}\rangle/M_\odot ]=9.84^{+0.48}_{-0.59}$, which is {an underestimation} limited by the narrow redshift bin, the {strong degeneracy with} the systematics, and the low-density galaxy sample. 
These shortfalls will be resolved for future MeerKLASS data to enable accurate measurements of the \hi\ density through stacking of intensity maps.
\end{abstract}

\section{Introduction}
A primary goal of observational cosmology is to map the distribution of the cosmic large-scale structure (LSS) throughout the evolutionary history of the Universe. 
To probe the initial conditions and the subsequent growth of the cosmic structure, different tracers covering a wide range of cosmological redshifts are needed. 
Among them, neutral hydrogen (\hi) intensity mapping \citep{1997ApJ...475..429M,2001JApA...22...21B,2004MNRAS.355.1339B,2008MNRAS.383.1195W,2008PhRvL.100i1303C} emerges as a unique probe. 
Instead of resolving the sources of \hi, intensity mapping aims to map the flux density of the 21\,cm emission line from the hyperfine transition of the \hi\ \citep{1970ITIM...19..200H} over large cosmological volumes. 
It can be used to probe the cosmic dark ages \citep{2007PhRvD..76h3005L}, the cosmic dawn, and the subsequent epoch of reionization \citep{2006PhR...433..181F}. 
After the cosmic reionization is complete at $z\lesssim 5.5$ \citep{2022MNRAS.514...55B,2022ApJ...932...76Z,2024MNRAS.533L..49Z,2024A&A...688L..26S}, the 21\,cm line traces primarily the dark matter halos and structures therein \citep{2018ApJ...866..135V}. 
As the 21\,cm line has a fixed rest frame frequency, intensity mapping surveys are spectroscopic in nature. 
Surveys in the near future with the Square Kilometre Array Observatory (SKAO) will be able to measure the \hi\ power spectrum and constrain the underlying cosmological model precisely, matching the precision of current optical galaxy surveys \citep{2020PASA...37....7S}. 
Moreover, it has the unique advantage of probing the high-redshift, post-reionization Universe $z\gtrsim 3.0$ with fine redshift resolution \citep{2023MNRAS.524.3724C}, and holds great synergy potential with the line-emission intensity mapping of other spectral lines~\citep{2022A&ARv..30....5B}.

Tremendous efforts have been made toward measuring the post-reionization \hi\ intensity mapping signal at large cosmological scales. 
Detections of the \hi\ clustering in cross-correlation with optical galaxies have been achieved, for example, using the Green Bank Telescope \citep{2013ApJ...763L..20M,2013MNRAS.434L..46S,2022MNRAS.510.3495W} and the Parkes telescope \citep{2018MNRAS.476.3382A}. 
There are numerous current and forthcoming experiments conducting post-reionization 21\,cm surveys, such as the Baryon Acoustic Oscillations from Integrated Neutral Gas Observations telescope (BINGO; \citealt{2022A&A...664A..14A}), the Canadian Hydrogen Intensity Mapping Experiment (CHIME; \citealt{2022ApJS..261...29C}), the Canadian Hydrogen Observatory and Radio-transient Detector (CHORD; \citealt{2019clrp.2020...28V}), the Five-hundred-meter Aperture Spherical radio Telescope (FAST; \citealt{2023ApJ...954..139L}), the Hydrogen Intensity and Real-Time Analysis Experiment (HIRAX; \citealt{2022JATIS...8a1019C}), the Tianlai array \citep{2021A&C....3400439Z}, and the upgraded Giant Metrewave Radio Telescope (uGMRT; \citealt{2022MNRAS.516.2851P}). 
In this paper, we focus on the progress made by the MeerKAT Large Area Synoptic Survey (MeerKLASS; \citealt{2016mks..confE..32S}) using the MeerKAT telescope. 
The MeerKAT telescope is the precursor to, and will be part of, the mid-frequency array of the SKAO (SKA-Mid). 
The MeerKLASS survey has produced 21\,cm intensity maps in the \textit{L}-band, which were used to detect the cross-power spectrum with optical galaxies \citep{2021MNRAS.505.3698W,2023MNRAS.518.6262C,2024arXiv241206750C}. 
The MeerKLASS \textit{L}-band deep-field observations, produced from 62\,hr of observation over 236\,$\rm deg^2$, represent the deepest single-dish \hi\ intensity maps to date, as presented in \cite{2025MNRAS.537.3632M}, hereafter \citetalias{2025MNRAS.537.3632M}.

The measurement of the 21\,cm signal in the intensity maps relies on the removal of foregrounds.
While the 21\,cm line is an emission-line signal, the foregrounds, such as Galactic synchrotron and extragalactic radio sources, are spectrally smooth and can therefore be separated using the technique of principal component analysis (PCA) with signal loss correction (e.g. \citealt{2023MNRAS.523.2453C}). 
As a result, the residual signal is susceptible to systematics that break the assumption that the foregrounds are spectrally smooth. 
The origins of such systematics can be polarisation leakage \citep{2014MNRAS.444.3183A,2020MNRAS.499..304C,2021MNRAS.504..208C}, chromaticity of the instrument beam \citep{2021MNRAS.506.5075M}, calibration errors (e.g. \citealt{2020MNRAS.494.5018H}), etc. 
These effects induce contamination of the signal in the measured summary statistics, which is at the same order of magnitude as the \hi\ signal for the MeerKLASS \textit{L}-band intensity maps (see Figures 14 and 20 of \citetalias{2025MNRAS.537.3632M}). 
Understanding and modelling the systematics are therefore important to enable the inference of the cosmological \hi\ signal, analogous to the marginalisation over nuisance parameters in optical galaxy surveys of LSS.

In this paper, we propose a novel way of modelling and constraining a type of multiplicative systematic induced by chromatic primary beam ripple, using the emission-line stacking of the 21\,cm intensity maps. 
Stacking the emission-line signal onto positions of optical galaxies is in itself a powerful way of probing cosmic \hi, and is one of the main scientific goals of interferometric MeerKAT observations. 
For example, the MeerKAT International GigaHertz Tiered Extragalactic Exploration survey (MIGHTEE; \citealt{2016mks..confE...6J,2021A&A...646A..35M}) has produced stacking measurements of \hi\ galaxies at $z\sim 0.37$ \citep{2022ApJ...935L..13S,2024MNRAS.529.4192S,2025arXiv250200110B} to probe scaling relations and dependencies of \hi\ mass on the LSS environment (see also \citealt{2021MNRAS.506.2753R,2021MNRAS.508.1195P,2022MNRAS.512.2697R,2022MNRAS.513.2168T,2023MNRAS.522.5308P,2023MNRAS.525..256P,2024MNRAS.534...76H,2024arXiv241114940T} for direct detections of \hi\ sources at $z\lesssim 0.01$). 
These stacking measurements utilise the information on source positions from external catalogues with maps in $\sim \,$arcsecond resolution made from interferometric observations (see also e.g. \citealt{2023ApJ...947...16A}).  Similarly, single dish intensity mapping observations can also be used for stacking, typically with maps in $\sim \,$arcminute resolution such as \hi\ stacking using the Parkes telescope \citep{2019MNRAS.489..385T,2020MNRAS.498.5916T} and CO emission-line stacking using the CO Mapping Array Project data \citep{2024ApJ...965....7D}.
{Detections of Ly$\alpha$ intensity mapping have also been achieved using the Hobby-Eberly Telescope Dark Energy Experiment observations~\citep{2022ApJ...929...90L, 2022ApJ...934L..26L}.}

The stacking measurement presented in \citetalias{2025MNRAS.537.3632M}, on the other hand, is made with intensity maps of $\sim 1\,$deg resolution. 
The resolution corresponds to large $\sim 30$ Mpc scales, over which we expect several galaxies contributing to the stacking measurements. Therefore, modelling clustering beyond Poisson statistics is required for a precise prediction (\citealt{2024PhRvD.109d3517B}; see also \citealt{2024MNRAS.535..826R} for the case of stacking Ly$\alpha$ emission).
Combined with the fact that the intensity map is affected by signal loss from foreground removal, forward modelling of the signal is required to infer the properties of the underlying \hi\ sources as well as the systematics in the data, which we aim to demonstrate in this work.

The modelling of the stacked signal can incorporate modelling of systematic effects in the data. 
The stacked signal can be averaged into a map of stacked emission in the angular plane. 
The excess emission in the centre region of the stacked map against the noise background can be used to describe the convolution of the instrument beam with the \hi\ signal. 
Similar analysis of systematics can be found in weak lensing, for example for cross-correlating the point spread function with galaxy shapes (e.g. \citealt{2023MNRAS.525.2441Z}). 
On the other hand, the stacked signal can also be averaged into a spectrum along the frequency direction. 
The stacked spectrum can then be used to examine the chromatic structure of the data that affects the two-point statistics, such as the structure seen in the line-of-sight power spectrum (e.g. \citealt{2022MNRAS.509.2048S}).
The complexity of the effects of the systematics requires detailed simulation and validation pipeline to study the viability of inference using emission-line stacking, which can be applied to the single dish \hi\ intensity mapping data using MeerKLASS and future SKAO.
This work lays the foundation for the pipeline and, for the first time, applies model inference to the emission-line stacking using MeerKLASS data.

The rest of the paper is organised as follows:
In \secref{sec:data}, we describe the specifications of the MeerKLASS \textit{L}-band deep-field data.
In \secref{sec:sim}, we present the simulation pipeline of stacked \hi\ signal. 
We validate the detectability of the stacked signal against the presence of thermal noise and foregrounds and establish the fact that the stacked signal needs to be forward-modelled in \secref{sec:validation}.
In \secref{sec:covest}, we discuss the method for covariance estimation.
We then describe the data analysis pipeline to measure the stacked signal in the data in \secref{sec:analysis}, which is an update of the one presented in \citetalias{2025MNRAS.537.3632M}. 
In \secref{sec:sys}, we present an analysis that pinpoints the nature of the systematics in the data. 
We then proceed to parameterise the systematics to enable the modelling of the stacked \hi\ signal, which we describe in \secref{sec:fit}. 
The modelling framework is used for parameter inference and the results are presented in \secref{sec:results}. 
We discuss the implications of our results for future \hi\ intensity mapping surveys in \secref{sec:discussion}, and conclude in \secref{sec:conclusion}. 
Throughout this paper, we assume a $\Lambda$ cold dark matter cosmology with the values of the model parameters reported in \citet{2020A&A...641A...6P}. 

\section{The MeerKLASS \textit{L}-band deep-field data}
\label{sec:data}
The MeerKLASS \textit{L}-band deep-field data is observed from 41 observation blocks with a total of 62\,hr of integration time per dish before flagging, spanning across 236$\,{\rm deg^2}$ in terms of sky area in ${\rm R.A.} \sim (330\degree, 360\degree)$ and ${\rm decl.}\sim (-36\degree, -25\degree)$. 
The details of the scanning strategy and observation time can be found in Table A1 of \citetalias{2025MNRAS.537.3632M}.

The \textit{L}-band data is observed from 900--1670\,MHz with a frequency channel width of $\delta \nu = 208.984375$\,kHz and a time resolution of $\delta t = 2$\,s. 
Noise diodes are fired to each receiver for 0.585\,s once every 19.5\,s for relative reference calibration at the level of time-ordered data (TOD). 
Before and after each scan of around 100 minutes, the telescope is pointed to track a nearby celestial point source, either PKS 1934-638 or Pictor A, as a bandpass and absolute flux calibrator.

\begin{figure}
    \centering
    \includegraphics[width=1.0\linewidth]{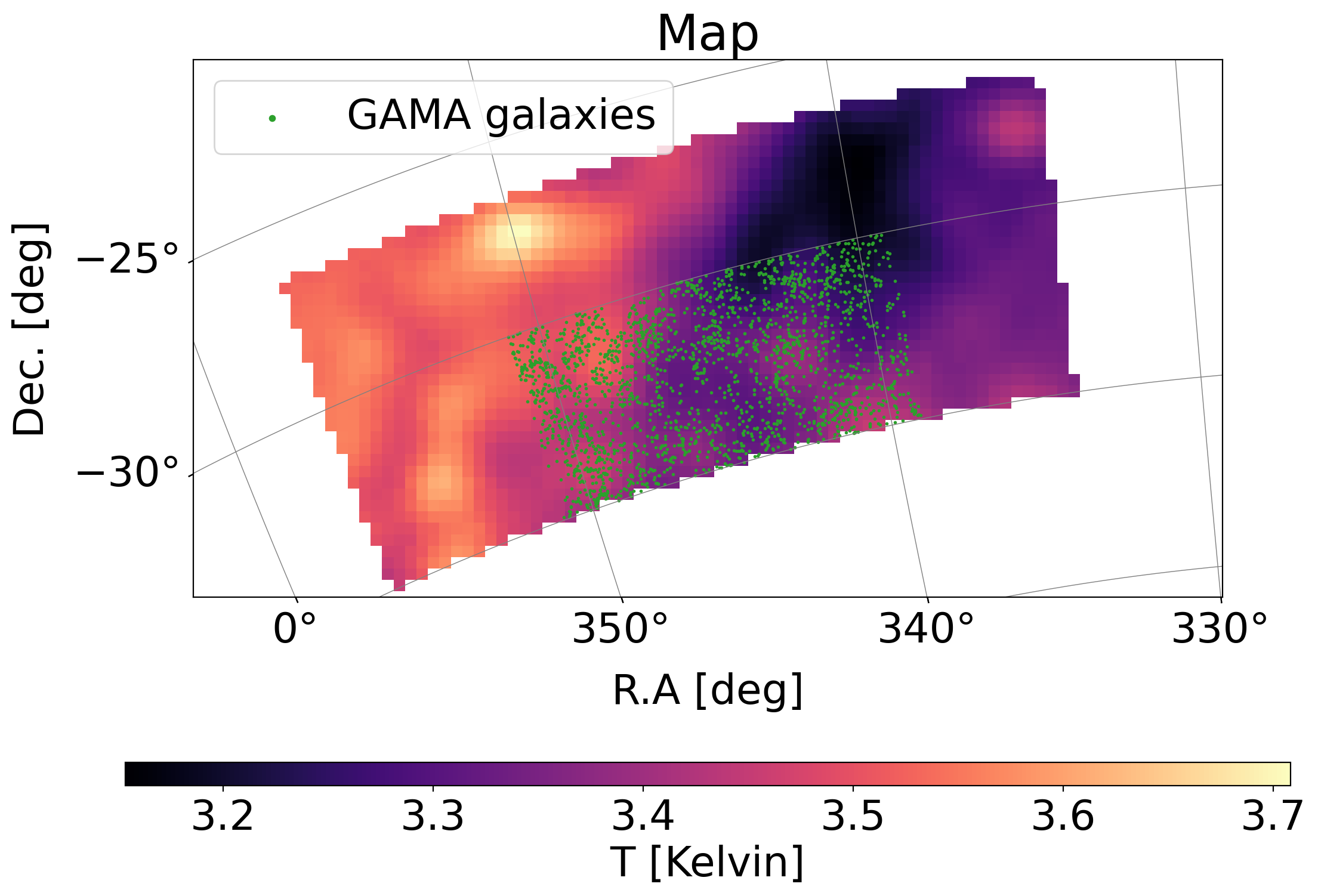}
    \includegraphics[width=1.0\linewidth]{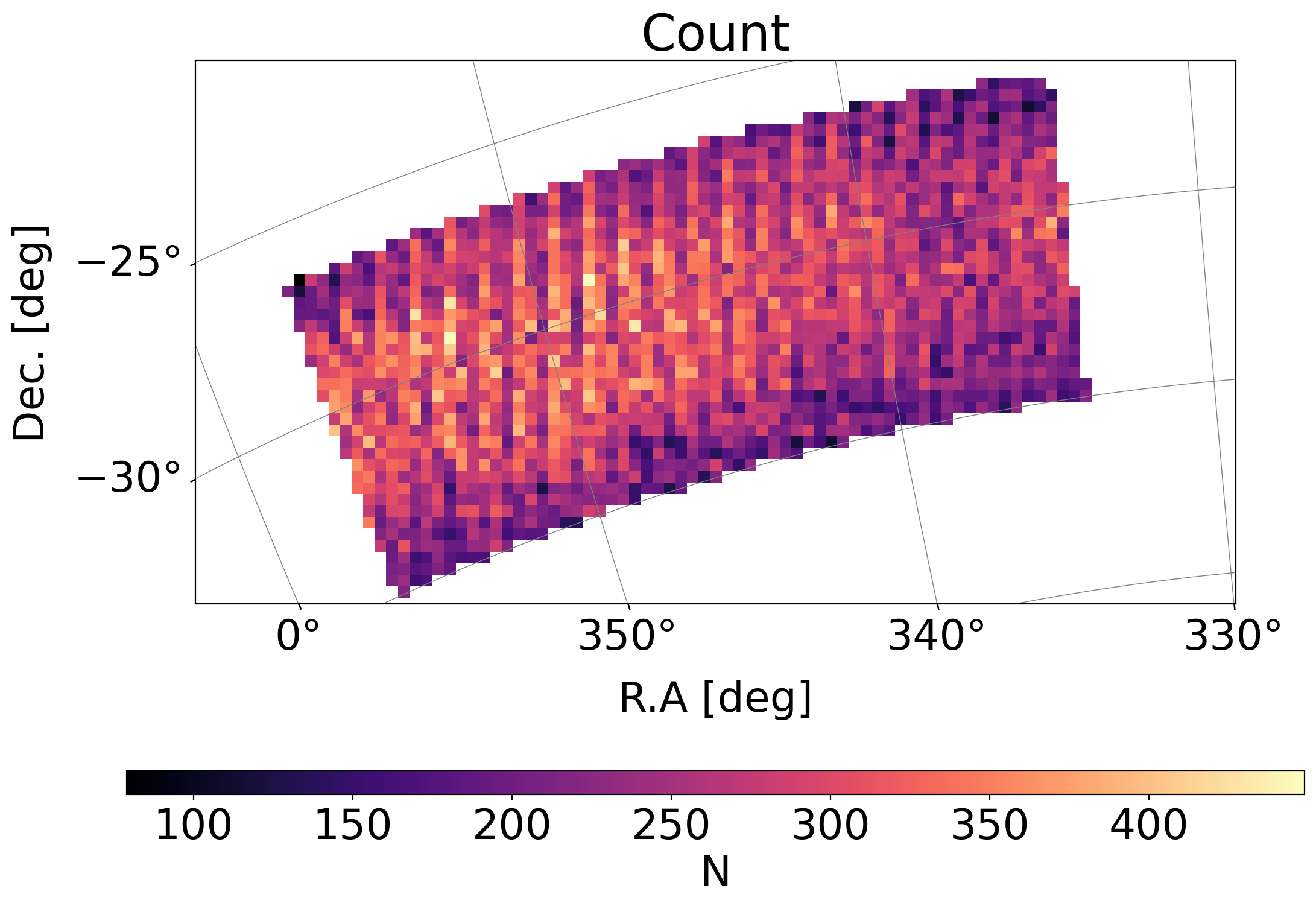}
    \caption{Top panel: the frequency-averaged intensity map from the MeerKLASS \textit{L}-band deep-field data reported in \citetalias{2025MNRAS.537.3632M}. The maps are trimmed so that only pixels within $334\degree<{\rm R.A.}<357\degree$ and $-35\degree<{\rm decl.}<-26.5\degree$ are kept. The green dots denote the positions of the overlapping GAMA galaxies within the redshift bin of the intensity mapping data. Bottom panel: the time-stamp counts of each pixel in the intensity map. }
    \label{fig:datamap}
\end{figure}

The data is then calibrated using the calibration pipeline \textsc{katcali}\footnote{\url{https://github.com/meerklass/katcali}}.
The details of the calibration process are described in \citet{2021MNRAS.505.3698W}, and we briefly summarise it below.
An initial radio frequency interference (RFI) flagging is applied to the raw data. 
{Removing channels with strong RFI contamination leaves the frequency ranges 971-1075\,MHz and 1305-1504\,MHz to be considered} (see Figure 5 of \citealt{2021MNRAS.505.3698W}).
The data is then modelled as a combination of several time- and frequency-dependent components,
\begin{equation}
    T_{\rm obs}(t,\nu) = g(t,\nu) [T_{\rm ps} + T_{\rm diffuse} + T_{\rm el} + T_{\rm diode} + T_{\rm rec}](t,\nu),
\label{eq:tobs}
\end{equation}
where $t$ is observation time, $\nu$ is the observing frequency, $g(t,\nu)$ is the gain, $T_{\rm ps}$ is the brightness temperature of the calibrator, $T_{\rm diffuse}$ is the diffuse foreground emission, $T_{\rm el}$ is the elevation dependent terrestrial emission from the Earth's atmosphere and ground spillover, $T_{\rm diode} $ is the noise diode signal, and the residual temperature from the receiver as well as modelling errors is absorbed into $T_{\rm rec}$. 
The modelling of each component of brightness temperature is described in Section 3 of \citet{2021MNRAS.505.3698W}.
The residual temperature $T_{\rm rec}$ is assumed to be slowly varying in time, and its time dependence can be described by a Legendre expansion up to the third-order Legendre polynomial.

The gain is then solved independently at each frequency for each feed of each dish using the reference model. 
The solution is assumed to be smooth in time, described by a Legendre expansion up to the fourth order. 
A Bayesian fitting framework is applied to solve for the gain and model temperature parameters. 
A more detailed description of the fitting and examination on the calibration quality can be found in \citet{2021MNRAS.505.3698W} and Section 2.4 of \citetalias{2025MNRAS.537.3632M}.

After applying the inverse of the gain solution, the dataset is then flagged again to remove outliers along the frequency direction. 
$T_{\rm el}$ and $T_{\rm rec}$ are then subtracted from the TOD, which is subsequently converted from polarization to Stokes \textit{I} intensity and gridded onto a sky map with azimuthal equal area projection. 
The angular {size} of the sky map pixels is chosen to be $0.3\,$deg.
Another round of RFI flagging is then applied to the scans by comparing the TOD with the median value along the R.A. direction in the sky map. 
The TOD after flagging are then averaged again into the final sky map.

\begin{figure}
    \centering
    \includegraphics[width=1.0\linewidth]{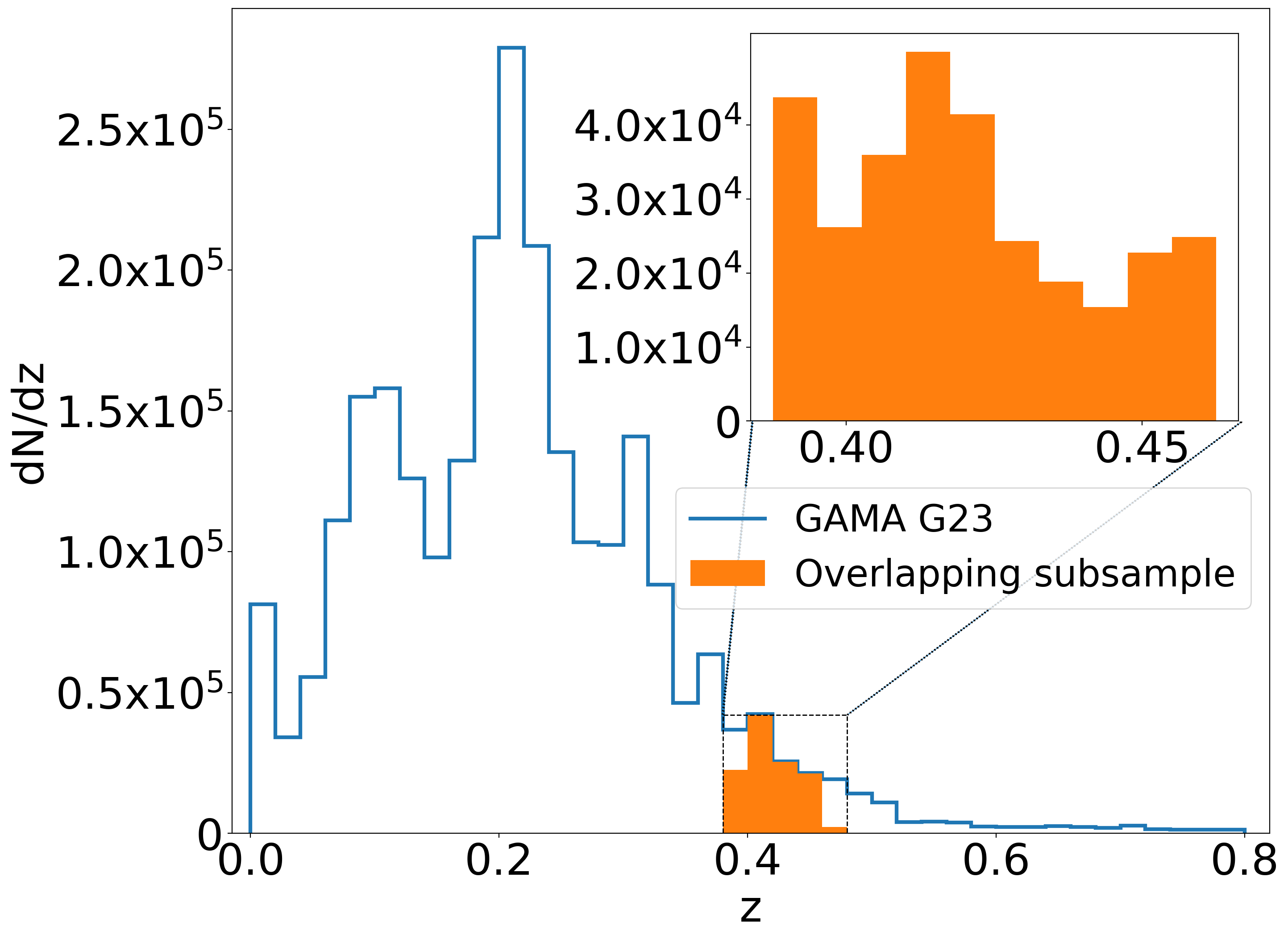}
    \caption{The redshift distribution of the GAMA G23 catalogue. The blue solid line denotes the distribution for the entire catalogue, whereas the orange filled region denotes the distribution of the subsample that overlaps with the intensity mapping data. The top-right corner shows the zoom-in of the overlapping redshift bin. Note that the values are different in the zoomed-in plot due to different choices of binning width in redshift.}
    \label{fig:dndz}
\end{figure}

In \autoref{eq:tobs}, the model components are mostly fixed from external measurements, for example, using the Python Sky Model \citep{2017MNRAS.469.2821T} for synchrotron emission in the modelling of the diffuse foregrounds. 
Any mismatch is then absorbed into $T_{\rm rec}$ and may induce nontrivial calibration errors. 
To further improve the gain solutions, the sky map obtained from the reference calibration is then passed back to \autoref{eq:tobs} to replace the original sky model of $T_{\rm diffuse}$.
The calibration process is then repeated to obtain an updated gain solution, which is then used to produce a new iteration of the final sky map.
This self-calibration step is iterated 5 times for convergence to produce the intensity maps used for subsequent data analysis.
Finally, by examining of flagging percentage of the data in each channel, the relatively clean sub-band of $971.15\,{\rm MHz}<\nu<1023.61\,{\rm MHz}$ is chosen for the data analysis, corresponding to $0.39\lesssim z \lesssim 0.46$ with an effective redshift $z=0.424$ {at the centre of the frequency sub-band}.
The frequency-averaged sky map after calibration, as reported in \citetalias{2025MNRAS.537.3632M} is shown in \autoref{fig:datamap} for illustration.
For reference, the hit counts, i.e. number of time stamps averaged in each pixel, are also shown.

To perform the stacking analysis, we use an overlapping spectroscopic galaxy catalogue from the Galaxy and Mass Assembly (GAMA) survey \citep{2009A&G....50e..12D,2011MNRAS.413..971D,2015MNRAS.452.2087L}. 
The specific region we use in this paper is the 23\,hr (G23) field \citep{2022MNRAS.513..439D} covering the area of $339\degree<{\rm R.A.}<351\degree$ and $-35\degree<{\rm decl.}<-30\degree$ as shown in the upper panel of \autoref{fig:datamap}.
{The galaxy sample has a magnitude limit of $i<19.2$ (for target selection, see \citealt{2010MNRAS.404...86B,2015MNRAS.452.2087L}).}

We note that, the GAMA Data Release 2 \citep{2015MNRAS.452.2087L} only provides complete survey information and value-added catalogues for part of the total survey region, namely the G09, G12, and G15 regions. 
For simplicity, we assume that the G23 region has uniform survey geometry and constant redshift distribution in the redshift range of our interest, following \citetalias{2025MNRAS.537.3632M}.
For reference, we show the redshift distribution of the GAMA G23 catalogue in \autoref{fig:dndz}. 
As shown, the redshift bin $0.39 \lesssim z \lesssim 0.46$ is at the tail of the redshift distribution.
The galaxy catalogue is therefore likely to be \hi\ incomplete, with only the most-massive star-forming galaxies being selected.
In total, $N_{\rm g}^{\rm GAMA}=2269$ galaxies are selected, resulting in a comoving number density of $\sim2.0\times 10^{-4}\,{\rm Mpc^{-3}}$.
As the \hi\ mass correlates with the stellar mass (e.g. \citealt{2021ApJ...918...53G}), it is expected that only some of the massive \hi\ galaxies are present in the catalogue, with a larger fraction of \hi\ sources missing.
The interpretation of stacking onto an incomplete catalogue is discussed later in \secref{sec:validation}.

\section{Simulation of stacked \hi\ signal}
\label{sec:sim}
In this section, we describe the simulation pipeline used in this paper. The simulation pipeline validates the viability of detecting a stacked signal using the MeerKLASS \textit{L}-band data which will be presented in \secref{sec:validation}, and also instructs us on the optimal way of performing the stacking analysis later in \secref{sec:analysis}.
\subsection{\althi\ emission}
\label{subsec:hisim}
For an \hi\ galaxy with total \hi\ mass $M_{\rm \hi}$, the total 21\,cm flux of the source is \citep{2017PASA...34...52M}
\begin{equation}
    S_{\rm \hi} = \frac{3 h_{\rm P} \nu_{21}A_{21}}{16\pi D_L^2}M_{\rm \hi},
\label{eq:hiflux}
\end{equation}
where $h_{\rm P}$ is the Planck constant, $\nu_{21}$ is the rest frequency of the 21\,cm line, $A_{21}$ is the spontaneous emission rate of the \hi\ atoms, and $D_L$ is the luminosity distance of the source. Throughout this section, we assume an effective redshift $z=0.424$ when calculating background quantities such as $D_L$ and the matter power spectrum discussed later.

The flux is distributed across the frequencies into an emission-line profile, due to the inner velocity dispersion of the source. The emission-line profile of the \hi\ flux density can be described by a busy function \citep{2014MNRAS.438.1176W}
\begin{equation}
    F_\nu(\Delta \nu) = \frac{a_\nu}{2} \times \Big( {\rm erf} \big[ b_\nu(w_\nu^2 - \Delta \nu^2) \big]+1  \Big)
    \times \Big( c_\nu\Delta\nu^2 + 1 \Big),
\label{eq:busyfunc}
\end{equation}
where ``erf'' denotes the Gauss error function, $a_\nu,b_\nu,c_\nu,w_\nu$ are parameters of the busy function, $\Delta \nu = \nu - \nu_0$ is the difference between the observed frequency $\nu$ and the central frequency $\nu_0$ of the \hi\ emission of the source, and the profile $F_\nu$ is the observed flux density of the source so that $\int {\rm d}\Delta\nu F_\nu = S_{\rm \hi }$.
The observed flux density is discretised at each frequency channel.
For a given busy function profile of source $i$, the flux density at a specific observing frequency $\nu$ is
\begin{equation}
    I_\nu^i(\nu-\nu_i) = \frac{1}{\delta\nu} \int_{\nu-\delta\nu/2}^{\nu+\delta\nu/2} {\rm d}\nu' F_\nu(\nu'-\nu_i),  
\end{equation}
where $\nu_i$ is the centre frequency of the \hi\ profile for source $i$.

The busy function encodes the velocity width of the \hi\ galaxy. 
In particular, $w_\nu$ controls the positions of the double peaks of the busy function and effectively describes the overall width of the profile. 
If the emission-line profile has a width of $w_\nu$ in the observed spectrum, the corresponding velocity width in the source rest frame is
\begin{equation}
    w_{V} = \frac{c}{\nu_{\rm obs}}w_\nu,
\label{eq:vandnu}
\end{equation}
where $c$ is the speed of light, and $\nu_{\rm obs}$ is the observing frequency.

We note that \autoref{eq:busyfunc} is a simplified version of the busy function, with the emission-line profile being symmetric along $\Delta\nu = 0$. 
In our case of stacking, the emission-line profiles are averaged across many sources, which will symmetrize the underlying \hi\ signal. 
Furthermore, as we discuss later in \secref{sec:validation}, the effect of clustering greatly stretches the stacked spectrum and wipes out information on the velocity width. 
As a result, the effect of assuming a simplified busy function is negligible.
For the same reason, we also do not consider the peculiar velocity of the observer and the impact of no Doppler correction on the map data.

Throughout this paper, \hi\ galaxies are treated as point sources in the angular plane, as the intensity maps are of $\sim 1\,$deg resolution. The observed 21\,cm intensity can then be described as
\begin{equation}
    I(l,m,\nu) = B(l,m,\nu)\, \circledast\, 
    \sum_i \bigg[
    \delta_{\rm D}(\bm{l}-\bm{l}_i)
    I_\nu^i(\nu-\nu_i)
    \bigg]
    ,
\label{eq:hifd}
\end{equation}
where $\bm{l} = (l,m)$ is the position on the sky, $B(l,m,\nu)$ is the beam of the instrument, $\circledast$ denotes convolution along the angular plane, $i$ loops over each source, and $\delta_{\rm D}$ is the Dirac delta function.
\autoref{eq:hifd} indicates that the simulation of \hi\ signal requires the position, the \hi\ mass, and the velocity width of the sources. 
We describe the routine for simulating each of these ingredients as follows.

We first determine the number density of \hi\ galaxies.
Given a population of \hi\ galaxies, the distribution of the \hi\ mass, the \hi\ mass function (HIMF), can be described by a Schechter function \citep{1976ApJ...203..297S}
\begin{equation}
\begin{split}
    \phi(M_{\rm \hi}) & = \frac{{\rm d}n}{{\rm dlog}_{10}(M_{\rm \hi})} \\ 
    & = {\rm ln}(10) \,\phi_*\,\bigg(\frac{M_{\rm \hi}}{M_*}\bigg)^{\alpha+1}\,
    {\rm exp}\bigg[-\frac{M_{\rm \hi}}{M_*} \bigg],
\end{split}
\label{eq:himf}
\end{equation}
where $n$ is the number density of \hi\ galaxies and $(\phi_*,M_*,\alpha)$ are the HIMF parameters.

In this paper, we adopt the values reported in \citet{2018MNRAS.477....2J} at $z\sim 0$.
It is expected that the values of the HIMF parameters evolve slowly over redshift \citep{2021MNRAS.501.4550X,6b89139973fc4d14ab13a986f2f93aac,2023MNRAS.522.5308P}, although the evolution may become significant for our data at $z\sim 0.4$ \citep{2022ApJ...940L..10B}.
Since the HIMF is not well understood beyond the local Universe, we resort to using the values at $z\sim 0$.
Note that future measurements of \hi\ stacking with Bayesian inference may help probe the HIMF over large samples of galaxies \citep{2020MNRAS.491.1227P,2025arXiv250111872W}.
For a given HIMF, we can calculate the number density $\bar{n}_{\rm \hi}$, and average density $\rho_{\rm \hi}$ of \hi\ sources that have \hi\ mass larger than $M_{\rm \hi}^{\rm min}$ as 
\begin{equation}
    \bar{n}_{\rm \hi} = \int_{M_{\rm \hi}^{\rm min}} {\rm dlog_{10}}(M_{\rm \hi}) \,\phi(M_{\rm \hi}),
\label{eq:nhi}
\end{equation}
\begin{equation}
    \bar{\rho}_{\rm \hi} = \int_{M_{\rm \hi}^{\rm min}} {\rm dlog_{10}}(M_{\rm \hi})\, M_{\rm \hi}\, \phi(M_{\rm \hi}).
\end{equation}
We find that for the HIMF parameter values in \citet{2018MNRAS.477....2J}, the \hi\ density $\bar{\rho}_{\rm \hi}$ is $97\%$ complete for $M_{\rm \hi}^{\rm min} = 10^8 M_\odot$, which we choose as the lower limit.
This gives a number density of $\bar{n}_{\rm \hi} = 0.031\,{\rm Mpc^{-3}}$ (note that this is much larger than the GAMA galaxy number density discussed in \secref{sec:data}).

Second, we generate the clustering of galaxy positions by simulating a lognormal realization of the galaxy overdensity field, $\delta_{\rm g}$, following a model power spectrum,
\begin{equation}
\begin{split}
    P_{\rm g} (\bm{k}) = &  V|\tilde{\delta}_{\rm g}(\bm{k})|^2 \\
    = & b_{\rm g}^2\,\Big( 1 + f \mu^2/b_{\rm g}\Big)^2 \,P_m(\bm{k})
\end{split}
\end{equation}
where $V$ is the survey volume, $\tilde{\delta}_{\rm g}(\bm{k})$ is the galaxy overdensity in Fourier space, $b_{\rm g}$ is the galaxy bias, $f$ is the growth rate, $\mu = k_\parallel/|\bm{k}|$, and $P_m$ is the matter power spectrum in real space.
We choose $b_{\rm g} = 1.9$ matching the auto-power spectrum of the GAMA galaxies as discussed in \citetalias{2025MNRAS.537.3632M}.
Note that redshift space distortions are applied to the galaxy power spectrum using the Kaiser effect \citep{1987MNRAS.227....1K} without the Finger-of-God effect, since the velocity dispersion is included in the simulation of the \hi\ profile.

The nonlinear matter power spectrum is calculated using \textsc{camb} \citep{2011ascl.soft02026L} with \textsc{halofit} \citep{2003MNRAS.341.1311S,2012ApJ...761..152T}.
A lognormal realization of the galaxy overdensity field is then generated using \textsc{powerbox} \citep{2018JOSS....3..850M} based on the formalism described in \citet{2011MNRAS.416.3017B}.
The lognormal simulation is motivated by its similarities to a Gaussian distribution as well as desirable physical properties, such as ensuring that the overdensity is always larger than -1 \citep{1991MNRAS.248....1C}.
The covariance of lognormal simulations is found to be more accurate than Gaussian realizations (e.g. \citealt{2011A&A...536A..85H}), and is sufficient for our signal-to-noise ratio in the data.
The overdensity field is then converted to the number density field,
\begin{equation}
    n_{\rm g}(\bm{x}) = \bar{n}_{\rm \hi}\,\big(\delta_{\rm g}(\bm{x})+1\big)\,W(\bm{x}),
\end{equation}
where the survey selection function $W(\bm{x})$ is 1 inside the 21\,cm survey volume and 0 otherwise in our case.
The galaxy positions are then Poisson sampled and projected onto the sky. 
Note that the comoving box used for generating galaxy positions is larger than the survey volume, as the beam smoothing requires information outside the survey area.

\begin{figure}
    \centering
    \includegraphics[width=1.0\linewidth]{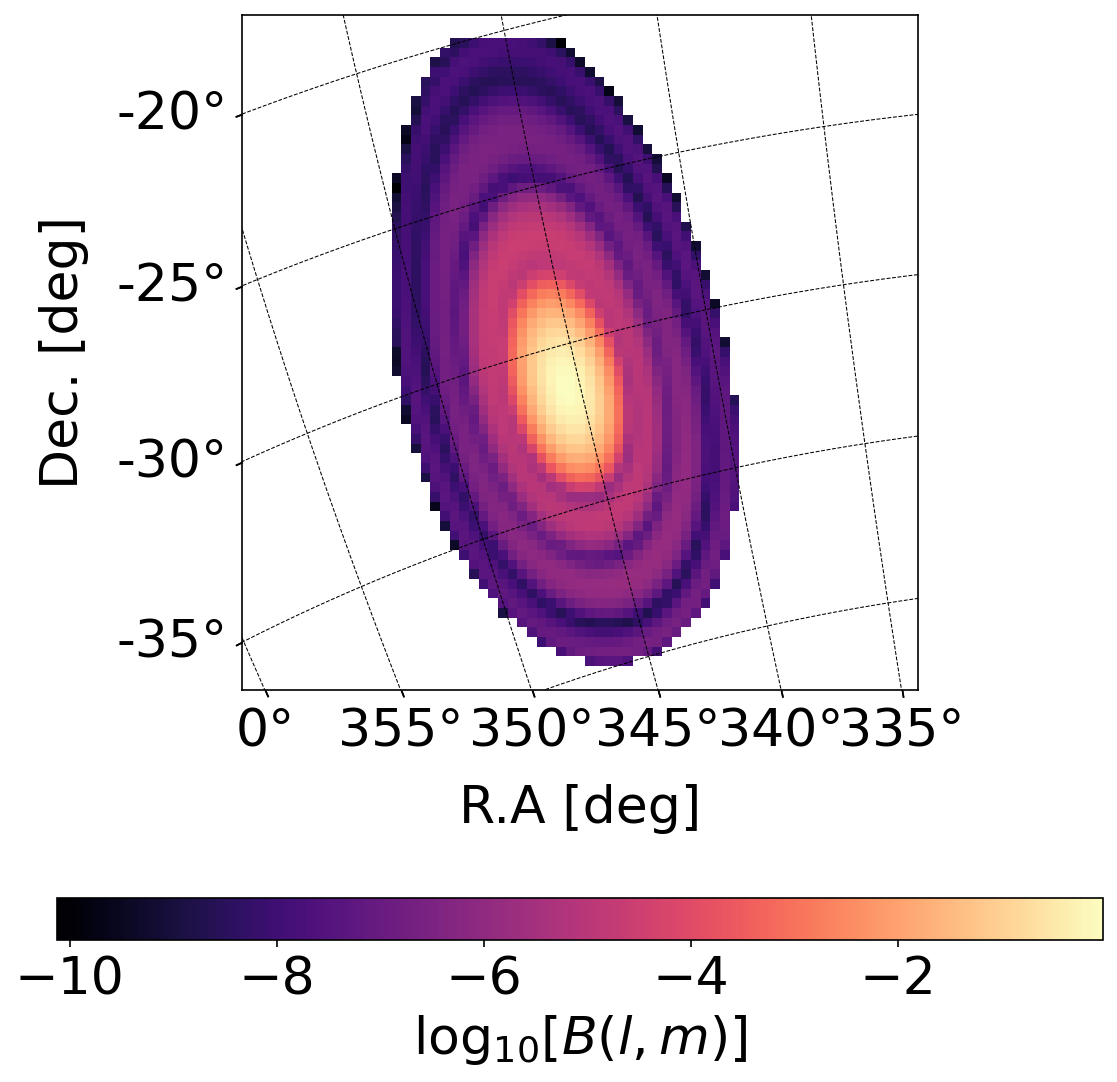}
    \caption{The \textsc{eidos} primary beam model \citep{2021MNRAS.502.2970A} for the MeerKAT telescope. For the visualization, the beam is pointed toward the centre of the survey area $\rm (RA,Dec) = (345.4\degree, -31.4\degree)$ at the centre of the frequency sub-band $\nu=997.38\,$MHz. Note that the apparent ellipticity and orientation of the beam are due to projecting the sky coordinates onto the Cartesian grids. The angular grid lines are shown as black dotted lines to further illustrate the projection.}
    \label{fig:beam}
\end{figure}

Third, we sample the input HIMF to assign \hi\ mass to each galaxy. 
{Note that the random assignment of} \hi\ {mass is uncorrelated with the galaxy positions. Therefore, the bias of the} \hi\ {map, as well as the bias of the stacking subsample described later, is the same as the input $b_{\rm g}=1.9$.}
The width of the emission profile is then calculated by first converting the \hi\ mass to a velocity dispersion through the Tully-Fisher (T-F) relation \citep{1977A&A....54..661T}. 
We choose the slope and the intercept of the T-F relation to be the measured values in \citet{2021MNRAS.508.1195P}. 
The intrinsic velocity dispersion is that projected to the line-of-sight direction by assuming a random inclination angle, $\theta_{\rm incl}$, so that
\begin{equation}
    w_v = \sin(\theta_{\rm incl}) v_{\rm T-F} \, ,
\end{equation}
where $v_{\rm T-F}$ is the velocity calculated from the T-F relation, and we randomly sample {the inclination from a uniform distribution} $\theta_{\rm incl}\in [0,2\pi]$ for each mock galaxy. 
The velocity dispersion $w_v$ is then converted to $w_\nu$ in frequency according to \autoref{eq:vandnu}. 
For the other two busy function parameters $b_\nu$ and $c_\nu$, we assume wide flat priors, so that $b_\nu\in [0.01,1]\,{\rm km^{-2} s^2}$ and $c_\nu\in [0.001,0.01]\,{\rm km^{-2} s^2}$, and sample random values of $b$ and $c$ for each source. 
Note that the amplitude parameter $a_\nu$ is not a free parameter and instead set by the shape of the profile and the total \hi\ mass of the galaxy.
The emission-line profile can then be calculated according to \hyperref[eq:hiflux]{Equations \ref{eq:hiflux}} and \ref{eq:busyfunc}.
{As we show later in \secref{sec:validation}, the stacked signal is dominated by the effects of double counting and clustering, and the shape of the emission-line profile has negligible effects in our modelling.}

\begin{figure}
    \centering
    \includegraphics[width=1.0\linewidth]{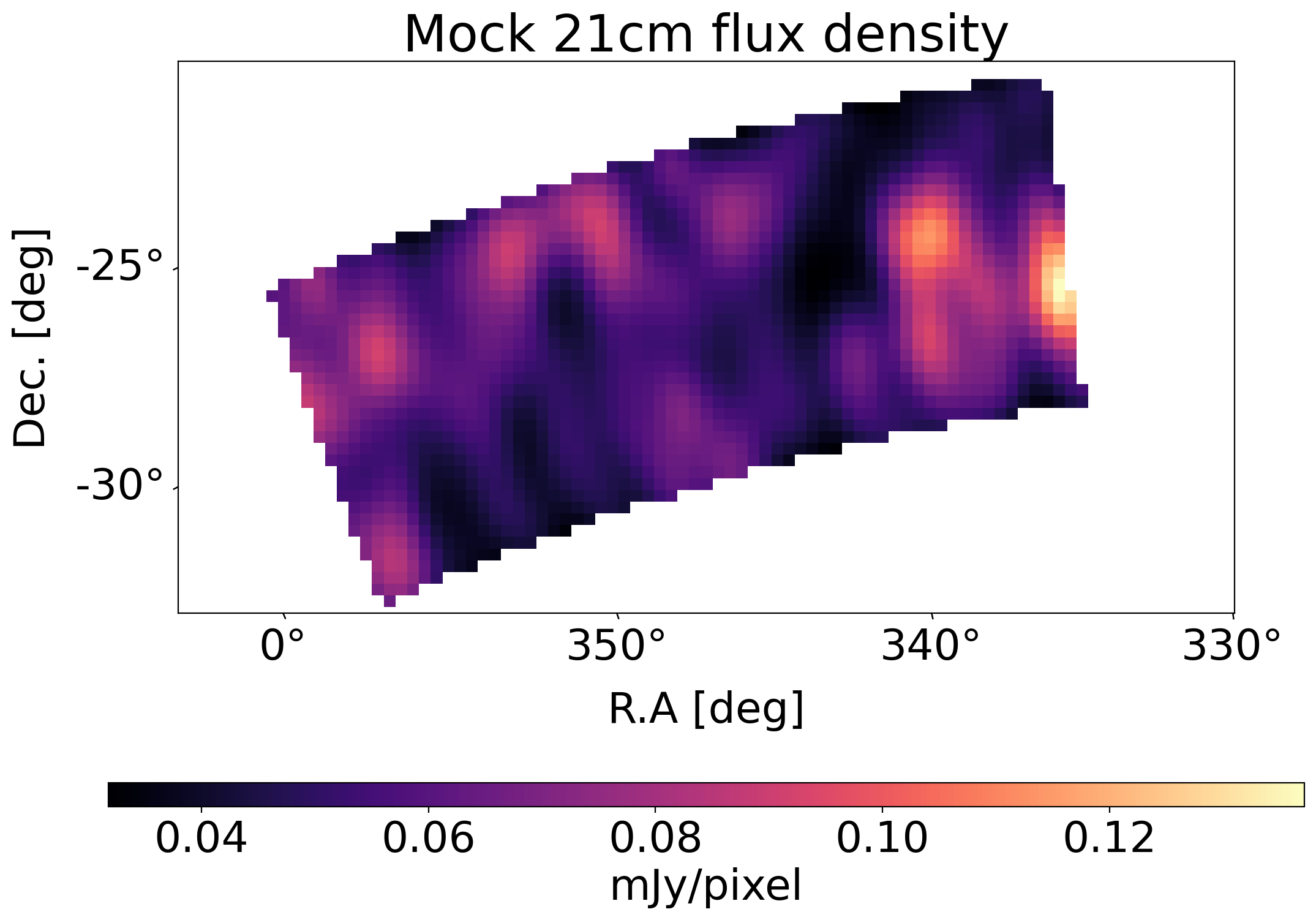}
    \includegraphics[width=1.0\linewidth]{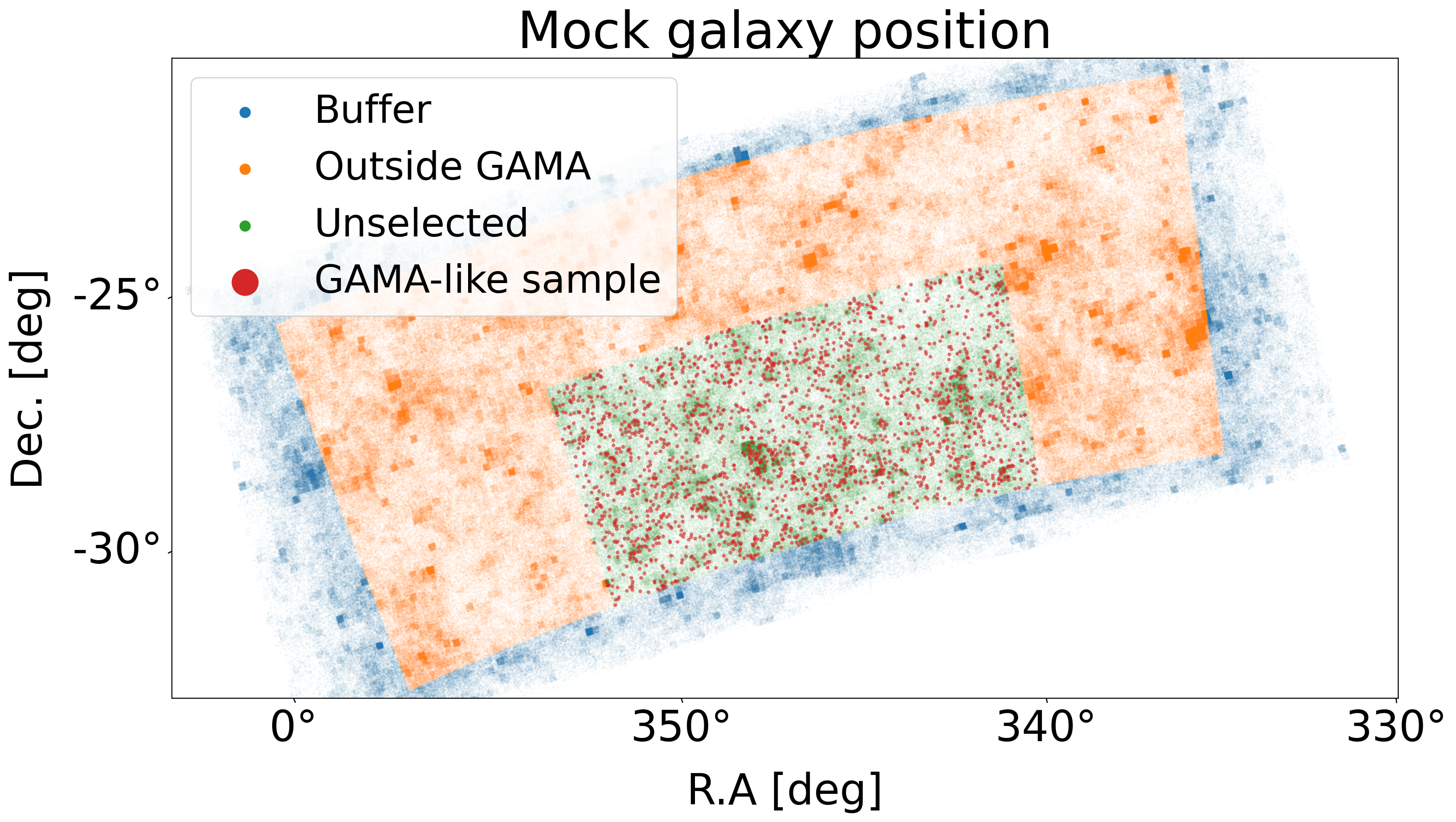}
    \caption{Upper panel: the frequency-averaged flux density map of mock \hi\ signal for one realization. The flux density is shown in the unit of Jy\,pixel$^{-1}$, where the pixel area is $0.3\times 0.3\,\deg^2$. Lower panel: the mock galaxy positions in one realization. Blue dots show galaxies that are generated in the comoving box that are outside the MeerKLASS survey area (``Buffer''). They are included in the \hi\ signal simulation as required by the beam smoothing. Orange dots show galaxies that are inside the MeerKLASS survey area but outside the GAMA region (``Outside GAMA''). Green dots show the galaxies inside the GAMA region that have relatively small \hi\ mass and are not included in the subsample for stacking (``Unselected''). The red dots show the galaxies selected for stacking (``GAMA-like sample''). Note the high number density of galaxies results in pixelated clumps seen in the panel. They correspond to the resolution over which the galaxy overdensity is generated, which is much smaller than the resolution of the 21\,cm maps. }
    \label{fig:mockillus}
\end{figure}

Finally, based on the emission-line profiles and the galaxy positions, an intensity map can be generated as described in \autoref{eq:hifd}. We use \textsc{eidos}, which is based on the astro-holographic measurement of the MeerKAT beam \citep{2021MNRAS.502.2970A}, to generate the input beam model $B(l,m,\nu)$ for the convolution that we show in \autoref{fig:beam}. 
An illustration of one realization of the simulated \hi\ intensity maps is shown in \autoref{fig:mockillus}.

After generating the \hi\ galaxies, we still need to select the subsample in the mock that resembles the GAMA galaxy catalogue for the stacking. 
As mentioned in \secref{sec:data}, the galaxy data catalogue is incomplete, with little information on derived quantities to be used for simulating the survey selection.
We assume that the GAMA galaxy catalogue to contain the most-massive \hi\ galaxies, and we select the $N_{\rm g}^{\rm GAMA} = 2269$ most-massive mock galaxies inside the GAMA survey area to be the mock subsample.
The process is illustrated in \autoref{fig:mockillus}.
Note that we can also use the number density of the data catalogue to recalculate a lower limit $M_{\rm \hi}^{\rm min, GAMA}$ by inverting \autoref{eq:nhi}, and randomly select among the mock \hi\ galaxies with \hi\ mass larger than the limit.
However, given the extremely small number density of the data, we find the lower limit $M_{\rm \hi}^{\rm min, GAMA} = 10^{10.235}\,M_\odot$, at the very tail of the HIMF distribution.
If we choose among mock galaxies with masses larger than $M_{\rm \hi}^{\rm min, GAMA}$, the number of selected galaxies will fluctuate a lot around the expected $N_{\rm g}^{\rm GAMA}$ due to sampling at the tail of the HIMF distribution.
To compensate for the low number of galaxies in the mock stacking subsample, we then still have to select the most-massive mock galaxies below $M_{\rm \hi}^{\rm min, GAMA}$, effectively performing the same selection as simply choosing the $N_{\rm g}^{\rm GAMA}$ most-massive mock galaxies.

As we discuss later, in the model fitting, we use the simplified assumption that the \hi\ emission comes entirely from the galaxy catalogue, and all galaxies in the catalogue have the same \hi\ mass.
This is because in the data analysis, we have no knowledge of the underlying distribution of \hi\ galaxies outside the GAMA galaxy sample.
To mimic this lack of information in the mock, for each realization, we generate an alternative \hi\ simulation following the simplified assumption.
In each realization of the mock, we use the original simulation of mock \hi\ galaxies, and calculate the total \hi\ mass within the GAMA survey region.
We then exclude all other galaxies from the simulations except the GAMA-like mock subsample (see \autoref{fig:mockillus}).
For the GAMA-like subsample, we then assign the same \hi\ mass to each galaxy and keep the total \hi\ mass within the GAMA region the same, so that
\begin{equation}
    M_{\rm \hi}^{\rm GAMA, alt} = \sum_i M_{\rm \hi}^{i} /N_{\rm g}^{\rm GAMA},
\end{equation}
where $i$ iterates over all mock galaxies in the original simulation that is inside the GAMA survey area.
We then use the updated mock galaxy catalogue and the \hi\ mass to re-assign random velocity profiles.
An alternative mock \hi\ map is then generated.

Comparing the original simulation with the full galaxy sample following an HIMF distribution and the simulation based on the simplified assumption\footnote{From now on, the ``simplified assumption'' always refers to the scenario where we only consider 21\,cm signals from the galaxy catalogue with the same \hi\ mass for all galaxies in the forward modelling, ignoring the incompleteness and the mass distribution.}, we can quantify the biasing of the stacked signal from the simplified assumption in \secref{sec:validation}, which we then use to forward model the signal for inference in \secref{sec:fit}.

The mock simulations are generated with multiple realizations. Foregrounds and thermal noise are then added to the mock \hi\ signal, which we discuss for the rest of this section.

\subsection{Foregrounds}
\label{subsec:fg}
In 21\,cm experiments, foregrounds originate from Galactic and extragalactic radio sources that have continuous spectra in frequency. 
For the \textit{L}-band observation with $\sim 1$\,deg angular resolution, it is expected that Galactic synchrotron radiation will dominate the foreground emission. 
For simplicity, we only consider synchrotron in the simulation. 
For the actual data, foreground removal is performed and validated by the null tests described in \citetalias{2025MNRAS.537.3632M}, and further validation tests are presented later in \secref{subsec:dataangular} and \secref{subsec:dataspectral}.
The foreground signal is therefore not included in the forward modelling for parameter fitting, and instead, the effects of PCA from the actual data are propagated into the mock \hi\ as we describe later in \secref{sec:fit}. 
Here, foreground simulation is only used in the mock to examine the effect of PCA foreground removal on the stacked spectrum qualitatively.

\begin{figure}
    \centering
    \includegraphics[width=1.0\linewidth]{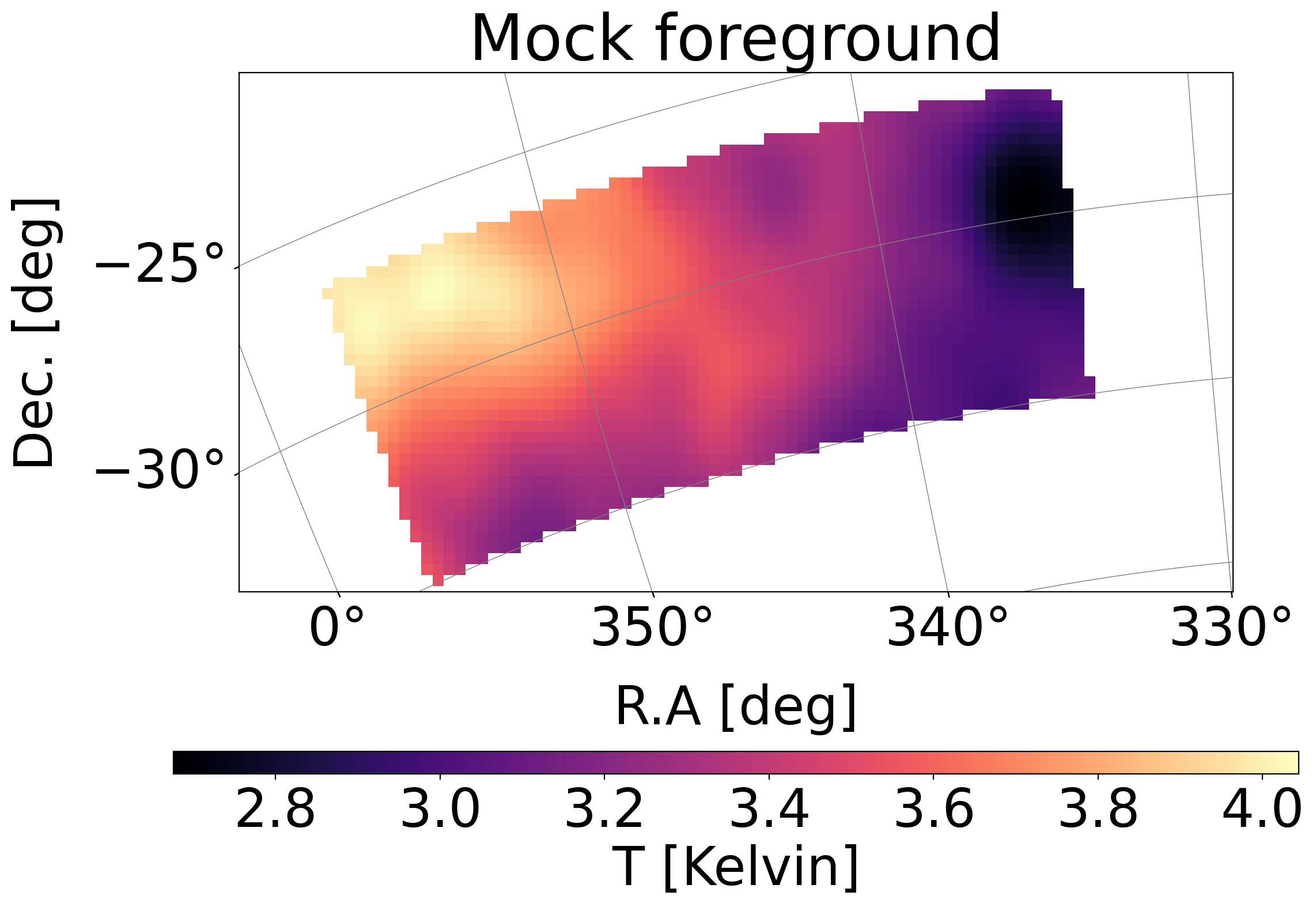}
    \caption{The frequency-averaged brightness temperature map for the mock foreground signal. }
    \label{fig:fgsim}
\end{figure}

We use the improved version of the 408\,MHz measurement of the synchrotron map \citep{1982A&AS...47....1H,2015MNRAS.451.4311R} as the template.
The synchrotron map is then extrapolated to the observing frequencies by assigning spectral indices to each pixel.
The spectral indices are calculated based on the sky model at 1.4\,GHz and 2.3\,GHz provided in the Global Sky Model \citep{2017MNRAS.464.3486Z}, based on the observations of \citet{2001A&A...376..861R} and \citet{1998MNRAS.297..977J}.
The resulting synchrotron maps are then smoothed by the primary beam using the \textsc{eidos} model.
An illustration of the mock foreground map is shown in \autoref{fig:fgsim}.

Note that, the assumption of a fixed spectral index for each pixel on the sky ignores the secondary curvature of the synchrotron spectrum (e.g. \citealt{2022MNRAS.509.4923I}). 
Consistency between low-frequency and \textit{L}-band measurements of Galactic foregrounds is also contested \citep{2024arXiv240906770W}.
In fact, this can be seen by visually comparing the maps shown in \autoref{fig:datamap} and \autoref{fig:fgsim}.
As mentioned, a more realistic foreground simulation is outside the scope of this work.

\subsection{Thermal noise}
\label{subsec:tn}
The radiometer equation states that for Stokes \textit{I} intensity at map level, the standard deviation of the thermal noise fluctuation is
\begin{equation}
    \sigma_{\rm N} = \frac{T_{\rm sys}(\nu)}{\sqrt{2\delta\nu \delta t N_{\rm hits}}},
\label{eq:sigmaN}
\end{equation}
where 
$\delta \nu$ is the frequency resolution, $\delta t$ is the time resolution, 
$N_{\rm hits}$ is the number of time stamps in the pixel, {and, following} \citetalias{2025MNRAS.537.3632M}, we model the temperature of the system as
\begin{equation}
    T_{\rm sys} = T_{\rm rx} + T_{\rm el} + T_{\rm CMB} + T_{\rm gal},
\end{equation}
where $T_{\rm rx}$ is the receiver temperature, $T_{\rm el}$ is the spillover introduced in \autoref{eq:tobs}, $T_{\rm CMB}$ is the CMB temperature, and $T_{\rm gal}$ is the temperature of Galactic synchrotron.
For the narrow frequency sub-band of our data, the system temperature is found to be near constant $T_{\rm sys} \approx 16\,$K (see Section 3.2 of \citetalias{2025MNRAS.537.3632M}).
In reality, the measured fluctuation from residual temperature in the calibration model is typically higher than expected.
We enlarge the system temperature by a factor of 1.2 so that $T_{\rm sys} = 19.2\,$K, following Section 2.4.3 of \citetalias{2025MNRAS.537.3632M}.
The thermal noise is then generated randomly following a Gaussian distribution at each pixel.

The different components of the mock observation are then added together to produce the mock intensity maps. 
For now, we leave out the component of the systematics caused by beam ripple as seen in the data, which will be discussed in detail later in \secref{sec:sys}.

\section{Viability of the stacking measurement}
\label{sec:validation}
In this section, we use the mock observations to validate the fact that the MeerKLASS \textit{L}-band intensity maps can be used to achieve a stacking detection.
From the mock stacked cubes, we can average the signal into the angular plane or into a spectrum, to inform us on the optimal region around a source that goes into the averaging for optimal signal-to-noise ratio, which can be used for the data.
We then discuss how the signal can be modelled and how the covariance of the stacked signal can be estimated. 
All results in this section are averaged over $100$ realizations.

\subsection{Double counting in stacked cubelets}
\label{subsec:stack}
To investigate the validity of the stacking measurement, we need to first define how a stacked signal is obtained from the intensity maps.
The stacked signal is calculated as
\begin{equation}
    I(\Delta\alpha,\Delta\phi,\Delta\nu) = \frac{\sum_i I(\alpha_i{+}\Delta\alpha,\,\phi_i{+}\Delta\phi,\,\nu_i{+}\Delta\nu) w_i}{\sum_i w_i},
\label{eq:stack}
\end{equation}
where $(\alpha_i,\phi_i)$ are the R.A. and decl. of the map pixel in which the $i^{\rm th}$ galaxy resides, $I$ is the flux density, and $w_i = w_{\hi}(\alpha_i{+}\Delta\alpha,\phi_i{+}\Delta\phi,\nu_i{+}\Delta\nu)$ is the weight of each pixel.
$(\Delta\alpha,\Delta\phi,\Delta\nu)$ is the position of the stacked signal relative to the source position in terms of R.A., decl., and frequency. 
Throughout this paper, we adopt inverse noise variance weighting so that $w_i = N_{\rm hits}^i$, as shown in the lower panel of \autoref{fig:datamap}.

It is common to express the stacked signal in terms of velocity instead of frequency. 
In our work, since we do not consider the Doppler correction, the peculiar velocity of the observer is not accounted for. 
Therefore, there is an ambiguity between the frequency offset $\Delta\nu$ and the velocity it should correspond to, so the stacked signal is not in actual velocity units. 
Nevertheless, when showing the results along the spectral direction, we also express $\Delta \nu$ in an effective velocity unit so that\footnote{We express velocity in terms of $v$ instead of $\Delta v$, since it is difficult to visually distinguish $\Delta v$ and frequency offset $\Delta \nu$.}
\begin{equation}
    v = \frac{c}{\nu_{\rm obs}} \Delta \nu,
\end{equation}
where $\nu_{\rm obs}=997.38\,$MHz is the central frequency of the frequency sub-band.

The frequency offset $\Delta \nu$ can also be transformed in to an approximate comoving scale, $\Delta D_c$, along the line of sight so that
\begin{equation}
    \Delta D_c \approx \frac{c \,\nu_{21}}{H \nu_{\rm obs}^2} \Delta \nu,
\end{equation}
where $H$ is the Hubble parameter at the observing redshift.
In our frequency sub-band, a frequency offset of $\Delta \nu = 1\,$MHz corresponds to $\Delta D_c \approx 5\,$Mpc.

\autoref{eq:stack} states that the stacked signal is a weighted average of the nearby volume of each source position.
We refer to the final stacked signal as the stacked cube, and the contribution of each source as the stacked cubelet. 
This expression also can be understood as equivalent to a cross-correlation function between the galaxy positions and the \hi\ line-intensity fluctuations.

The stacked cube of flux density can then be summed along the frequency direction into integrated flux so that
\begin{equation}
    \bar{F}(\Delta\alpha,\Delta\phi) = \sum_{\Delta\nu_i} I(\Delta\alpha,\Delta\phi,\Delta\nu_i) W(\Delta\nu_i) \delta\nu,
\label{eq:angular}
\end{equation}
where $\delta\nu$ is again the frequency channel bandwidth, $\Delta\nu_i$ loops over each frequency interval, and $W(\Delta\nu_i)$ is the selection function.
In this paper, we choose 
\begin{equation}
    W(\Delta\nu) = 1,\,{\rm for\,} |\Delta\nu|<3.5\,{\rm MHz},
\label{eq:velrange}
\end{equation}
which corresponds to $|v|\lesssim 1000\,$km\,s$^{-1}$.
The choice of summing within the $\pm$1000\,km\,s$^{-1}$ range is justified in \secref{subsec:expsnr}, and we note that the stacked image is only used for visual checks in the data analysis later.

As we will show in \secref{subsec:pca}, the structure of the \hi\ emission in the stacked image follows closely the shape of the beam.
It is therefore also useful to check the polar average of the stacked image, so that
\begin{equation}
\begin{split}
    \bar{F}_{\rm 1D}(\Delta\theta_i) = & \sum_{\Delta\alpha,\Delta\phi} \bar{F}(\Delta\alpha,\Delta\phi) W(\sqrt{\Delta\alpha^2 + \Delta\phi^2};\Delta\theta_i) \\ & / \sum_{\Delta\alpha,\Delta\phi }W(\sqrt{\Delta\alpha^2 + \Delta\phi^2};\Delta\theta_i)\, ,
\end{split}
\label{eq:angular1d}
\end{equation}
where $W(\sqrt{\Delta\alpha^2 + \Delta\phi^2};\Delta\theta_i)$ is a selection function that selects pixels that are inside the $i^{\rm th}$ angular annulus bin.
In this work, we choose seven annulus bins to be linearly spaced between 0 and 3.5\,deg.
We can also compute the polar average of the primary beam, by simply replacing the stacked image $\bar{F}(\Delta\alpha,\Delta\phi)$ with the frequency-averaged primary beam $B(l,m)$ in \autoref{eq:angular1d}.

Alternatively, the stacked cube can be summed along the angular plane into a spectrum so that
\begin{equation}
    \bar{I}(\Delta \nu) = \sum_{i,j} I(\Delta\alpha_i,\Delta\phi_j,\Delta\nu) W(\Delta\alpha_i,\Delta\phi_j), 
\label{eq:spectral}
\end{equation}
where $i,j$ loops over each pixel along the angular plane, and $W(\Delta\alpha_i,\Delta\phi_j)$ is the selection function.
Since \autoref{eq:stack} normalises over the sum of weights, the stacked spectrum can be understood as the \hi\ {signal for the average galaxy.}
In this paper, for visualization, we consider the range ($|\Delta\alpha| < 3\,$deg, $|\Delta\phi| < 3\,$deg), and for the spectrum, we consider 
\begin{equation}
    W(\Delta\alpha_i,\Delta\phi_j)= 1,\,{\rm for\,} \sqrt{\Delta\alpha^2+\Delta\phi^2}<1.2\,{\rm deg},
\label{eq:angrange}
\end{equation}
based on the mock discussed later in \secref{subsec:expsnr}.
Furthermore, in the stacked spectrum, we choose the binning in frequency to be 3 times the frequency channel bandwidth.

It is worth pointing out that, at the angular scales of the MeerKAT beam $\sim 1\,$deg corresponding to $\sim 30\,$Mpc, it is not evident that a stacked signal can be measured.
A simple way to see this is to imagine two extreme scenarios: one where the beam is uniform across the sky and one where the beam is infinitely small.
If the beam is uniform across the full sky, for any arbitrary position as the centre of the stacking, the stacked cubelet will always simply return the average \hi\ flux.
If the beam is infinitely small, for stacking on the positions of \hi\ galaxies, the stacked cubelet will give an excess signal at the centre angular pixel and no signal anywhere else.
In general, due to beam smoothing, the stacked cube will always contain an excess signal at the centre of the cubelet from the source we are stacking on, and an extra signal from double counting the other sources near the target stacking source.
{In terms of a two-point correlation function, the effect of the beam can be understood as a mixture between the one-halo and the two-halo terms.}
``Two-halo'' term corresponds to the correlation between \hi\ galaxies that are not in the same parent dark matter halo, whereas the one-halo term describes the inner-halo correlation between \hi\ galaxies of the same halo (see \citealt{2002PhR...372....1C} for a review and e.g. \citealt{2021MNRAS.502.5259C} for \hi\ halo models).
{A cubelet at a given separation to the detected galaxy, which would contain its \hi\ signal and therefore correspond to the one-halo term, would also include signal from a different, close-by galaxy, which is a clustering signal.}
{The $\sim 30\,$Mpc scale of the beam means that the mixture happens at small inner-halo scales and out to large physical scales, corresponding to the two-halo correlation.}

\begin{figure}
    \centering
    \includegraphics[width=1.0\linewidth]{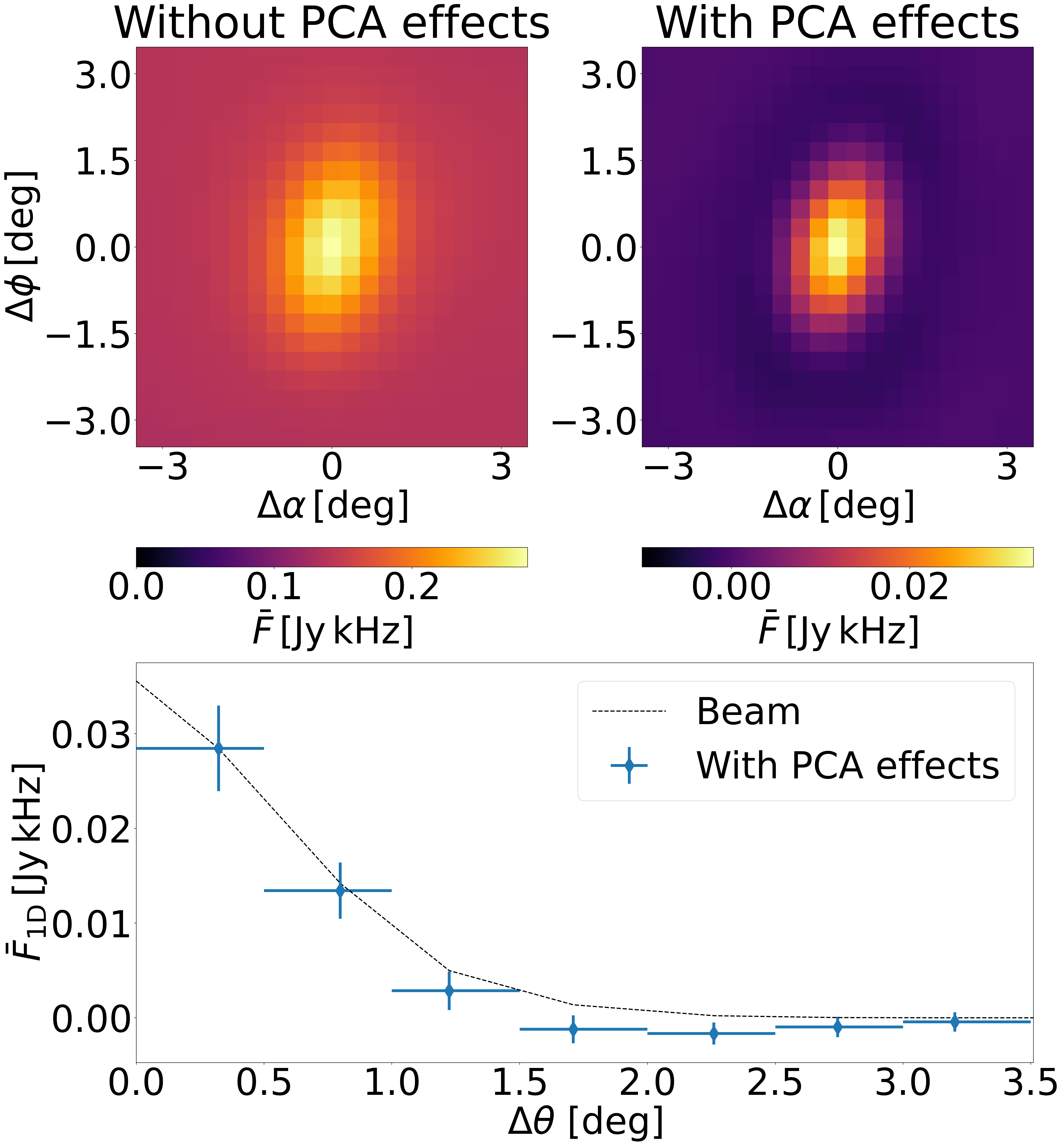}
    \caption{The angular stacking signal from \hi\ only mocks (no foregrounds and thermal noise), averaged over 100 realizations. The stacked image is calculated according to \autoref{eq:angular} and averaged into a 1D profile according to \autoref{eq:angular1d}. The upper-left panel shows the stacked image without any PCA cleaning effects, and the upper-right panel shows the case with PCA cleaning of 10 modes. The excess signal at the centre corresponds to the primary beam. The apparent ellipticity and orientation of the excess signal around the centre are due to projection, as explained in \autoref{fig:beam}. Note that the pixels away from the centre region also have a nonzero emission in the left panel. In the \hi\ only case, the first 10 eigenmodes contain much more \hi\ signal as there is no foreground to be removed, and therefore, the signal loss is more severe. The lower panel shows the 1D polar average of the stacked image for the PCA cleaned case (``With PCA effects''). The error bars are the standard deviations of the stacked profile among the realizations. For illustration, the primary beam profile matching the amplitude of the first $\Delta\theta$ bin is plotted in black dashed line (``Beam'').}
    \label{fig:angularpurehi}
\end{figure}

We illustrate this effect of double counting, using the angular stacking signal following \autoref{eq:angular}. 
As shown in the left panel of \autoref{fig:angularpurehi}, the stacked \hi\ flux shows a consistent positive signal at $\rm \sim 0.15\,Jy\,kHz$ level, regardless of how far away the angular position is from the centre.
This indicates a severe amount of double counting from the beam smoothing.
Despite the double counting, the excess emission around the centre of the image is still clearly visible, and its profile corresponds to the shape of the beam. 
This suggests that stacking the intensity maps onto the GAMA galaxy positions produces a detectable signal.

\begin{figure}
    \centering
    \includegraphics[width=1.0\linewidth]{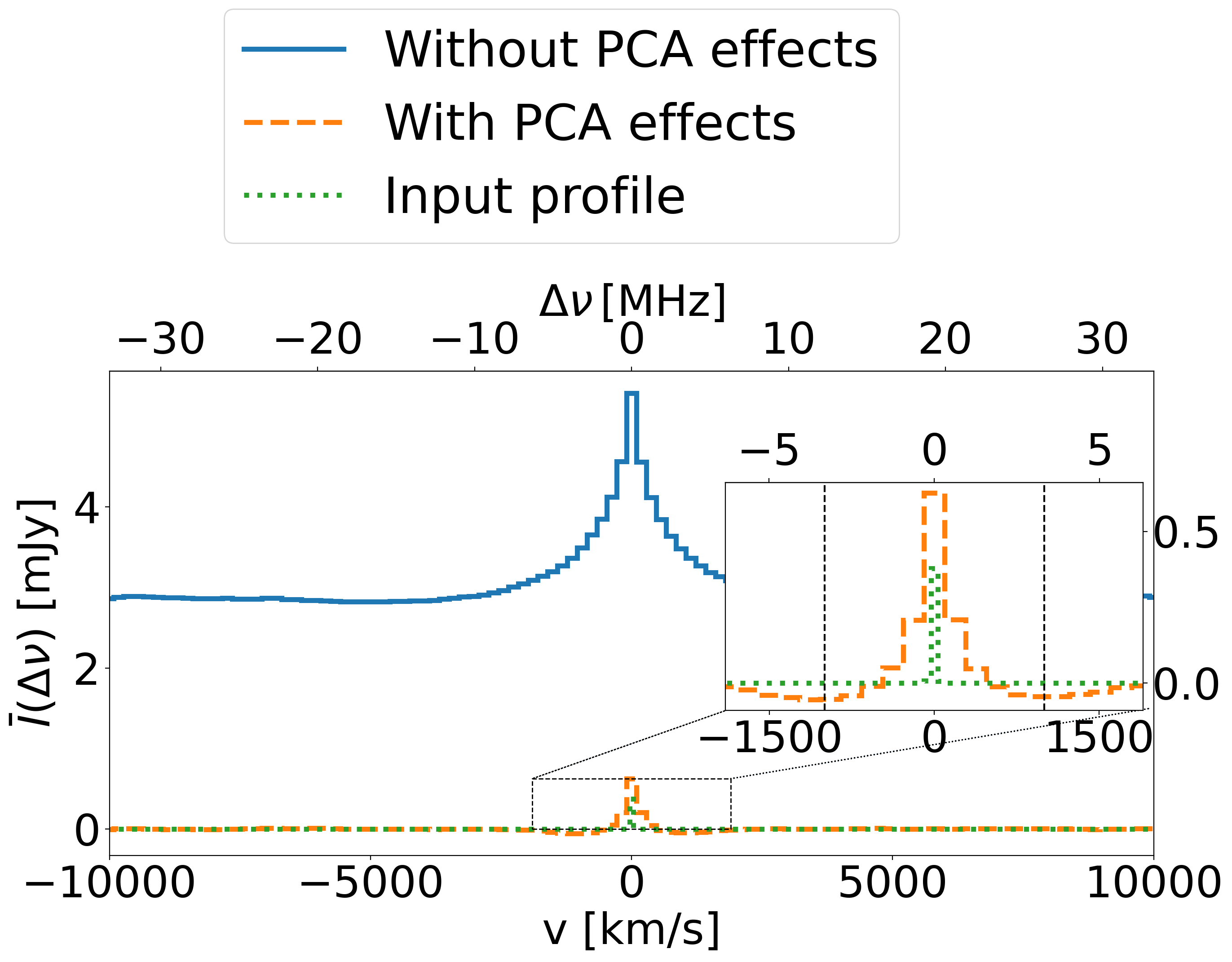}
    \caption{The spectral stacking signal from \hi\ only mocks, averaged over 100 realisations. The stacked spectrum is calculated according to \autoref{eq:spectral}. The stacked \hi\ signal without any PCA cleaning (``Without PCA effects''), with PCA cleaning of 10 modes (``With PCA effects''), and the average of the input \hi\ profiles of the GAMA-like galaxies (``Input profile'') are shown. 
    The zoomed-in panel shows the stacked spectrum after PCA for $|\Delta v|<1600\,$km\,s$^{-1}$, compared against the average of the input \hi\ galaxies.
    The stacked spectrum is much larger than the input profile due to the beam smoothing, which causes double counting of \hi\ sources. The black dashed lines show the $|\Delta v|=1000\,$km\,s$^{-1}$ boundary within which the stacked angular image is produced. In the \hi\ only case with no foreground and noise, the first 10 eigenmodes of PCA contain much more \hi\ signal as there is no foreground to be removed, and therefore, the signal loss is more severe compared to the full mock case shown later in \autoref{fig:spectralmock}.}
    \label{fig:spectralpurehi}
\end{figure}

The double counting is further demonstrated in \autoref{fig:spectralpurehi}. The \hi\ signal spectrum (``Without PCA effects'') shows a plateau of emission-line signal, indicating the existence of double counting.
Comparing the \hi\ signal spectrum with the average input \hi\ profile of the galaxy subsample, one can see that in addition to a signal plateau, the peak around the centre frequency is also amplified with a greatly enlarged velocity width.
If the excess signal is only a result of beam smoothing and summing over the angular plane, there is no correlation along the frequency direction.
The widened velocity profile of the stacked spectrum suggests that the stacked cubelets are correlated at these frequency scales.
This points to the fact that, the \hi\ signal is correlated at $\Delta\nu < 5\,$MHz frequency intervals and $\sim 30\,$Mpc transverse scales.
The $\Delta\nu < 5\,$MHz frequency scales correspond to a line-of-sight comoving distance of $\lesssim 25\,$Mpc.
This is an effect of clustering of the \hi\ signal.
Similar features of clustering in the stacked signal have also been found in the context of CO line-intensity mapping \citep{2025arXiv250321743D}.

{We emphasise that the effects of double counting and clustering are distinctive. The double counting is due to the high number density of} \hi\ {emitters compared to the angular size of the beam. One can imagine a case where there is no clustering, and the number density of} \hi\ {sources is extremely low within the size of the primary beam. In that case, along the line of sight of any source, there will be very little leakage from other sources as they are sufficiently away. Hence, there will not be a positive plateau in the stacked signal. On the other hand, even if the average number density of} \hi\ {sources is low, there will always be pairs of sources close to each other if there is clustering. Therefore, the widening of the stacked spectrum at small separations will be present as a result of the clustering.}

\subsection{Effects of PCA}
\label{subsec:pca}
We have established that for the MeerKLASS \textit{L}-band intensity maps, a stacking signal can be measured with the caveat of the stacked signal having a large contribution due to beam smoothing and source clustering.
The additional contribution is then partially removed due to the procedure of foreground cleaning, which removes the mean of the signal and causes signal loss.
Therefore, it is important to validate that the excess signal in the central region of the stacked cube is robust against the PCA.

To perform the PCA cleaning, we first calculate the weighted average of the maps at each channel, and obtain the mean centred intensity mapping data matrix $\mathbf{M}_{ip}$, where $i$ iterates over each frequency and $p$ iterates over each map pixel. We can then compute the frequency-frequency covariance matrix so that
\begin{equation}
    \mathbf{C}_{ij} = \Big(\sum_{p} w_{ip} w_{jp} \mathbf{M}_{ip}\mathbf{M}_{jp}\Big) / \Big(\sum_p  w_{ip} w_{jp}\Big).
\end{equation}
Eigendecomposition of the frequency-frequency covariance matrix is then performed to find its eigenvalues and eigenvectors. 
The eigenvectors $\vec{u}_{i}$ are then sorted based on the values of their corresponding eigenvalues from largest to smallest.
Choosing a $N_{\rm fg}$ number of modes, we can define a mixing matrix $\mathbf{A}$ so that 
\begin{equation}
    \mathbf{A} = [\vec{u}_{1},\vec{u}_{2},...,\vec{u}_{N_{\rm fg}}],
\end{equation}
from which we can then define a cleaning matrix $\mathbf{R}$
\begin{equation}
    \mathbf{R} = \mathbf{I} - \mathbf{A}\mathbf{A}^{\rm \mathbf{T}},
\label{eq:rpca}
\end{equation}
where $\mathbf{I}$ is the identity matrix and $^{\rm \mathbf{T}}$ denotes transpose of a matrix.
The residual data matrix after cleaning is then
\begin{equation}
    \mathbf{M}^{\rm res}_{ip} = \sum_j \mathbf{R}_{ij} \mathbf{M}_{jp}.
\label{eq:mapres}
\end{equation}
The residual data $\mathbf{M}^{\rm res}$ can then be used to perform the stacking analysis.

The choice of the number of modes to be removed, $N_{\rm fg}$, is subtle and depends on the level of contamination in the data (see, e.g. \citealt{2023MNRAS.518.6262C}).
Throughout this paper, we choose $N_{\rm fg}=10$ as suggested in \citetalias{2025MNRAS.537.3632M}.
This choice is to be consistent with the level of foreground removal needed in the data, as discussed in \citetalias{2025MNRAS.537.3632M}.

We first discuss the case where only \hi\ signal is considered. We present the angular stacking image 
for \hi\ only mocks after PCA cleaning in the right panel of \autoref{fig:angularpurehi}.
As can be seen, the excess signal in the central region of the stacked image is still present after PCA cleaning.
Significant signal loss can be found, as is evident in the decrease in the amplitude of the emission signal with respect to the case before PCA cleaning.
The region of excess emission still follows closely the shape of the primary beam, suggesting that the excess is indeed caused by the source in the GAMA-like subsample (recall \autoref{fig:mockillus}).
In the bottom panel of \autoref{fig:angularpurehi}, we show the polar average of the stacked image after PCA cleaning according to \autoref{eq:angular1d}.
As we can see, within the primary beam of $\sim 1\,$deg, the excess emission closely follows the beam.
The emission then decreases as the angular distance increases and reaches zero.

The removal of the extra emission away from the cubelet centre results in the removal of the plateau in the stacked spectrum, as shown in the orange dashed line (``With PCA effects'') in \autoref{fig:spectralpurehi}.
For large $\Delta\nu$, the signal is consistent with zero.
The amplitude of the central peak is much smaller compared to the spectrum before PCA, while still being much larger than the input.
The PCA cleaning also results in a slightly negative amplitude at $|\Delta v|\sim 1000\,$km\,s$^{-1}$.
The removal of relatively large line-of-sight modes, together with the subtraction of the mean, results in the negative amplitude (see also Fig. 18 of \citealt{2023ApJ...947...16A}).
The complication of these effects requires forward modelling to describe.

\subsection{Expected detection significance}
\label{subsec:expsnr}
Based on the viability of detecting a stacked signal around the central region of the stacked cube, we then include foreground signal and thermal noise and perform the PCA subtraction to see if the excess emission can be detected given the depth of the MeerKLASS \textit{L}-band survey.

\begin{figure}
    \centering
    \includegraphics[width=1.0\linewidth]{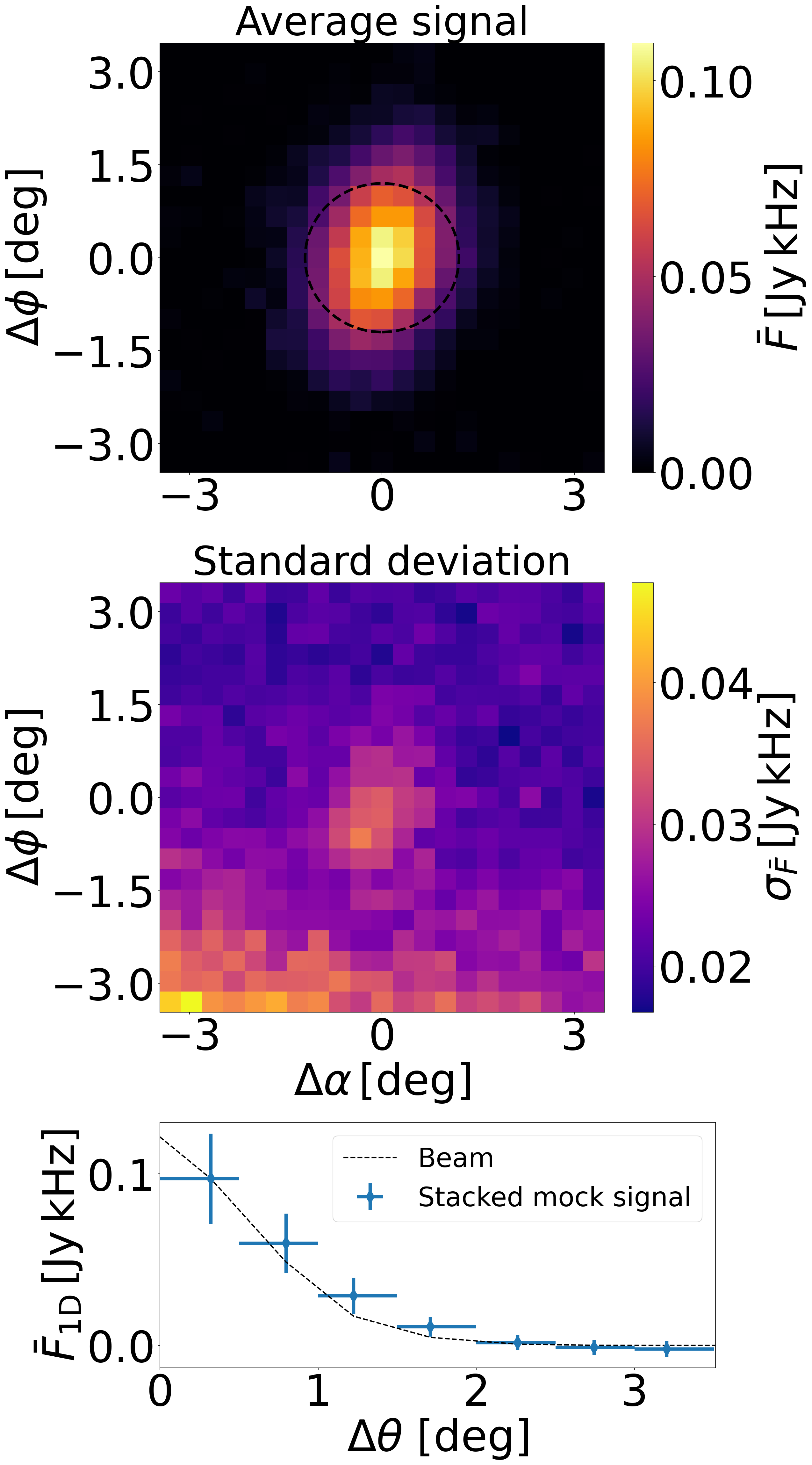}
    \caption{The angular stacking signal from full mocks (including foreground and noise simulation), averaged over 100 realisations. The top panel shows the average of the stacked images across the realisations, and the central panel shows the standard deviation of the realisations. The dashed circle in the top panel denotes the 1.2\,deg boundary within which the signal is summed into the stacked spectrum.  Note the differences in the colour scale in the two panels.
    The lower panel shows the 1D polar average of the stacked image for the full mock (``Stacked mock signal''). The error bars are the standard deviations of the stacked profile among the realisations. For illustration, the primary beam profile matching the amplitude of the first $\Delta\theta$ bin is plotted in black dashed line (``Beam'').
    }
    \label{fig:angularmock}
\end{figure}

In the top panel of \autoref{fig:angularmock}, we show the angular stacked image averaged from 100 independent realisations. 
Comparing the stacked image with the one in \autoref{fig:angularpurehi}, we can see that the excess signal remains, with the background becoming noisy due to the presence of thermal noise. 
Notably, the signal loss is much less severe.
This is because in the \hi\ only case, the first 10 eigenmodes contain much more \hi\ signal as there is no foreground to be removed.
When foregrounds and thermal noise are included, the PCA first cleans the smooth foregrounds and the same number of modes $N_{\rm fg}=10$ removes much less \hi\ signal.
The mocks can also be used to calculate the signal variance by taking the standard deviation among the realisations, as we show in the bottom panel of \autoref{fig:angularmock}.
Overall, the stacked signal is around a factor of 3 larger than the standard deviation, suggesting that the stacked image can indeed be detected.
Note that, the bottom half, and especially the bottom-left part, of the stacked image has a higher level of variance.
This can be traced back to the survey area shown in \autoref{fig:datamap}.
When stacking near the boundary of the GAMA region, the upper half of the cubelet is always sampled by the 21\,cm intensity maps, whereas the region lower than the GAMA galaxies is not covered by the MeerKLASS survey area.
The bottom half of the stacked image is therefore less sampled and has a higher noise level.
The polar average of the stacked image, shown in the bottom panel of \autoref{fig:angularmock}, is consistent with the beam profile.
We find that for the stacked image, the region within 1.2\,deg of the centre of the cube has a clear excess, which is the criterion we use for calculating the stacked spectrum in \autoref{eq:angrange}.

\begin{figure}
    \centering
    \includegraphics[width=1.0\linewidth]{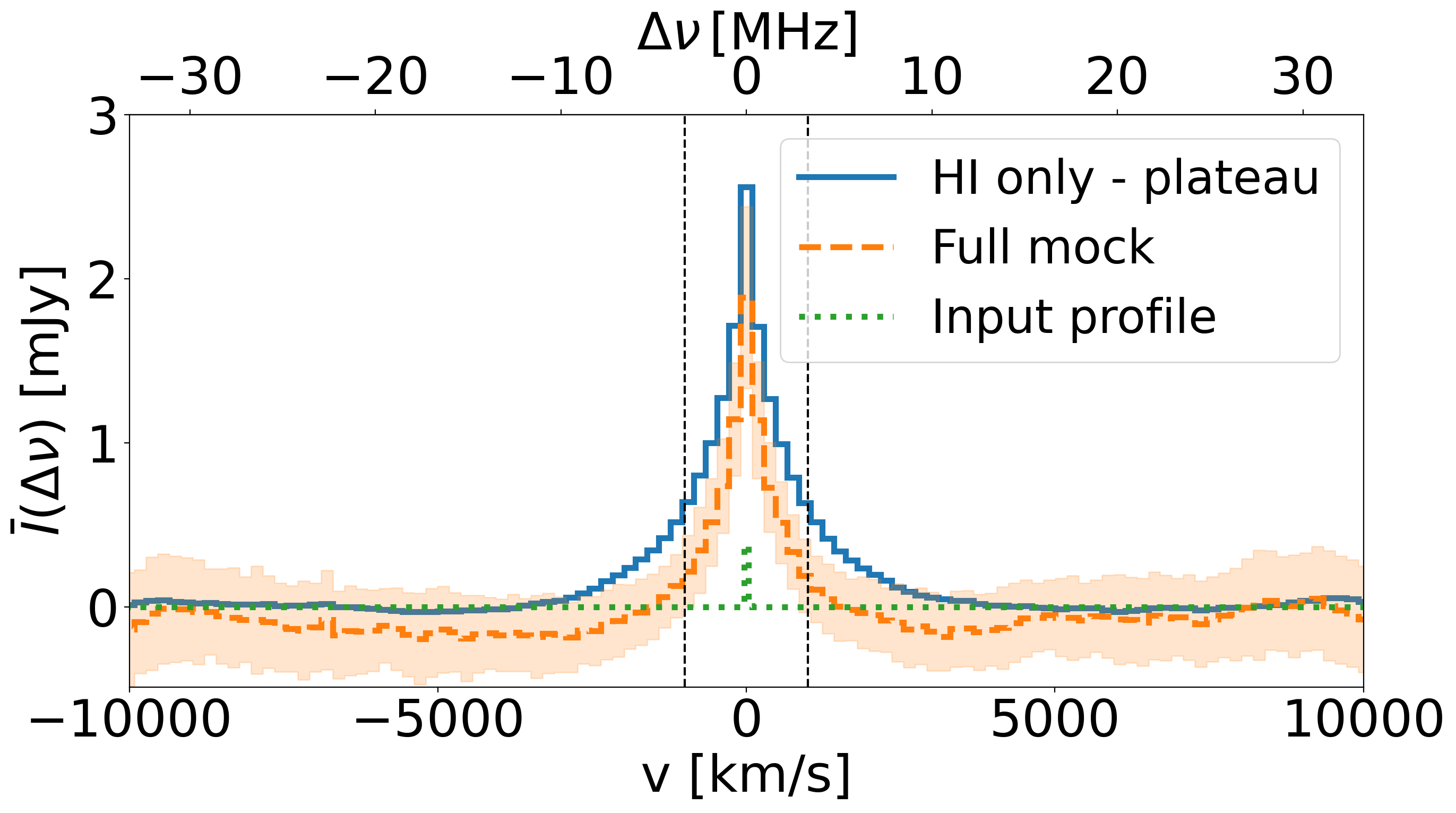}
    \caption{The spectral stacking signal from full mocks, averaged over 100 realisations. The \hi\ only signal without any PCA cleaning is shown in the blue solid line, with the plateau from double counting subtracted (``HI only $-$ plateau''). The averaged spectrum of the full mocks with PCA cleaning of 10 modes is shown in the yellow dashed line. The shaded region corresponds to the standard deviation among the realisations. The average of the input \hi\ profiles of the GAMA-like galaxies (``Input profile'') is also shown for reference.}
    \label{fig:spectralmock}
\end{figure}

We then proceed to calculate the stacked spectrum as shown in \autoref{fig:spectralmock}.
To compare against the signal without PCA, we show the \hi\ only simulation without PCA cleaning, and subtract out the plateau from the double counting (``HI only - plateau'') for comparison.
The averaged stacked spectrum (``Full mock'') shows similar amplitude compared to the input, with a visible amount of signal loss due to PCA cleaning of foregrounds.
Again, the stacked spectrum contains a large component of beam smoothing and clustering; it is much larger than the input \hi\ profile.
Overall, it is expected that the central peak can be detected with high statistical significance $> 3 \sigma$.
The peak of the stacked spectrum extends across $|\Delta v| \lesssim 1000\,$km\,s$^{-1}$, and we therefore choose 1000\,km\,s$^{-1}$ as the upper limit for the frequency channels in calculating the stacked image in \autoref{eq:velrange}.

The overall detection significance of the stacked signal is harder to calculate than simply comparing the variance with the signal.
This is because the signal is correlated between different angular positions and frequencies, as one pixel in the intensity map will be averaged into different voxels in the stacked cube.
To quantify the overall significance, the full covariance matrix is needed, which will be discussed in \secref{sec:covest}.

\subsection{Biasing from simplified forward modelling}
\label{subsec:biasing}
As mentioned in \secref{sec:sim}, in the actual data analysis, we would not have information on the underlying distribution of the \hi\ mass of the sources.
The detection in our data, while statistically significant, as shown later in \secref{sec:analysis}, is not enough to allow a large number of parameters to be constrained.
Therefore, as a simplification, we need to assume that all \hi\ density resides within the GAMA galaxy subsample, and the galaxies have the same \hi\ mass when using forward modelling to model the signal.
This leads to a mismatch between the interpreted model and the underlying truth.
We examine the impact of this biasing using the mocks with the following steps.

The alternative \hi\ map, described in \secref{subsec:hisim}, is applied with the \emph{original} PCA cleaning matrix from the full simulation.
This is because in the data analysis, the PCA cleaning matrix is obtained based on the data itself to ensure the same level of signal loss in the data as well as forward modelling.
The cleaned \hi\ map is then passed to the stacking pipeline to generate the stacked cube.
The resulting stacked signal is the ``forward modelling'' case, since the simplified assumption is applied to the simulation while keeping the overall \hi\ density unchanged.
For simplicity, we only show the stacked spectrum, since later in \secref{sec:covest} we demonstrate that the stacked spectrum is more robust for covariance estimation.

\begin{figure}
    \centering
    \includegraphics[width=1.0\linewidth]{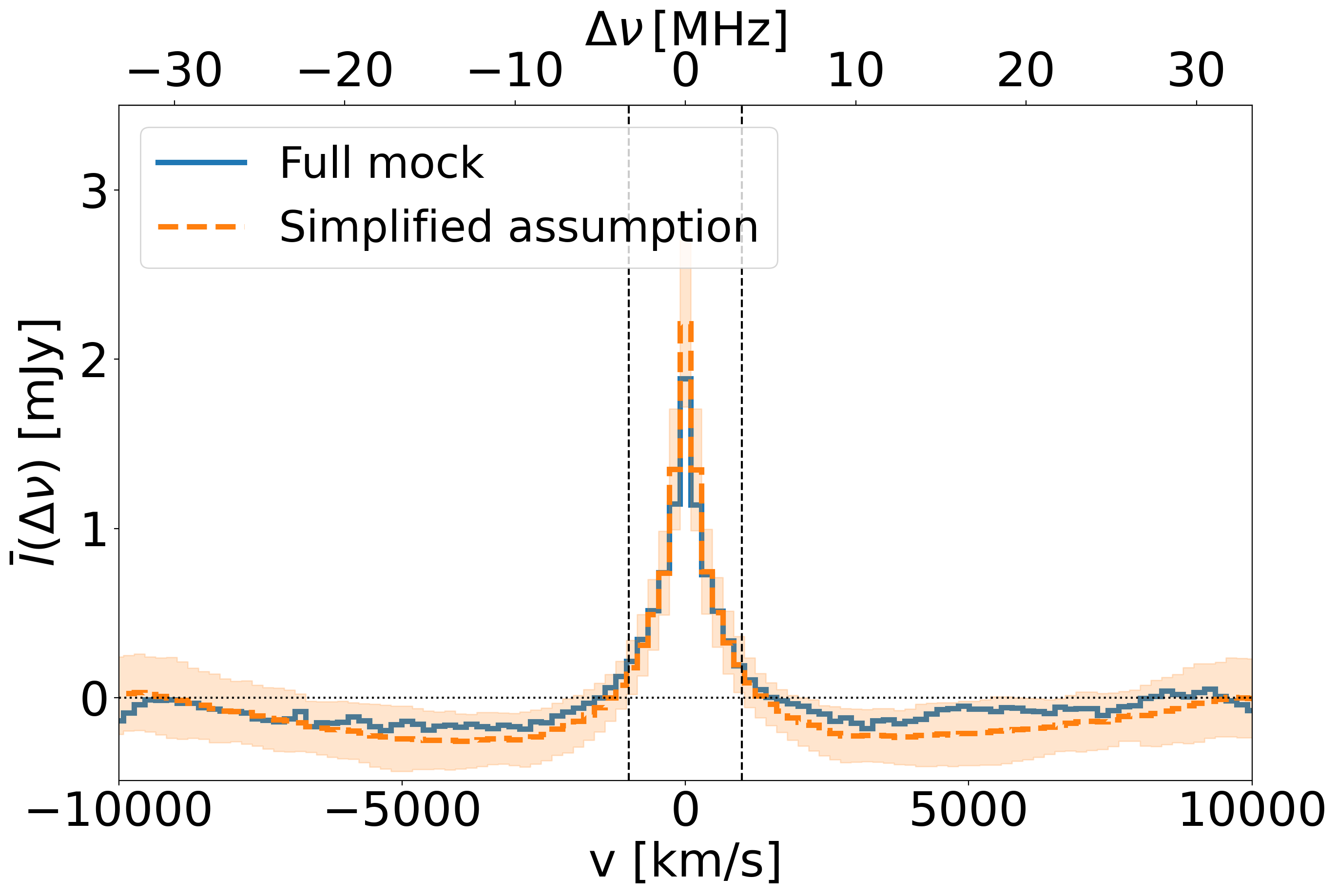}
    \caption{The stacked spectrum of the \hi\ signal from simplified assumption (``Simplified assumption''), compared against the true \hi\ signal (``Full mock''). The results shown are averaged across 100 realisations. The shaded region around the dashed line shows the standard deviation for the simplified assumption case among the realisations.}
    \label{fig:spectralassume}
\end{figure}

The comparison between the original mock and the simplified assumption is shown in \autoref{fig:spectralassume}.
The simplified assumption closely follows the true \hi\ signal, with the differences between the two smaller than the standard deviation across the realisations.
The \hi\ signal from the simplified assumption has a higher central peak with a slightly narrower width, due to far fewer sources which have much higher \hi\ mass than the full mock.
Integrating the spectrum between $|v|<2500\,$km\,s$^{-1}$ where the peak resides, we find that the differences in the integrated flux are smaller than $5\%$.
We deduce that, due to the large physical scale of the beam, the clustering signature of the \hi\ flux density in the simplified assumption is similar to the full mock.
Even though there are much fewer sources resulting in a more extreme distribution of \hi\ mass, after beam smoothing {the aggregate \hi\ signal over a resolution element is similar.}

In conclusion, the simplified assumption only slightly biases the modelled \hi\ signal, and is sufficiently accurate for the purpose of this work. We note that, in this case, the inferred \hi\ mass per source is not the actual mass, but should be interpreted as the total \hi\ mass over the number of galaxies in the stacking subsample.

\section{Covariance estimation}
\label{sec:covest}
\subsection{Mock covariance}
\label{subsec:mockcov}
The covariance of the stacking measurement consists of two uncorrelated components, which are the signal covariance and the noise covariance.
Understanding the contribution of both components is necessary to fully quantify the detection significance and perform model inference.
In this section, we first use the realisations to calculate the covariance, and then discuss the method for covariance estimation from one realisation, which is then used later for the data analysis.

\begin{figure*}
    \centering
    \includegraphics[width=1.0\linewidth]{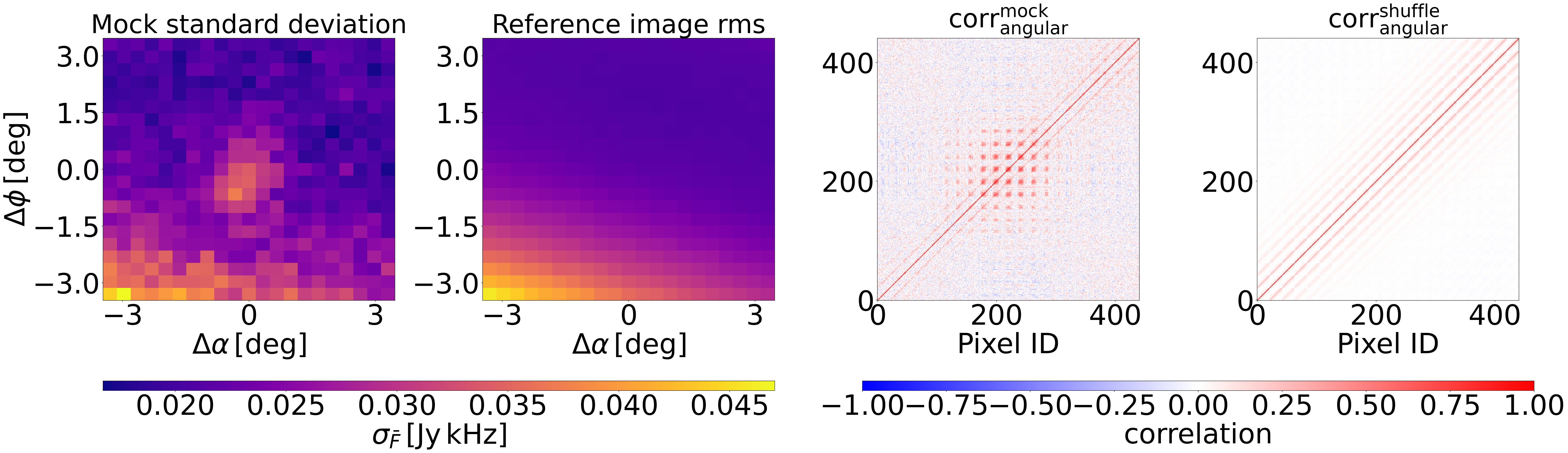}
    \caption{Left panel: the standard deviation of the angular stacked image across the mock realisations, which is also the square root of the diagonal elements of the mock covariance matrix $\mathbf{C}^{\rm mock}_{\rm angular}$. Centre-left panel: the reference image standard deviation among the random shuffles, averaged across all mock realisations, which is also the square root of the diagonal elements of the estimated covariance from shuffling $\bar{\mathbf{C}}^{\rm shuffle}_{\rm angular}$. Centre-right panel: the correlation matrix from mock covariance. The indexing of the array iterates from left to right, and then bottom to top of the stacked image. Right panel: the correlation matrix from covariance estimation using shuffling.}
    \label{fig:angularmockcov}
\end{figure*}

To calculate the covariance of the stacked cube, in each realisation, we start from the PCA cleaning matrix in \autoref{eq:rpca}, and apply the cleaning separately to the \hi\ signal, foregrounds, and noise.
We find that the residual foregrounds are negligible compared to the level of the \hi\ signal.
The residual \hi\ signal and noise are then used to calculate the stacked cube respectively, producing 100 realisations of stacked cubes for both components.
The covariance can then be calculated from the realisations,
\begin{equation}
    \mathbf{C}^{\rm mock}_{ij} = \sum_{\rm n}\frac{(I^{\rm n}_i - \bar{I}_i)(I^{\rm n}_j - \bar{I}_j)}{N_r-1},
\label{eq:covmock}
\end{equation}
where $i,j$ denote two voxels in the stacked cube, $n$ iterates over the realisations, $I^n_i = I^n(\Delta\alpha_i,\Delta\phi_i,\Delta\nu_i)$ is the stacked signal for one realisation, $\bar{I}$ is the average of the stacked cube across the realisations and $N_r$ is the number of mock realisations.
For any covariance matrix $\mathbf{C}$, we can also compute the correlation matrix,
\begin{equation}
    {\rm corr}_{ij} = \frac{\mathbf{C}_{ij}} {\sqrt{\mathbf{C}_{ii}}\sqrt{\mathbf{C}_{jj}}},
\end{equation}
which informs us on the correlation between different data points of a measurement.

The covariance obtained from \autoref{eq:covmock} is for the 3D stacked cube. 
In practice, we are only interested in the stacked image and the stacked spectrum.
The covariance for the averaged image/spectrum can be calculated by simply substituting the 3D signal $I^n$ with the averaged signal in \autoref{eq:covmock}.

\subsection{Covariance estimation using random shuffling}
\label{subsec:shuffle}

In reality, we only have access to the data without much information on the underlying model.
The ``true covariance'' from \autoref{eq:covmock}, built from specific choices of model parameters, does not necessarily reflect the observation.
Moreover, the observation data contain multiplicative systematics that affect the covariance of the data, which we will discuss in detail later in \secref{sec:sys}.
We therefore need a method of covariance estimation from the data itself, which we can then compare against the true covariance using the mocks.
In this work, we explore using random shuffles of galaxy positions for covariance estimation.
We generate realisations of random galaxy positions following the same procedure and clustering statistics of the galaxy catalogue, but based on realisations of mock dark matter uncorrelated with the \hi\ signal.
Such a random shuffling of galaxy positions should have the same statistical significance as the true catalogue, but produce a stacking signal consistent with zero.
Therefore, it is common to use such galaxy shuffling as a null detection test in the data analysis, as shown in \citetalias{2025MNRAS.537.3632M}.
In this work, we refer to the stacked image/spectrum using a random shuffle as the ``reference image/spectrum''.
A covariance can be then calculated based on multiple realisations of the shuffling \emph{in a single mock},
\begin{equation}
    \hat{\mathbf{C}}^{\rm shuffle}_{ij} = \sum_{\rm n}\frac{(I^{\rm n,shuffle}_i - \bar{I}^{\rm shuffle}_i)(I^{\rm n,shuffle}_j - \bar{I}^{\rm shuffle}_j)}{N_{\rm shuffle}-1},
\label{eq:covshuffle}
\end{equation}
where $i,j$ denote two voxels in the stacked cube, $\rm n$ iterates over the realisations of the shuffling of galaxy positions, $I^{\rm n,shuffle}$ is the stacked signal over the shuffled galaxy positions for one shuffle, $\bar{I}^{\rm shuffle}$ is the averaged stacked signal over all shuffles, and $N_{\rm shuffle}$ is the number of shuffling in total.
In this work, we choose $N_{\rm shuffle} = 400$ and find that convergence has been reached.
From \autoref{eq:covshuffle}, it is also easy to see that the standard deviation of the reference stacked signal is simply the diagonal elements of $\hat{\mathbf{C}}^{\rm shuffle}$.
Using the 400 realisations of shuffled galaxy positions, we can then estimate the covariance for each mock observation, and compute the average of the estimated covariance across all mock realisations.
We denote the average of $\hat{\mathbf{C}}^{\rm shuffle}$ \emph{across all mock realisations} as $\bar{\mathbf{C}}^{\rm shuffle}$.
For simplicity, we do not show the results for the entire 3D cube, but instead average the signal into the stacked image and the stacked spectrum for examination.

\begin{figure*}
    \centering
    \includegraphics[width=1.0\linewidth]{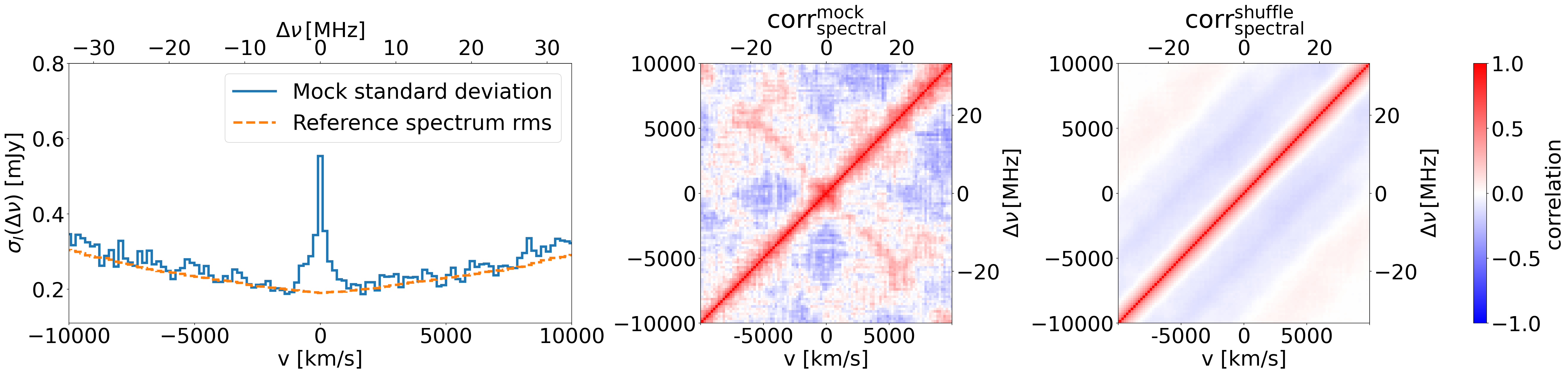}
    \caption{Left panel: the blue solid line shows the standard deviation of the spectral stacked spectrum across the mock realisations (``Mock standard deviation''), which is also the square root of the diagonal elements of the mock covariance matrix $\mathbf{C}^{\rm mock}_{\rm spectral}$. The orange dashed line shows the average of the standard deviation of the randomly shuffled reference spectrum across the realisations (``Reference spectrum rms''). Centre-right panel: the correlation matrix from mock covariance. Right panel: the correlation matrix from covariance estimation using random shuffling.}
    \label{fig:spectralmockcov}
\end{figure*}

The comparison between the true mock covariance and the random shuffling estimate is shown in  \autoref{fig:angularmockcov}.
From the variance of the angular stacking shown in the left panels of \autoref{fig:angularmockcov}, we can see that the random shuffling captures accurately the overall noise level of the signal.
In particular, the higher noise variance at the lower half of the image, which is due to less sampling as discussed in \secref{subsec:expsnr}, is also reflected in the variance of the reference image.
The random shuffling does not capture the \hi\ signal variance at the centre of the image though, which is expected, since the shuffling does not contain the excess \hi\ signal at the centre of the stacked cube.
Apart from the lack of signal variance, it is also expected that the correlation of the signal due to the primary beam is also missing from the shuffling estimate.
In the right panels of \autoref{fig:angularmockcov}, we can see that the true correlation matrix contains strong correlation between pixels near the centre of the stacked image.
This correlation is not being captured by the random shuffling.

We then examine the differences between the true mock covariance and the shuffling estimate in the stacked spectrum in \autoref{fig:spectralmockcov}.
The overall amplitude of the variance from the reference spectrum matches closely the true mock variance, as seen in the left panel.
Similar to the stacked image, the shuffling estimate does not capture the variance of the \hi\ signal near the centre $\Delta \nu \sim 0$.
Comparing the correlation between different velocities shown in the right panel of \autoref{fig:spectralmockcov}, we can see that the correlation between nearby channels is largely consistent between the truth and the shuffling estimate.
The largest inconsistency comes from the antidiagonal direction, i.e. correlation between positive and negative values of $\Delta \nu$.
This inconsistency again stems from signal covariance, as the same \hi\ sources are averaged into the spectrum multiple times at different $\Delta \nu$ due to the double counting.
The PCA cleaning, while removing the amplitude of the plateau as discussed in \secref{subsec:pca}, does not fully remove the correlation induced by the plateau.

In conclusion, we find that the random shuffling can be used to estimate the overall amplitude of the variance of the stacked signal.
Due to not including the excess \hi\ signal at the centre of the stacked cube, the shuffling estimate does not include the \hi\ signal covariance nor its corresponding correlation near the centre of the stacked image.
In the stacked spectrum, the correlation is largely consistent, but the \hi\ signal variance around the centre is still missing from the shuffling estimate.
From now on, we focus on the stacked spectrum for covariance estimation.

\subsection{Mock-corrected covariance estimation}
\label{subsec:mockcovest}
Based on the conclusions reached in \secref{subsec:shuffle}, we can see that an accurate covariance estimation can be obtained in the stacked spectrum, where we have summed over the angular pixels.
There are two problems that remain, which are the missing \hi\ signal variance around $\Delta \nu \sim 0$, and the fact that the antidiagonal direction of the correlation matrix is not captured by the shuffling estimate.

We first discuss how to mitigate the problem of antidiagonal correlation.
Note that, if only the upper-right quadrant of the correlation matrix is considered, i.e. $\Delta \nu \geq 0$, then the antidiagonal direction is naturally excluded.
This prompts the usage of symmetrised stacking \citep{2022MNRAS.514.4205S} so that
\begin{equation}
    \bar{I}_{\rm sym}(\Delta \nu) = \big(\bar{I}(\Delta \nu)+\bar{I}(-\Delta \nu)\big)/2, 
\label{eq:symstack}
\end{equation}
where for the averaged spectrum $\bar{I}_{\rm sym}$ at $\Delta \nu$, both the $\Delta \nu$ and $-\Delta \nu$ of the original stacked spectrum are included.
Since the stacked \hi\ signal is on average symmetric along $\Delta\nu=0$ as seen in \autoref{fig:spectralmock}, we expect no loss of information from the symmetrisation.
In our case where the angular plane is collapsed into the spectrum, \autoref{eq:symstack} is equivalent to the 180$\degree$ rotational symmetry along the zero-velocity axis described in \citet{2022MNRAS.514.4205S}.
In our case, this does not increase the signal-to-noise ratio, as suggested in \citet{2022MNRAS.514.4205S}, since in our case, there is significant double counting, and there is no meaningful increase in the number of sampling at each $\Delta\nu$ from symmetrisation.

\begin{figure}
    \centering
    \includegraphics[width=1.0\linewidth]{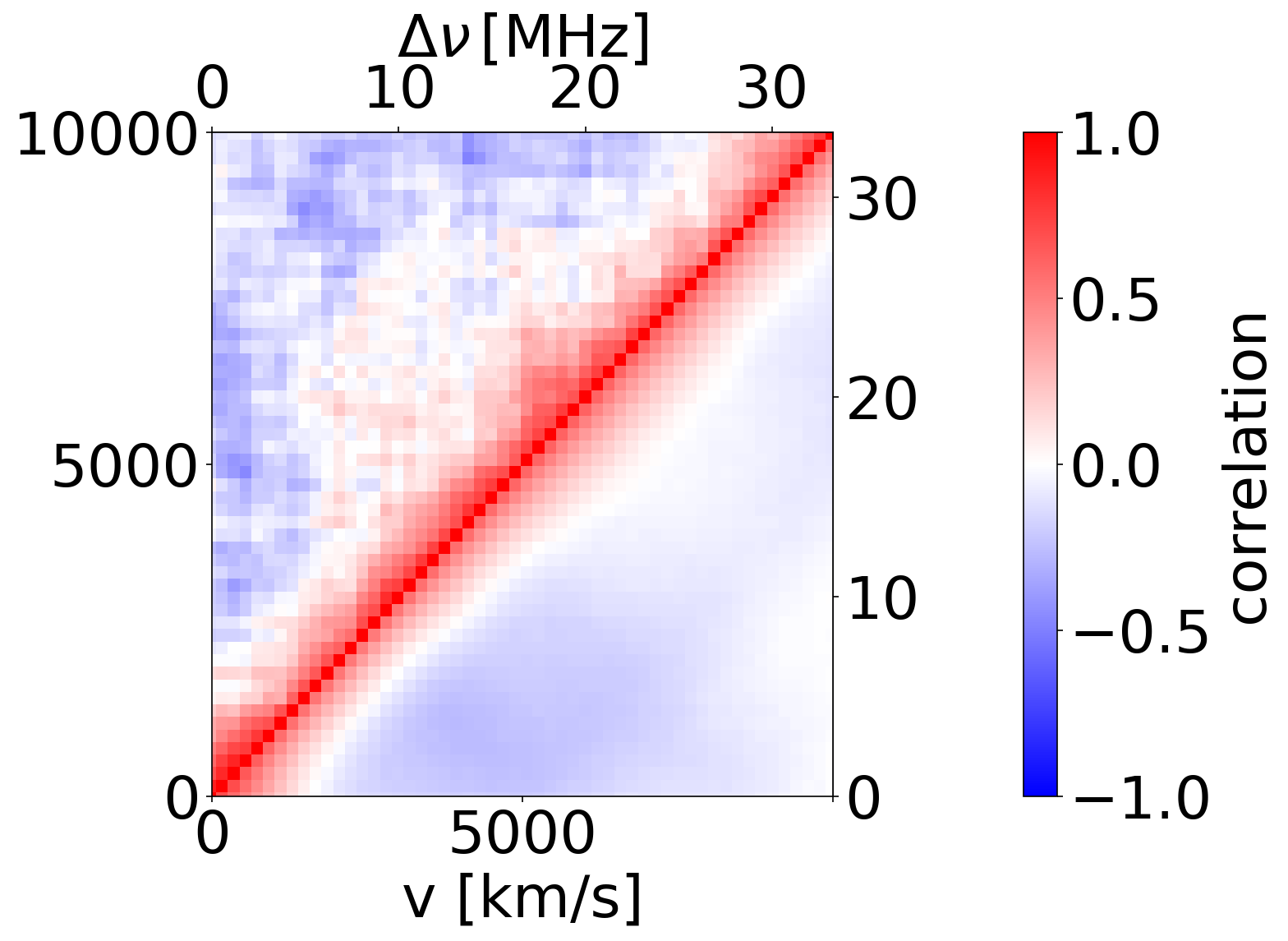}
    \caption{The comparison between the correlation matrix of the symmetrised stacked spectrum using the true mock covariance and the shuffling estimate. The upper triangle of the matrix shows values from the true mock covariance, and the lower triangle shows values from the shuffling.}
    \label{fig:corrmix}
\end{figure}

Using the symmetrised stacked spectrum, we recalculate the mock stacked signal in all realisations, the mock covariance, and the shuffling estimate.
The resulting correlation matrix is shown in \autoref{fig:corrmix}.
Since the spectrum has been symmetrised, we only need to consider the range $\Delta\nu\geq0$.
Comparing the results from \autoref{fig:corrmix} and \autoref{fig:spectralmockcov}, one can see that the correlation matrix from the shuffling estimate becomes more consistent with the underlying truth when the spectrum is symmetrised.
The correlation matrix is highly consistent for small values of $|\Delta \nu|$, where small inconsistencies still persist at intermediate values.

We now turn to the fact that we need to correct for the missing \hi\ signal covariance.
We construct a mock-informed correction of the covariance estimation so that
\begin{equation}
    \hat{\mathbf{C}}^{\rm data}_{ij} = \vec{r}_{i}\vec{r}_{j}\hat{\mathbf{C}}^{\rm data,shuffle}_{ij},
\label{eq:covcorrected}
\end{equation}
where $\hat{\mathbf{C}}^{\rm data,shuffle}$ is the shuffling estimate described in \autoref{eq:covshuffle}, using the actual data. 
$\vec{r}$ is a vector of the ratio between the square root of the diagonal elements of mock covariance and the shuffling estimate in the mock so that
\begin{equation}
    \vec{r}_i = \sqrt{\mathbf{C}^{\rm mock}_{ii} / \hat{\mathbf{C}}^{\rm mock, shuffle}_{ii}}.
\end{equation}
It is easy to see that $\hat{\mathbf{C}}^{\rm data,shuffle}$ and the final $\hat{\mathbf{C}}^{\rm data}$ share the same correlation matrix.
For the mock simulations, \autoref{eq:covcorrected} returns the true mock covariance when averaged over all realisations, since in this case, ``data'' in $\hat{\mathbf{C}}^{\rm data,shuffle}$ simply stands for one realisation of the mock.

In short, \autoref{eq:covcorrected} describes the covariance estimation from the stacked intensity maps in two steps: First, use the symmetrised stacking and randomly shuffled galaxy positions to calculate an initial estimate $\hat{\mathbf{C}}^{\rm data,shuffle}_{ij}$.
Second, use the mock realisations to obtain a correction of the amplitude from the shuffling to the true covariance.
This method has several desired advantages.
The amplitude of the covariance does not rely on the amplitude obtained in the mock, but instead depends on the data itself through $\hat{\mathbf{C}}^{\rm data,shuffle}$.
The correlation matrix also follows the structure of the data.
Even if there are systematics not considered in the mock, for example, the multiplicative systematics that we discuss in \secref{sec:sys}, the correlation matrix will follow the data affected by the systematics instead of following the mock, since the correlation matrix follows $\hat{\mathbf{C}}^{\rm data,shuffle}$ (although the correlation will be slightly distorted; A simple analytical derivation is presented in \appref{apdx:covanalytic}).
The downside of \autoref{eq:covcorrected} is that the fractional correction of missing \hi\ covariance is based on the mock, where the amplitude of the \hi\ signal and the noise are likely to be different from the actual data.
We come back to the effect of covariance estimation later in \secref{sec:results}.

\section{Stacking measurement}
\label{sec:analysis}
\begin{figure*}
    \centering
    \includegraphics[width=1.0\linewidth]{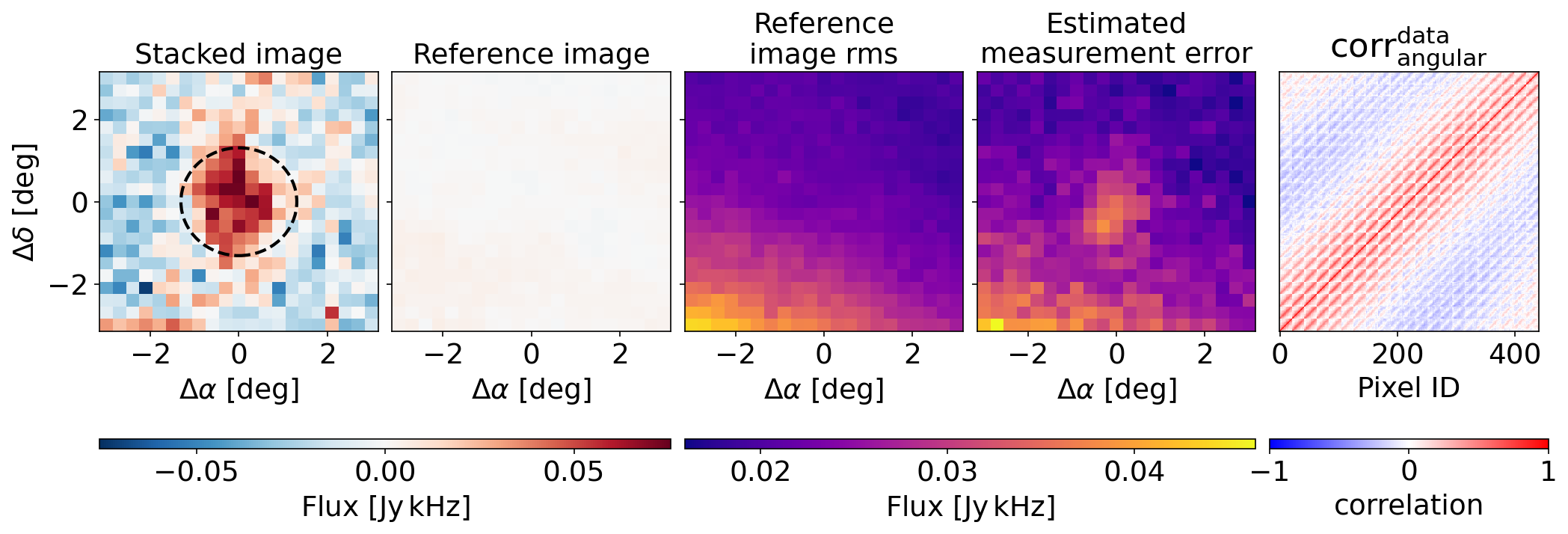}
    \caption{Left panel: the measured angular stacked image. The dashed circle shows the $1.2\,$deg boundary within which we use to calculate the stacked spectrum. Centre-left panel: the average of the reference image across all random shuffles. Centre panel: the reference image standard deviation among the random shuffles. Central-right panel: the measurement error at each pixel in the stacked image, which is also the square root of the diagonal elements of the estimated covariance matrix. Note that the colour scale for the centre and centre-right panels are different from the colour scale for the left and centre-left panels. Right panel: the estimated correlation matrix from mock covariance. The indexing of the array iterates from left to right, and then bottom to top of the stacked image. }
    \label{fig:angularstackdata}
\end{figure*}

\begin{figure}
    \centering
    \includegraphics[width=1.0\linewidth]{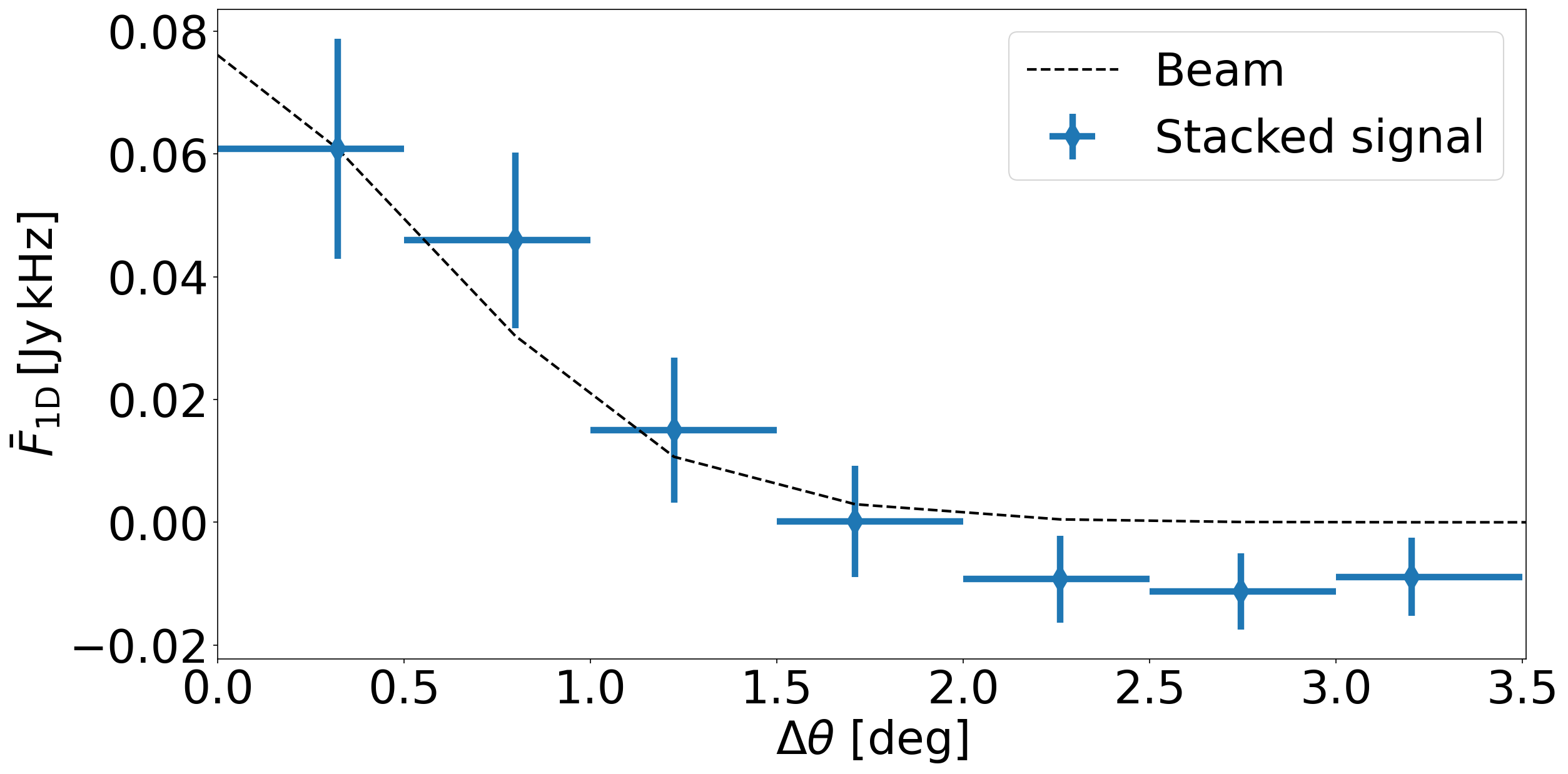}
    \caption{The 1D polar average of the stacked image (``Stacked signal''). The error bars are the standard deviations of the stacked profile among the realisations. For illustration, the primary beam profile matching the amplitude of the first $\Delta\theta$ bin is plotted in black dashed line (``Beam''). }
    \label{fig:angularstackdata1d}
\end{figure}

In this section, we present the measurement of the stacked signal using the MeerKLASS \textit{L}-band deep-field intensity maps.
The intensity maps are cleaned using PCA, removing 10 modes following \citetalias{2025MNRAS.537.3632M}.
We note the difference between the foreground cleaning in \citetalias{2025MNRAS.537.3632M} and this work is that we no longer perform the reconvolution (see Section 4.1 of \citetalias{2025MNRAS.537.3632M} and the discussion in \citet{2024arXiv241206750C}).
We will come back to the effect of reconvolution and its relation to the systematics in \secref{sec:sys}.
We perform the stacked signal estimation following \autoref{eq:stack}, and then collapse the stacked cube into the stacked image and the stacked spectrum following \autoref{eq:angular} and \autoref{eq:spectral}.
The stacked spectrum is subsequently symmetrised.
We then use random shuffling of galaxy positions to perform covariance estimation following \autoref{eq:covcorrected}.

\subsection{Stacked image}
\label{subsec:dataangular}

In \autoref{fig:angularstackdata}, we present the stacked image of the MeerKLASS \textit{L}-band deep-field data over the GAMA galaxies.
The central area of the stacked image shows a clear excess of \hi\ signal, corresponding to the structure of the primary beam blurred by the random thermal noise.
The amplitude of the peak flux is $\sim 0.7\, \rm Jy\,kHz$, slightly smaller than the expected signal level at $\sim 1\rm \,Jy\,kHz$ in our mock simulation shown in \secref{subsec:expsnr}.
This is a combination of the fact that our \hi\ model in the mock simulation is not based on the $z\sim 0.4$ redshift range, and that the data itself possess different foregrounds and systematics that change the PCA cleaning matrix with the same number of modes removed.

The reference image from random shuffling reveals an average that is consistent with zero.
It suggests that there is no overall foreground residual in the data.
Since foreground residuals are uncorrelated with the \hi\ data, any foreground residual would have similar structures in the stacked image and in the reference image.
The reference image, however, does not show any statistically significant structure of excess emission.
The null test using the reference image suggests that the excess in the stacking signal indeed originates from the \hi\ emission of the GAMA galaxies.

Using the random shuffling, we can estimate the covariance of the measurement.
The resulting measurement errors and the correlation matrix are shown in \autoref{fig:angularstackdata}.
The estimated measurement error is consistent with the mock variance we find in \autoref{fig:angularmock}.
It also correctly reproduces the higher variance at the bottom half of the image as we see in the mock.
The correlation matrix, on the other hand, is inconsistent with the mock, as can be seen by the comparison between the rightmost panels of \autoref{fig:angularstackdata} and \autoref{fig:angularmockcov}.
In the mock simulation, the reference image is dominated by noise, which is not related to the primary beam, and therefore there is an inconsistency between the true mock covariance and the estimate as we have shown in \autoref{fig:angularmockcov}.
In the data, however, the estimated covariance produces a strong correlation between the central pixels, seemingly suggesting a convolution between thermal noise and the primary beam, which is not possible.
As we explain later in \secref{sec:sys}, this is due to the chromaticity of the beam affecting the PCA, which then is applied to the entire data vector, which affects the noise as well.

\begin{figure*}
    \centering
    \includegraphics[width=1.0\linewidth]{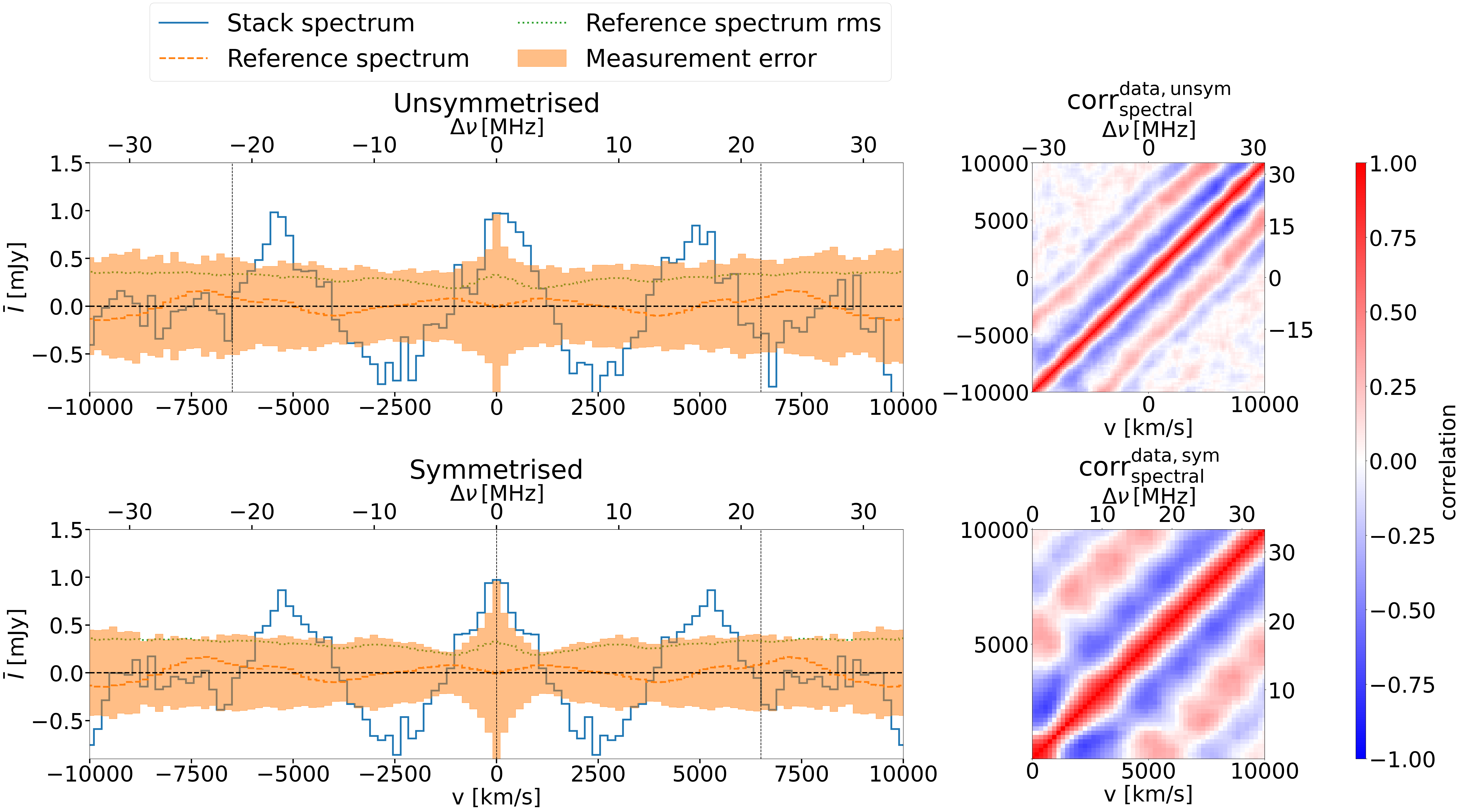}
    \caption{Upper-left panel: the measurement of the stacked spectrum of the MeerKLASS \textit{L}-band deep-field data over the GAMA galaxies (``Stacked spectrum''). The orange dashed line shows the average of the reference spectrum over the random shuffles (``Reference spectrum''). The green dotted line shows the standard deviation of the reference spectrum among the random shuffles (``Reference spectrum rms''). The shaded region shows the estimated measurement error (``Measurement error''), which is also the square root of the diagonal elements of the estimated covariance matrix. The vertical dashed lines show the $|v|<6500\,$km\,s$^{-1}$ boundary within which we use to calculate the detection significance. Upper-right panel: the estimated correlation matrix. Lower-left panel: the same as the upper left panel, but the stacked spectrum is symmetrised. The vertical dashed lines show the $0<v<6500\,$km\,s$^{-1}$ boundary within which we use to perform model inference. Lower-right panel: the estimated correlation matrix for the symmetrised spectrum.}
    \label{fig:spectralstackdata}
\end{figure*}

For reference, we calculate the 1D polar average of the stacked data image and show the results in \autoref{fig:angularstackdata1d}.
The stacked emission follows the attenuation of the primary beam, with most of the detection significance lying within the $1.2\,$deg range.
Therefore, we use the central ($1.2\,$deg)$^2$ area of the stacked image to calculate the detection significance.
Given a data vector $\vec{d}$ and its covariance matrix $\mathbf{C}$, the detection significance can be quantified as
\begin{equation}
    \chi^2 = \vec{d}^{\rm \, \: T} \,\mathbf{C}^{-1}\vec{d}.
\end{equation}
Using the central ($1.2\,$deg)$^2$ area and the estimated covariance, we find a detection significance of 8.66$\sigma$.
The high detection significance suggests that the \hi\ signal is dominant compared to the noise level in the MeerKLASS \textit{L}-band deep-field intensity maps.

\subsection{Stacked spectrum}
\label{subsec:dataspectral}

In \autoref{fig:spectralstackdata}, we present the stacked spectrum of the MeerKLASS \textit{L}-band deep-field data over the GAMA galaxies.
The unsymmetrised stacked spectrum in the upper panel shows clear detection of excess signal relative to the noisy background and the reference spectrum.
The peak amplitude of the excess signal is at $\sim 1\,$mJy which is around the expected level from the mock simulation as shown in \autoref{fig:spectralmock}.
The width of the central peak is $\sim 2000\,$km\,s$^{-1}$, similar to the mock signal.
However, the stacked spectrum also exhibits a clear structure of systematics, shown by the repeated peaks at $\sim -5000\,$km\,s$^{-1}$ and $\sim 5000\,$km\,s$^{-1}$ that are statistically significant against null detection.
Correspondingly, at $\sim \pm 2500\,$km\,s$^{-1}$ there are negative amplitude troughs in the spectrum, clearly reflecting an oscillating structure with an oscillating period of $\sim 5000\,$km\,s$^{-1}$, or equivalently $\sim 15\,$MHz.

The detection of the stacked spectrum indeed originates from the \hi\ signal, as the reference spectrum is consistent with zero.
The standard deviation of the stacked spectrum is at $\sim 0.3 - 0.4\,$mJy, also consistent with the noise level found in the mock simulation, as shown in \autoref{fig:spectralmockcov}.
Similar to the angular stacked image, we find that the estimated covariance matrix does not follow the structure seen in the mock simulation.
Instead, the covariance also shows structures indicating the existence of a component of systematics.

As we have discussed in \secref{sec:covest}, in order to eliminate the need to evaluate correlations between positive and negative values of $\Delta \nu$, we need to use the symmetrised staked spectrum which we show in the lower panel of \autoref{fig:spectralstackdata}.
The symmetrised spectrum and the estimated covariance show similar oscillating structures as seen in the unsymmetrised spectrum.

Using the estimated covariance, we can quantify the detection significance. 
We restrict the stacked spectrum to $|v|<6500\,$km\,s$^{-1}$ for the unsymmetrised spectrum, as there is no signal outside this range, and the detection significance is found to be 7.45$\sigma$.
The significance is slightly lower than the one in the stacked image.
It is worth noting that, a large contribution to the significance is from the peaks and troughs outside the $v \sim 0$ region from the oscillating systematics.
For the symmetrised spectrum, we only need the $v >0$ region and apply the same $v<6500\,$km\,s$^{-1}$ cut.
The detection significance is found to be 5.29$\sigma$.
The decrease is due to the averaging of the spectrum, which symmetrises the spectrum.
If we limit the velocity range to $v<1500\,$km\,s$^{-1}$ where only the central peak is included, the detection significance decreases to $2.43\sigma$.
It suggests that a large part of the information contained in the stacked spectrum is on the systematics in the data.

The detection of the stacked spectrum indicates the depth of the MeerKLASS \textit{L}-band deep-field data.
The peak amplitude of the systematics at $|v| \sim 5000\,$km\,s$^{-1}$ is similar to the \hi\ signal at $v\sim 0$, suggesting that the \hi\ signal and the effect of systematics are around the same order of magnitude.
A similar conclusion can also be found in the auto-power spectrum of the data as shown in Figure 14 of \citetalias{2025MNRAS.537.3632M}.
While the presence of the systematics is still large, it is no longer dominant in the observed signal.
If we can understand the origin of the oscillating systematics, we can parameterise and model its effects, allowing for the inference of the systematics as well as the \hi\ model.

\section{Nature of systematics}
\label{sec:sys}
In this section, we demonstrate evidence that the oscillating systematics seen in the data are a convolutional effect caused by the diffraction of the secondary reflector of the MeerKAT telescope, which affects the chromaticity of the primary beam.
We then investigate the parameterisation and the modelling of the stacked \hi\ signal.

\subsection{Additive and multiplicative systematics}
\label{subsec:addandmulti}
In general, for a data vector of summary statistics $\vec{d}$, the effect of systematics can be written as two components (e.g. \citealt{2021MNRAS.503.5061W}),
\begin{equation}
    \vec{d} = \sum_i \mathbf{S}_{\rm M}^i \vec{d}_i + \vec{S}_{\rm A},
\end{equation}
where $i = \{{\rm \hi,\,n,\,fg}\}$ represents different components of the signal including the \hi, noise and foregrounds, respectively, $\vec{d}_i$ is the underlying uncontaminated signal, $\mathbf{S}_{\rm M}^i$ is the multiplicative systematics matrix, and the additive systematics $\vec{S}_{\rm A}$ is added as an additional component.

The systematics have different origins for different tracers of the LSS.
For example, in galaxy clustering surveys, the additive systematics can be induced by interlopers (e.g. \citealt{2016PASJ...68...12P}), and the multiplicative systematics can be induced by source blending (e.g. \citealt{2021NatRP...3..712M}).
Intensity mapping surveys can be contaminated by a number of sources of systematics.
Residual RFI contamination can be present in the data \citep{2025MNRAS.536.1035E}, leading to additive systematics.
Calibration errors due to insufficient modelling of the sky lead to multiplicative systematics as well as additive residual foreground leakage \citep{2016MNRAS.461.3135B}.
It is therefore important to first determine the type of systematics that contributes the most to the MeerKLASS data.

We note that, since the foregrounds and RFI are not of cosmological origin, the additive systematics should not correlate with the positions of the GAMA galaxies.
If additive systematics have significant contributions to the stacked signal, the oscillating features seen in \autoref{fig:spectralstackdata} should be present both in the signal and the reference spectrum.
This is not the case, as the reference spectrum and the reference image are consistent with null detection.
Therefore, the oscillating systematics must be a multiplicative component applied to the signal data vector.

\subsection{Evidence of convolutional systematics}
\label{subsec:convsys}
The nature of the multiplicative systematics can be split into two categories.
Effects such as bandpass errors are multiplied to the data vector at each pixel without convolving the signal.
On the other hand, effects such as beam chromaticity convolve the data vector in the angular plane and also create structures in the spectral direction.
The distinction between the two effects can be seen in the covariance of the data,
\begin{equation}
\begin{split}
    \mathbf{C} &= \langle \vec{d}\,\vec{d}^{\rm \:T} \rangle 
    = \sum_i \mathbf{S}_{\rm M}^i \langle \vec{d}_i\,\vec{d}^{\rm \:T}_i \rangle  (\mathbf{S}_{\rm M}^i)^{\rm T}\\
    &= \sum_i \mathbf{S}_{\rm M}^i \mathbf{C}_i (\mathbf{S}_{\rm M}^i)^{\rm T},
\end{split}
\end{equation}
where we have omitted the additive systematics.

For a data component $\vec{d}_i$ and a multiplicative systematics matrix $\mathbf{S}_i$, effects such as bandpass errors lead to a diagonal $\mathbf{S}_i$ matrix whereas convolutional effects lead to a nondiagonal matrix.
It is easy to see that, when $\mathbf{S}_i$ is diagonal, $\mathbf{S}_{\rm M}^i \mathbf{C}_i (\mathbf{S}_{\rm M}^i)^{\rm T}$ produces the same correlation matrix as the underlying covariance $\mathbf{C}_i $.
As a result, we expect that the estimated correlation matrix from the data, shown in \autoref{fig:angularstackdata} and \autoref{fig:spectralstackdata}, should be similar to the correlation matrix in the mock simulations.
As discussed in \secref{sec:analysis}, this is not the case, as the effects of the systematics can be clearly seen in the correlation matrices.
Therefore, the systematics must be a convolutional effect on the data.
This also aligns with the oscillation feature seen in the stacked spectrum, which requires a convolution of the oscillation feature with the \hi\ emission-line peaks along the spectral direction.

The existence of convolutional systematics hints toward the connection between the oscillations in the stacked spectrum and the instrument beam.
Beam chromaticity is known to be a limiting factor for intensity mapping surveys (e.g. \citealt{2024arXiv241209527S}), introducing contamination in the foreground removal procedure \citep{2021MNRAS.506.5075M,2022MNRAS.509.2048S}.
To confirm the connection, we examine the stacked signal with reconvolved intensity maps.
Instead of directly stacking on the data, we first perform a convolution of the intensity map at each frequency channel so that
\begin{equation}
    I_{\rm reconv}(l,m,\nu) = B_{\rm reconv}(l,m,\nu) \circledast I(l,m,\nu),
\end{equation}
where the reconvolution kernel, $B_{\rm reconv}(l,m,\nu)$, deconvolves a frequency-dependent Gaussian beam and then convolves the map to a common Gaussian kernel so that
\begin{equation}
    B_{\rm reconv}(l,m,\nu) = {\rm exp}\bigg[-\frac{1}{2}  \frac{\theta^2}{\gamma\sigma_{\rm max}^2 - \sigma(\nu)^2} \bigg],
\end{equation}
where $\sigma(\nu)$ is the beam size of the MeerKAT telescope assuming a Gaussian beam at each frequency, $\sigma_{\rm max}$ is the maximum beam size in the frequency sub-band, and $\gamma$ is a scaling factor to scale down the final resolution.
We follow \citetalias{2025MNRAS.537.3632M} to calculate the reconvolution kernel, and perform the PCA cleaning and stacking with the reconvolved maps.

\begin{figure}
    \centering
    \includegraphics[width=1.0\linewidth]{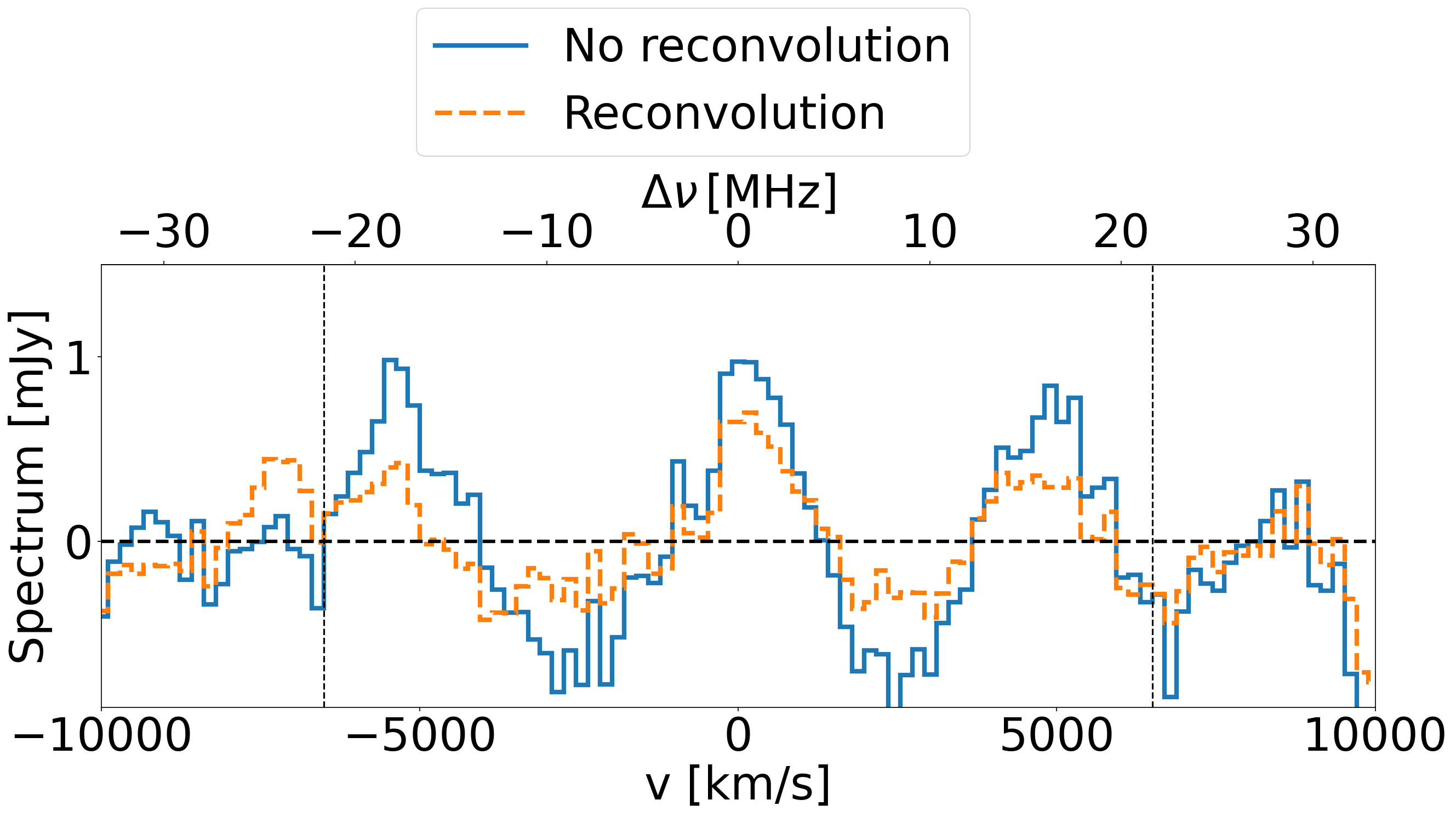}
    \caption{The comparison between the stacked spectra with and without the reconvolution. The blue solid line shows the stacked spectrum without the reconvolution (``No reconvolution''). The orange dashed line shows the spectrum with the reconvolution (``Reconvolution''). The vertical dashed lines show the $|v|<6500\,$km\,s$^{-1}$ boundary.}
    \label{fig:reconvolve}
\end{figure}

In \autoref{fig:reconvolve}, we show the comparison of the stacked spectra with and without the reconvolution.
For simplicity, we only showcase the unsymmetrised spectrum.
When reconvolved, the stacked signal exhibits an overall decrease in amplitude.
This is expected, since smoothing the maps to a lower resolution attenuates fluctuations at small scales.
More importantly, there is a visible decrease in the amplitude of the systematics relative to the central peak.
The oscillation structure is also less localised, as seen in the $v\sim -5000\,$km\,s$^{-1}$ region.
The fact that resmoothing the maps affects the oscillation features of the systematics suggests the systematics are related to the chromaticity of the primary beam.
Reconvolution partially eliminates the frequency-dependency of the map resolution, which is the incentive of performing the reconvolution in the power spectrum analysis.
The stacking measurement suggests that while there is a small effect of mitigating the systematics from the reconvolution, the systematics are still significant and become harder to describe as the oscillations are less localised (see also Appendix B of \citealt{2021MNRAS.506.5075M}).

\subsection{Beam oscillations}
\label{subsec:beamosci}
In \citet{2021MNRAS.506.5075M}, it is found that the contamination after foreground cleaning can be caused by the frequency ripple in the primary beam of the MeerKAT telescope.
The primary beam size of MeerKAT oscillates in frequency, due to the diffractive interference between the secondary and the primary reflector of the dish \citep{6410347}.
Measurements of the beam in \citet{2021MNRAS.502.2970A} show that it leads to a small modulation in the supposedly smooth dependency on frequency.
It is then natural to speculate that the ripple in the beam leads to the oscillations in the stacked spectrum.

\begin{figure}
    \centering
    \includegraphics[width=1.0\linewidth]{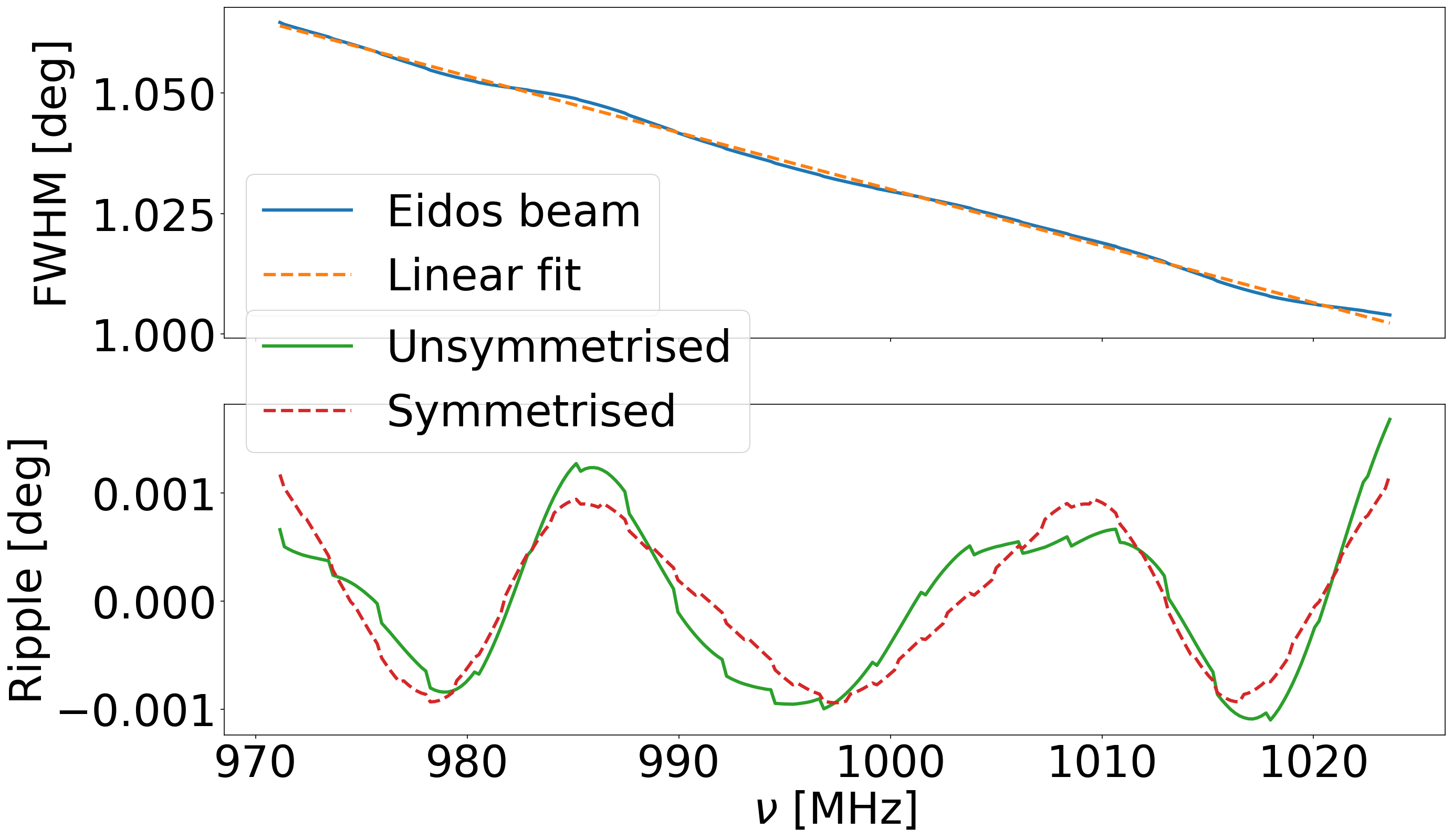}
    \caption{Upper panel: the blue solid line shows the FWHM of the MeerKAT primary beam across the frequency sub-band from the \textsc{eidos} model (``Eidos beam''). The orange dashed line shows a linear best fit of the FWHM (``Linear fit''). Lower panel: the green solid line shows the differences between the FWHM and its linear fit (``Unsymmetrised''), which represents the ripple in the primary beam. The red dashed line shows the symmetrised beam ripple (``Symmetrised'') according to \autoref{eq:symbeam}, which is used for forward modelling.}
    \label{fig:beamosci}
\end{figure}

In \autoref{fig:beamosci}, we use the \textsc{eidos} beam model to calculate the ripple in the primary beam.
Following the procedure in \citet{2021MNRAS.502.2970A}, we first calculate the beam area using only the primary beam, which we choose to be where the beam is larger than 0.1.
We then use the primary beam area to calculate the effective FWHM of the primary beam.
The primary beam FWHM indeed exhibits a small ripple, as seen in the upper panel of \autoref{fig:beamosci}.
We then fit the smooth component using a linear best fit, and subtract the smooth part out to obtain the ripple.
The ripple shown in the lower panel exhibits an oscillating feature, similar to the one seen in the stacked spectrum.

\begin{figure}
    \centering
    \includegraphics[width=1.0\linewidth]{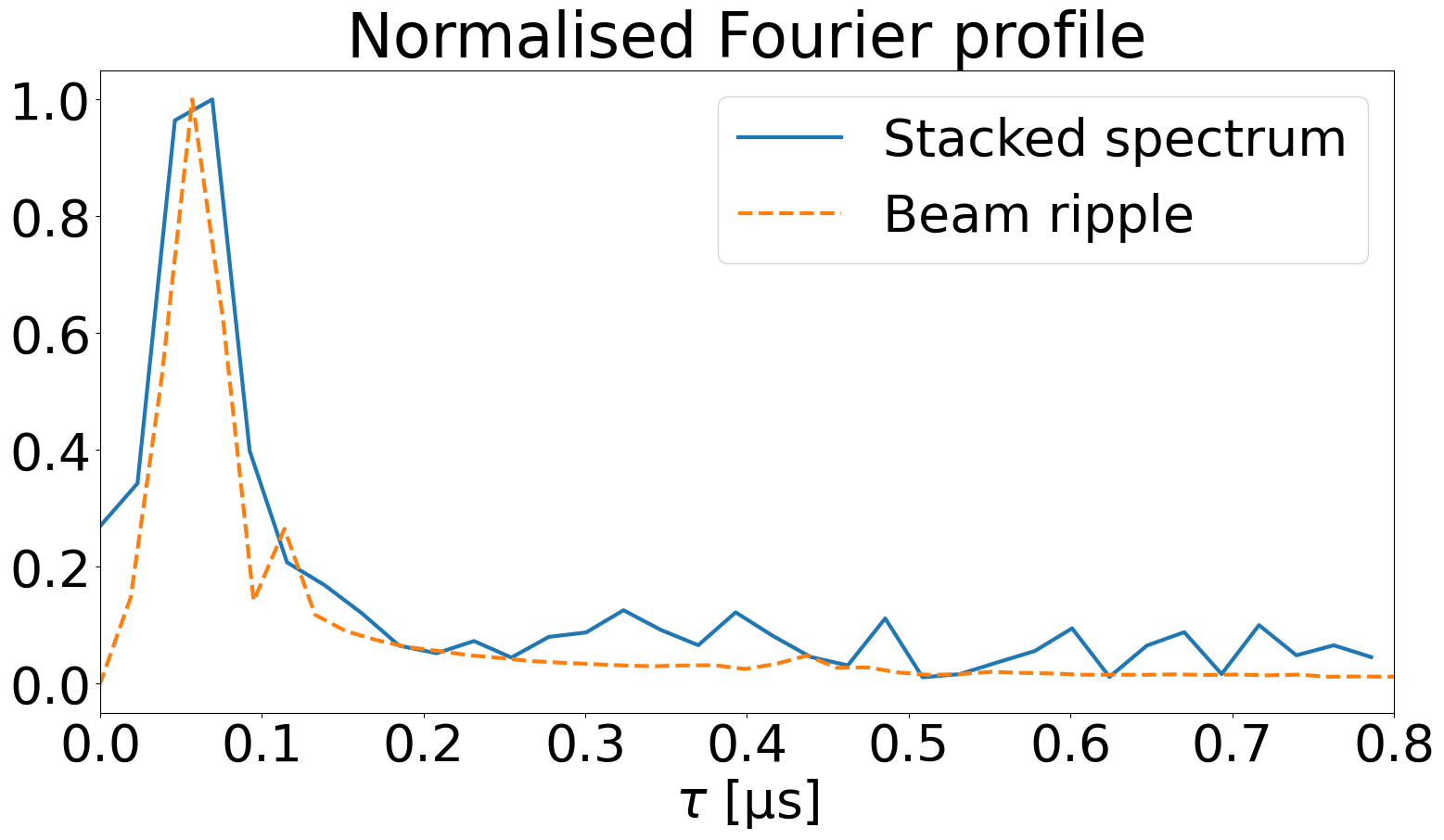}
    \caption{The blue solid line shows the Fourier transformed profile of the unsymmetrised stacked spectrum (``Stacked spectrum''). The orange dashed line shows the Fourier profile of the beam ripple (``Beam ripple''). The profiles are normalised so that the maximum amplitude is 1.}
    \label{fig:fourierosci}
\end{figure}

To further illustrate the connection, we show that the characteristic frequency scale of the oscillation and the beam ripple is the same.
We use the Fourier transform along the frequency direction, defined as
\begin{equation}
    \tilde{f}(\tau) = \int {\rm d}\nu \,{\rm exp}[-2\pi i \tau \nu] f(\nu),
\end{equation}
where $f(\nu)$ is a function of frequency, and $\tau$ is the Fourier pair of $\nu$.
We then perform the Fourier transform for the beam ripple shown in \autoref{fig:beamosci} and the unsymmetrised stacked profile shown in \autoref{fig:spectralstackdata}.
The transformed property is then rescaled, so that the maximum amplitude in Fourier space is 1.
The results are shown in \autoref{fig:fourierosci}.
The beam ripple and the stacked profile share the same peak position centred around $\sim 0.05\,\mu$s, giving an oscillating frequency of $\sim 20\,$MHz.
Both the position of the peak and the width of the peak overlap closely, providing strong evidence that the beam ripple and the oscillating systematics have the same origin.
The Fourier transform of the stacked profile is similar to the line-of-sight power spectrum often used in intensity mapping studies.
For example, in \citet{2021MNRAS.506.5075M}, it is also found that insufficient cleaning of foregrounds in the presence of beam ripple can lead to a peak in the line-of-sight power spectrum.

In conclusion, the oscillating systematics originate from the chromatic ripple in the primary beam of the MeerKAT telescope.
We note that, however, the interaction between the beam ripple and the stacked signal is not direct.
In \secref{sec:validation}, we use the \textsc{eidos} beam, which includes the beam ripple, and we find no systematics in the stacked signal in the mock.
If the systematics in the data are simply caused by the beam convolution, the noise-dominated reference spectrum of the data should not have the systematics.
Yet, in the shuffling covariance of the data, we observe the oscillating systematics when visualising the correlation matrix in \autoref{fig:spectralstackdata}.
It is worth noting that the beam model affects the modelling of the sky signal for calibration, as described in \secref{sec:data}.
The calibration solution is then affected by the systematics, leading to chromatic calibration errors that are commonly seen in the case of imperfect sky model (e.g. \citealt{2016MNRAS.461.3135B,2020MNRAS.494.5018H}).
The calibration errors serve as a multiplicative effect on the sky signal, which then affects the PCA cleaning of the data.
The PCA cleaning matrix, as described in \autoref{eq:rpca}, is then applied to the entire data vector including the noise.
In conclusion, the systematic oscillation in the stacked spectrum is introduced by the PCA, whose cleaning matrix is modulated by the MeerKAT beam ripple.
An end-to-end study of the systematics from the calibration of the time-ordered data is beyond the scope of this work.
Instead, we use the connection between the beam ripple and the systematics as a starting point for forward modelling the signal.

\section{Model fitting}
\label{sec:fit}
In this section, we describe the model inference framework we use in this work to constrain the systematics and the \hi\ signal.

\subsection{Forward modelling}
\label{subsec:forward}
As we have extensively discussed in \secref{subsec:biasing}, in the forward modelling we use the simplified assumption and distribute the total \hi\ mass inside the GAMA survey region to the GAMA galaxies.
Therefore, we can simulate the \hi\ signal with one free parameter $\bar{M}_{\hi}$ following the procedure described in \secref{subsec:hisim}.
We briefly review the procedure below.

First, we generate the positions of galaxies using the lognormal simulation routine.
The number density is set so that the expected number of galaxies within the GAMA region is equal to the catalogue.
Note that in each realisation, the number of galaxies is not equal to the number of galaxies in the GAMA catalogue due to assigning Poisson random to the mock dark matter field.
This is desired, as the modelling is supposed to reflect the Poisson fluctuations.

We then assign a uniform \hi\ mass, $\bar{M}_{\hi}$, to each galaxy and generate the \hi\ profile.
The \hi\ signal is then convolved with the beam model to produce the \hi\ map.
We then apply a modelling of the oscillating systematics, $f_{\rm sys}(\nu)$, to the map so that in each pixel the \hi\ signal is convolved with the systematics,
\begin{equation}
    I_{\rm sys}(\alpha,\phi,\nu) = \sum_{\Delta\nu_i=\nu-\nu_0}^{\nu_1-\nu} f_{\rm sys}(\Delta\nu_i + \nu_0)I_{\rm \hi}(\alpha,\phi,\nu-\Delta\nu_i),
\end{equation}
where $[\nu_0,\nu_1]$ are the lower and upper limits of the frequency sub-band, $I_{\rm \hi}$ is the \hi\ map before applying systematics, and $\Delta\nu_i$ iterates over a step size of frequency channel bandwidth.
The parameterisation of $f_{\rm sys}(\nu)$ is discussed later in detail in \secref{subsec:beampars}.

The \hi\ signal with systematics is then cleaned by applying the PCA cleaning matrix \emph{calculated from data}.
This is to ensure that the model has the same level of signal loss as the data.
As we have shown in \secref{subsec:pca}, for the \hi-only mock and the full mock, the level of signal loss from PCA is very different due to the change in the removed modes.
Since we do not have prior information on the exact level of the oscillating systematics, and the fact that the foreground model and the data have a mismatch, the PCA cleaning matrix from the mock simulation is different from the data, leading to different signal loss properties.
Therefore, applying the PCA cleaning matrix from data is important to keep the signal loss consistent.

The \hi\ map and the GAMA-like catalogue are then used to perform stacking, using the same weighting as the data.
The stacked signal is then used for model fitting.

\subsection{Parameterising the beam oscillation}
\label{subsec:beampars}
We first describe the parameterisation of the beam oscillation, which we use to describe the systematics.
As shown in \autoref{fig:fourierosci}, the main feature of the beam oscillation is the peak structure in Fourier space.
The systematics in Fourier space can then be parameterised to reflect the peak structure, so that
\begin{equation}
    |\tilde{f}_{\rm sys}(\tau)|^2 =  {\rm exp}\big[-\frac{(\tau-\frac{1}{\nu_{\rm sys}})^2}{2 \sigma_\tau ^2}\big],
\end{equation}
where $1/\nu_{\rm sys}$ is the position of the peak corresponding to an oscillating frequency of $\nu_{\rm sys}$ and $\sigma_\tau$ is the width of the peak.

After specifying the parameters $\nu_{\rm sys}$ and $\sigma_\tau$, we then perform an inverse Fourier transform of $\tilde{f}_{\rm sys}(\tau)$. 
$\tilde{f}_{\rm sys}(\tau)$ is assumed to be real-valued, which leads to a symmetric function in real space so that
\begin{equation}
    f(\nu_0 + \Delta \nu) = f(\nu_1 - \Delta\nu),
\label{eq:symbeam}
\end{equation}
where $\nu_0$ and $\nu_1$ are the lower and upper limits of the frequency sub-band, respectively.
Subsequently, the beam ripple discussed in \secref{subsec:beamosci} can be symmetrised in the same way. 

The function in real space is then mean-subtracted, and then rescaled so that the standard deviation of the function in the frequency sub-band is 1.
We then multiply another free factor, $A_{\rm sys}$, so that
\begin{equation}
    {\rm std}(f_{\rm sys}) = A_{\rm sys},
\end{equation}
to denote the amplitude of the systematics.

\begin{figure}
    \centering
    \includegraphics[width=1.0\linewidth]{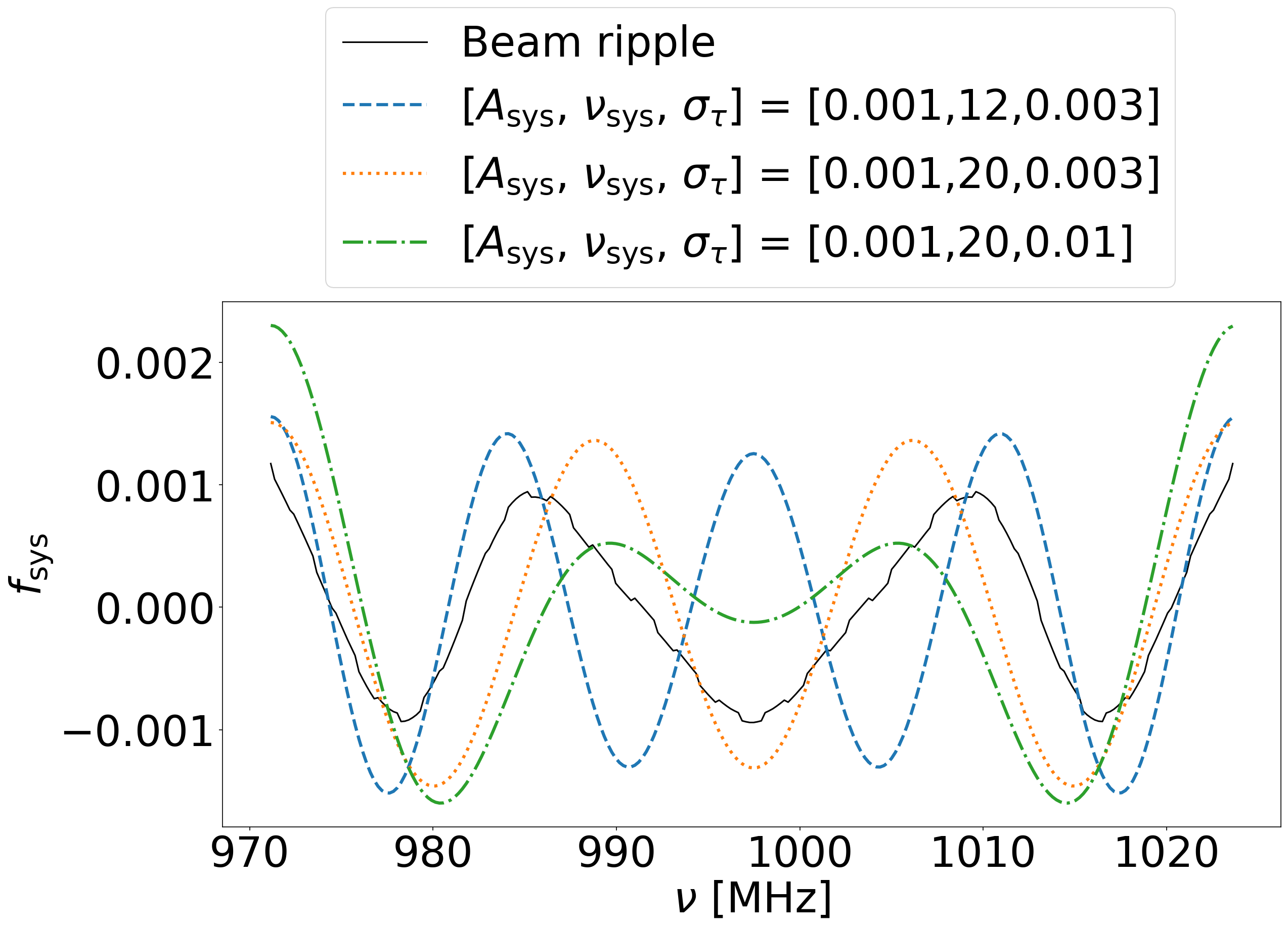}
    \caption{An illustration of the parameterisation of the systematics $f_{\rm sys}$. The black solid line shows the beam ripple. The other three lines show three different sets of parameters for the systematics listed in the figure.}
    \label{fig:oscipars}
\end{figure}

In short, we use three parameters, $\{A_{\rm sys}, \nu_{\rm sys}, \sigma_\tau\}$ to model the systematics to describe the amplitude, characteristic frequency and the shape of the systematics respectively.
An illustration of the parameterisation is shown in \autoref{fig:oscipars}.

In the fitting, to examine the impact of the parameterisation, we adopt three different scenarios.
First, we only vary $A_{\rm sys}$, and keep the shape of the $\tilde{f}_{\rm sys}(\tau)$ fixed to the beam oscillation.
Second, we fix $\sigma_\tau$ to be very small and vary $A_{\rm sys}$ and $\nu_{\rm sys}$, so that the oscillating frequency varies while $f_{\rm sys}$ has the shape of a sine function.
Finally, we allow all three parameters to vary to explore the full parameter space.

We emphasise that, the forward modelling routine used in this work is simplified.
For example, we fix cosmological parameters and other \hi-related parameters and only keep $\bar{M}_{\hi}$ as a free parameter.
The assumption that the systematics $f_{\rm sys}$ directly convolve with the \hi\ signal along the frequency direction is an effective approach to model a much more complex phenomenon that interacts with the data at the level of the time-ordered data.
Given that our detection significance in the symmetrised spectrum is $\sim 5.5\sigma$, we are limited by the number of parameters that we can vary to achieve sensible constraints.
A more detailed study is left for the future MeerKLASS data analysis, where the depth of both the \hi\ intensity maps and the galaxy catalogue will be larger.

\subsection{Importance nested sampling}
\label{subsec:sampling}
The forward modelled \hi\ signal produces a stacked spectrum, $\bar{I}_{\rm model}$ to be used to fit against the data.
Using the estimated covariance $\mathbf{C}$ and the measured symmetrised spectrum $\bar{I}_{\rm data}$, relative changes in the likelihood $\mathcal{L}$ can be calculated as 
\begin{equation}
    \Delta {\rm log}\mathcal{L} = -\frac{1}{2}(\vec{I}_{\rm model}-\vec{I}_{\rm data})\mathbf{C}^{-1}(\vec{I}_{\rm model}-\vec{I}_{\rm data})^{\rm T},
\end{equation}
where log stands for natural logarithm, $\vec{I}_{\rm model}$ is the model vector for the model spectrum at each $\Delta \nu$ and  $\vec{I}_{\rm data}$ is the data vector.

The likelihood is then used to perform Bayesian inference using importance nested sampling (INS; e.g. \citealt{2019OJAp....2E..10F}).
INS explores the prior volume of the parameter space, identifies the region with the highest likelihood, and sample with the posterior for Bayesian inference.
In practice, the algorithm starts with a number of points randomly distributed in the prior volume, finding the live points with the highest likelihood.
Based on the positions of the live points, it estimates a boundary around each point to sample within, and identifies a new iteration of the random points with higher minimum likelihood.
Iterating this process will result in convergence so that the boundary no longer changes for each point.
We use \textsc{nautilus}\footnote{\url{https://nautilus-sampler.readthedocs.io/}}~\citep{2023MNRAS.525.3181L} for performing the sampling.

\begin{table}
    \centering
    \begin{tabular}{cccc}
        ${\rm log}_{10}[\bar{M}_{\rm \hi}/M_\odot]$ & $A_{\rm sys}$ & $\nu_{\rm sys}\,$[MHz] & $\sigma_\tau$\,[$\mu$s] \\ \hline
        [7.0,14.0] & [0.0,1.0] & [10,32] & [0.003,0.04] \\
    \end{tabular}
    \caption{The priors we use for the parameters in this work. All priors are flat priors}
    \label{tab:priors}
\end{table}

The sampling depends on the priors we set for the parameters. Given our lack of understanding of the model parameters, we adopt wide flat priors, which we list in \autoref{tab:priors}.
The prior for ${\rm log}_{10}[\bar{M}_{\rm \hi}/M_\odot] \in [7.0,14.0]$ translates to an effective \hi\ density of $\Omega_{\rm \hi} \in [1.5\times 10^{-8},0.155]$.
Measurements of \hi\ density at various redshifts give the \hi\ density to be $\sim 5\times 10^{-4}$ (see, e.g. Figure 14 of \citealt{2019MNRAS.489.1619H} and references therein), and our prior is significantly looser than the current constraints from observations.
The amplitude of the systematics is sampled from $A_{\rm sys}\in [0,1]$, since values outside this range give unphysical negative values.
The oscillating frequency is sampled from $\nu_{\rm sys}\in [10,32]\,$MHz.
Higher values of the oscillation frequency will not be captured in our $\Delta \nu \lesssim 20\,$MHz range for the stacked spectrum.
Lower values, on the other hand, lead to rapid oscillations that are unphysical given the physical distance between the primary and secondary reflector of the MeerKAT telescope.
Finally, the width of the peak in the Fourier space is sampled from $\sigma_\tau\in [0.003,0.04]\,\mu$s.
Values smaller than 0.003\,$\mu$s are below the resolution of the frequency sub-band we use.
Values larger than 0.04\,$\mu$s will result in a wide peak, so that $f_{\rm sys}$ in frequency space is almost completely flat, which is unphysical.

For each fitting, 2000 live points are used for sampling.
A weighted sample of parameter values is then returned for Bayesian inference, which we discuss in the next section.

\begingroup
\setlength{\tabcolsep}{10pt} 
\renewcommand{\arraystretch}{1.5} 
\begin{table*}[!htbp]
    \centering
    \begin{tabular}{c|cccc|c}
         & $\rm log_{10}[\bar{M}_{HI}/M_\odot]$ & $A_{\rm sys}$ & $\nu_{\rm sys}\,$[MHz] & $\sigma_\tau\,$[$\mu$s] & log$\mathcal{Z}$ \\ \hline
        [$\bar{M}_{\rm HI}$,$A_{\rm sys}$,$\nu_{\rm sys}$,$\sigma_\tau$] & ${9.84}_{-0.59}^{+0.48}(10.24)$ & ${0.50}_{-0.33}^{+0.33}(0.63)$ & ${17.90}_{-4.27}^{+6.53}(16.17)$ & ${0.02}_{-0.01}^{+0.01}(0.004)$ & -14.08 \\
        $[\bar{M}_{\rm HI},A_{\rm sys},\nu_{\rm sys}]$ & ${9.99}_{-0.47}^{+0.44}(10.77)$ & ${0.51}_{-0.34}^{+0.33}(0.10)$ & ${17.40}_{-2.11}^{+2.66}(19.49)$ & / & -13.85 \\
        $[\bar{M}_{\rm HI},A_{\rm sys}]$ & ${10.17}_{-0.56}^{+0.45}(10.17)$ & ${0.50}_{-0.32}^{+0.34}(0.87)$ & / & / & -13.92 \\ \hline
        Shuffling Cov & ${9.93}_{-1.05}^{+0.57}(10.92)$ & ${0.43}_{-0.31}^{+0.37}(0.10)$ & ${17.84}_{-4.11}^{+6.41}(19.30)$ & ${0.02}_{-0.01}^{+0.01}(0.003)$ & -21.31 \\
    \end{tabular}
    \caption{The 68\,\% intervals of the model parameters from fitting the stacked spectrum of the MeerKLASS \textit{L}-band deep-field data onto the GAMA galaxies. Three different parametrisations of the systematics are considered, which vary the full shape of the oscillations (``[$\bar{M}_{\rm HI}$,$A_{\rm sys}$,$\nu_{\rm sys}$,$\sigma_\tau$]''), only the amplitude and the frequency (``[$\bar{M}_{\rm HI}$,$A_{\rm sys}$,$\nu_{\rm sys}$]''), and only the frequency (``[$\bar{M}_{\rm HI}$,$A_{\rm sys}$]''). For reference, the maximum a posteriori estimations of each parameter are listed in the brackets. The last column lists the Bayesian evidence log$\mathcal{Z}$ of each fitting. The last row shows the results using the shuffling estimate of the covariance (``Shuffling Cov'') instead of the corrected covariance estimate for the full shape case. }
    \label{tab:results}
\end{table*}
\endgroup

\section{Results from forward modelling}
\label{sec:results}

In this section, we present the main results of this paper, obtained from the Bayesian inference of the stacked spectrum.
As discussed in \secref{subsec:beampars}, we consider three different parametrisations of the systematics, varying the full shape of the oscillations, only the amplitude and the frequency, and only the amplitude.
In each case, we also treat the \hi\ mass of the galaxies $\bar{M}_{\rm \hi}$ as a free parameter for sampling.
From now on, for simplicity, we will denote each case using their respective number of free parameters, for example the two-parameter case refers to varying $[\bar{M}_{\rm \hi},A_{\rm sys}]$ and fixing the shape of beam oscillation.
Later in \secref{subsec:diffcov}, we discuss the fitting results with the full shape of the systematics but using a shuffling covariance instead of the corrected covariance, which we denote as the shuffling covariance case.

\subsection{Parameter constraints}
\label{subsec:constraint}

\begin{figure*}
    \centering
    \includegraphics[width=1.0\linewidth]{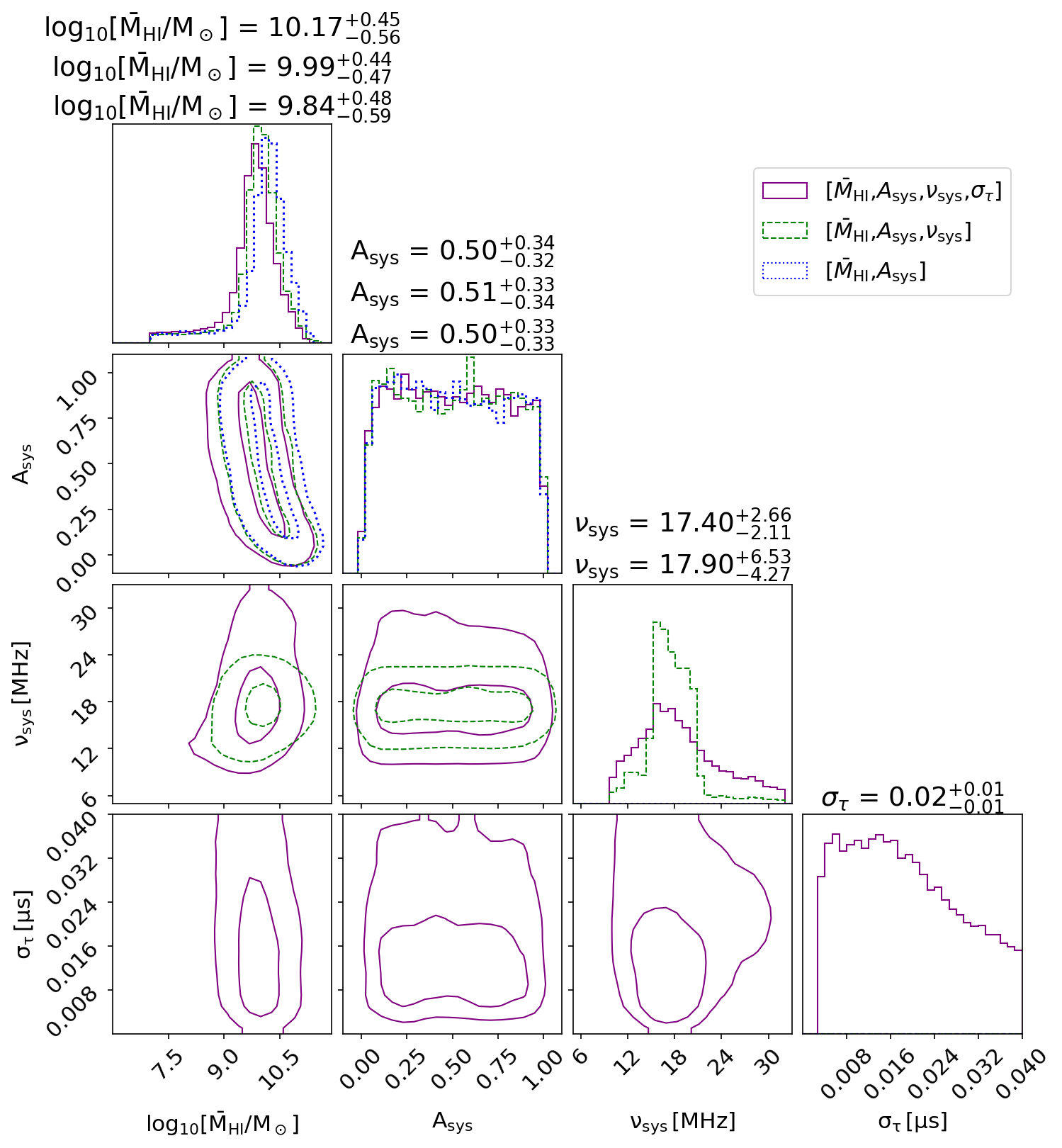}
    \caption{The posterior distribution of the parameter fitting results of this work. The histogram plots show the posterior distribution of each model parameter when marginalising over all other parameters. The contour plots show the marginalised 2D posterior distribution of each parameter pair. The outer contour denotes the $2\sigma$ confidence region, whereas the inner contour denotes the $1\sigma$ confidence region. Three scenarios are shown, with the shape of systematics fixed (``[$\bar{M}_{\rm HI}$,$A_{\rm sys}$] ''), varying the amplitude and the oscillation frequency (``[$\bar{M}_{\rm HI}$,$A_{\rm sys}$,$\nu_{\rm sys}$]'' ), and varying the full shape of the systematics ([$\bar{M}_{\rm HI}$,$A_{\rm sys}$,$\nu_{\rm sys}$,$\sigma_\tau$]). The title of each histogram plot shows the median and the $68\,\%$ interval. From top to bottom, the results shown are for the two-parameter, three-parameter, four-parameter cases, respectively.}
    \label{fig:posterior}
\end{figure*}

Using the posterior obtained from INS, we compute the 1$\sigma$ confidence interval, i.e. the 16\%, 50\%, and the 84\% percentiles, of each parameter and show the results in \autoref{tab:results}.
The posterior is also used to visualise the 2D posterior of the parameters as shown in \autoref{fig:posterior}.
For all cases of model fitting, we obtain consistent constraints on $\bar{M}_{\rm \hi}$ with the results from all three scenarios, with their differences smaller than the $1\sigma$ confidence interval.
As the number of free parameters increases for describing the systematics, the amplitude of the \hi\ emission, $\bar{M}_{\rm \hi}$, decreases.
The consistent shifts of the amplitude of \hi\ suggests that the modelling of the systematics may impact the inference of the \hi\ signal.

The amplitude of the systematics, $A_{\rm sys}$, is not constrained as the 68\,\% interval of the posterior extends to almost the entire prior volume.
While we have not achieved a constraint on the amplitude of the systematics, we note that $A_{\rm sys}$ is likely higher than expected.
For example, for the four-parameter case, the 5\,\% percentile of the posterior gives $A_{\rm sys}>6.71\,\%$.
The beam oscillation, on the other hand, is $\sim 0.1\,\%$ as shown in \autoref{fig:beamosci}.
This confirms our previous discussion in \secref{subsec:beamosci}, that the way that systematics are coupled to the data is not a trivial convolution of the primary beam and worth further investigation in future work.

Comparing the two-parameter and the three-parameter case, we can see that varying the oscillation frequency does not lead to larger measurement error but in fact a smaller $68\,\%$ interval for $\bar{M}_{\rm \hi}$.
This suggests that, while the beam oscillation does match the systematics in the stacked spectrum well, the exact oscillation frequency may be different from the measurement from the \textsc{eidos} beam.
The improvement in the fitting can also be seen in the small increase in Bayesian evidence log$\mathcal{Z}$.
The slight mismatch is expected, since the frequency of the beam ripple of the telescope has a dependency on elevation, and the effective frequency of the oscillation may be different from the measurements using one night of tracking observation in  \citet{2021MNRAS.502.2970A}.

When we vary the full shape of the systematics function in the four-parameter case, it can be seen that the constraints on the oscillation frequency $\nu_{\rm sys}$ degrade significantly.
When $\sigma_\tau$ is fixed so that the systematics follow the shape of a sine function, the constraint on $\nu_{\rm sys}$ gives $\nu_{\rm sys} = {17.40}_{-2.11}^{+2.66}\,$MHz.
In the four-parameter case, however, the measurement error increases by a factor of $\sim 2$ which gives $\nu_{\rm sys}={17.90}_{-4.27}^{+6.53}\,$MHz.
$\sigma_\tau$ is not well constrained, as the $68\,\%$ interval occupies a large part of the prior volume.
The increase in the errors is likely due to the fact that the stacked spectrum is measured in relatively low frequency resolution, and therefore cannot be used to describe the shape of the oscillations in detail.
There is also a decrease in Bayesian evidence log$\mathcal{Z}$.
This suggests that a small value of $\sigma_\tau$, which makes the oscillations following a sine function, is good enough for modelling the stacked signal.
Nevertheless, we adopt the more conservative estimation in the four-parameter case as our final results.
The measured oscillation frequency $\nu_{\rm sys}\approx 18\,$MHz is consistent with the $\sim 20\,$MHz beam ripple discussed in the literature.

We now further examine the constraints of the model parameters in terms of the degeneracy between parameters, as shown in \autoref{fig:posterior}.
For the \hi\ amplitude $\bar{M}_{\rm \hi}$, we can see that the distribution of the 1D posterior is well constrained, with an extended tail at the lower end of the distribution.
We discuss the implications of the $\bar{M}_{\rm \hi}$ posterior in more detail later in \secref{subsec:projection}.
The amplitude of the systematics, $A_{\rm sys}$, is indeed not constrained, as the posterior simply extends throughout the flat prior.
When the three-parameter model is considered, the oscillation frequency $\nu_{\rm sys}$ is well constrained.
In the four-parameter case, on the other hand, the tails of the posterior reaches the physically driven prior.
This is caused by the posterior of $\sigma_\tau$ not being constrained. 

Finally, we comment on the fact that the estimated $\bar{M}_{\rm \hi}$ is much lower than expected.
Note that, as we have discussed in \secref{subsec:biasing}, the values of $\bar{M}_{\rm \hi}$ should not be interpreted as the average \hi\ mass of the GAMA galaxies, but as the total \hi\ mass in the GAMA survey region distributed among the GAMA sample.
For $\Omega_{\hi} \sim 0.5\times10^{-3}$, we expect $\bar{M}_{\rm \hi} \sim 10^{11}M_\odot$, and our estimation is an order of magnitude lower than expected.
This suggests that there may be issues in the model fitting, and while there is a tentative measurement of the \hi\ density, the estimation is likely to be biased.
For the rest of this section, we examine issues of covariance estimation and parameter degeneracy that contribute to the underestimation.

\subsection{Impact of covariance estimation}
\label{subsec:diffcov}
In \secref{sec:covest}, we discussed in detail how we obtain the covariance estimation using the random shuffling of galaxy positions.
The covariance estimation, while corrected for signal covariance, leads to a distortion compared to the true covariance, which may impact the inference of the systematics, as shown in \appref{apdx:covanalytic}.
While a more accurate covariance estimation is beyond the scope of this work, we can use the shuffling covariance without the correction to perform the sampling and compare the results to understand the impact of covariance estimation.

In the bottom row of \autoref{tab:results}, we show the four-parameter case with shuffling covariance.
Note that, without the signal covariance correction, the measurement error of the stacked spectrum is lower, as seen in \autoref{fig:spectralstackdata}.
However, the resulting measurement error for $\bar{M}_{\rm \hi}$ becomes larger, and the Bayesian evidence log$\mathcal{Z}$ decreases significantly compared to the corrected covariance.
This suggests that indeed a correction to the shuffling covariance is needed.
The constraints on the oscillation frequency, on the other hand, are robust against the choice of systematics.
This is expected, as the primary feature of the stacked spectrum is the oscillation, so that the information in the stacked spectrum mostly goes to constraining $\nu_{\rm sys}$.
Furthermore, the measured maximum a posteriori (MAP) values of the systematics are lower when the shuffling covariance is used, while the \hi\ amplitude is larger.

\begin{figure*}
    \centering
    \includegraphics[width=1.0\linewidth]{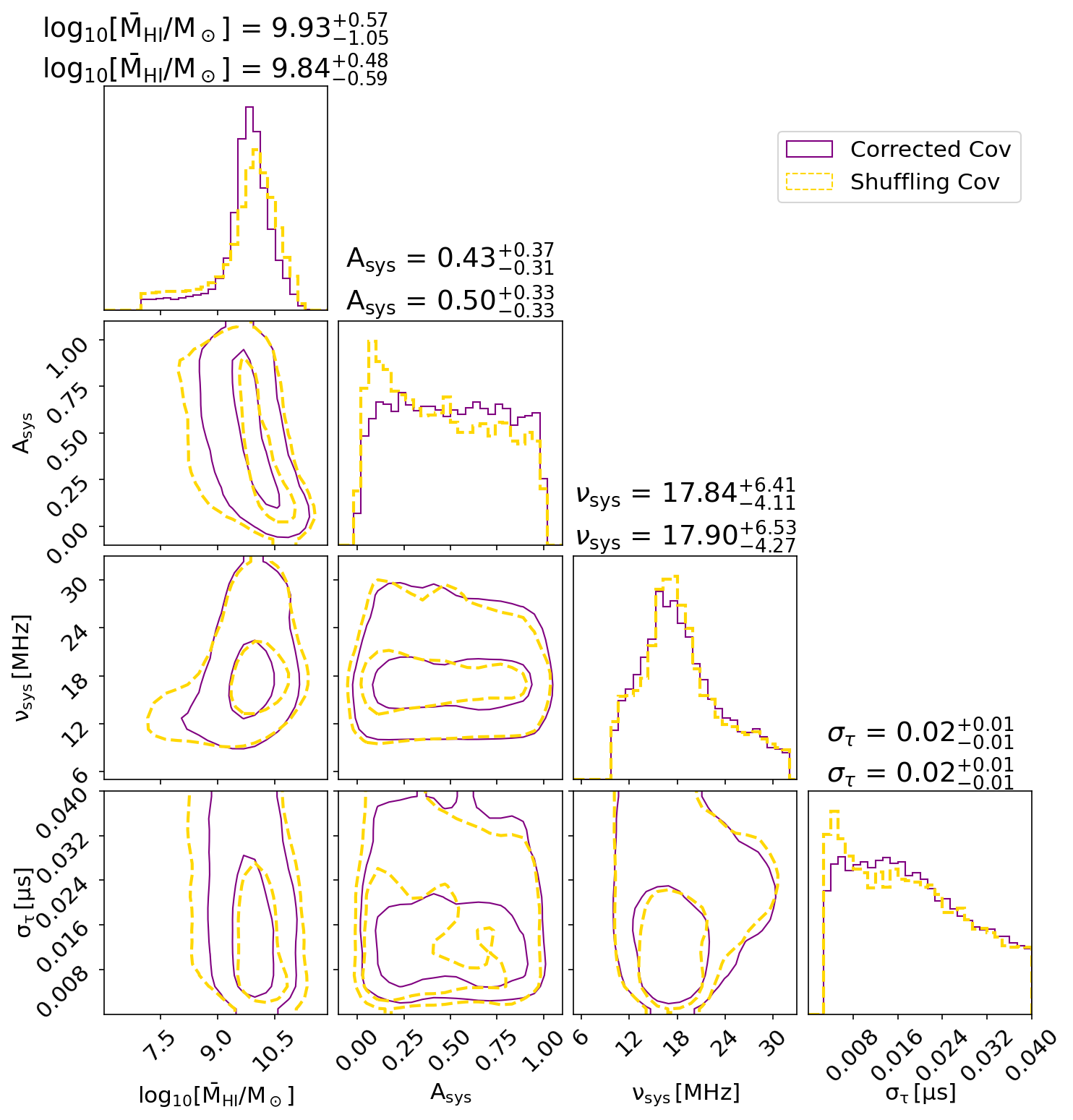}
    \caption{The same as \autoref{fig:posterior}, but with cases of corrected covariance and shuffling covariance for the four-parameter model. From top to bottom, the titles shown are for the shuffling covariance and the corrected covariance, respectively.}
    \label{fig:postdiffcov}
\end{figure*}

To further illustrate the effect of covariance estimation, we show the 2D posterior distribution of the model parameters for the shuffling covariance compared against the corrected covariance in \autoref{fig:postdiffcov}.
Notably, the posterior for $A_{\rm sys}$ changes significantly, leading to a peak at $A_{\rm sys}\sim0.15$.
As the amplitude of the systematics decreases, we note that there is a turn in the 2D posterior distribution of $A_{\rm sys} - \bar{M}_{\rm \hi}$,  leading to a higher estimation of $\bar{M}_{\rm \hi}$.
Similarly, a peak around small values of $\sigma_\tau$ also appears when shuffling covariance is used.

The comparison between the two choices of covariance shows that the constraints on systematics are affected by the covariance estimation.
When the amplitude parameter reaches small values $A_{\rm sys}\lesssim 0.1$, there is a stronger anticorrelation between the systematics amplitude and the \hi\ mass.
We note that this leads to a larger estimation of  $\bar{M}_{\rm \hi}$, which suggests that covariance estimation impacts the underestimation of the \hi\ density.

\subsection{Projection effects}
\label{subsec:projection}

\begin{figure}
    \centering
    \includegraphics[width=1.0\linewidth]{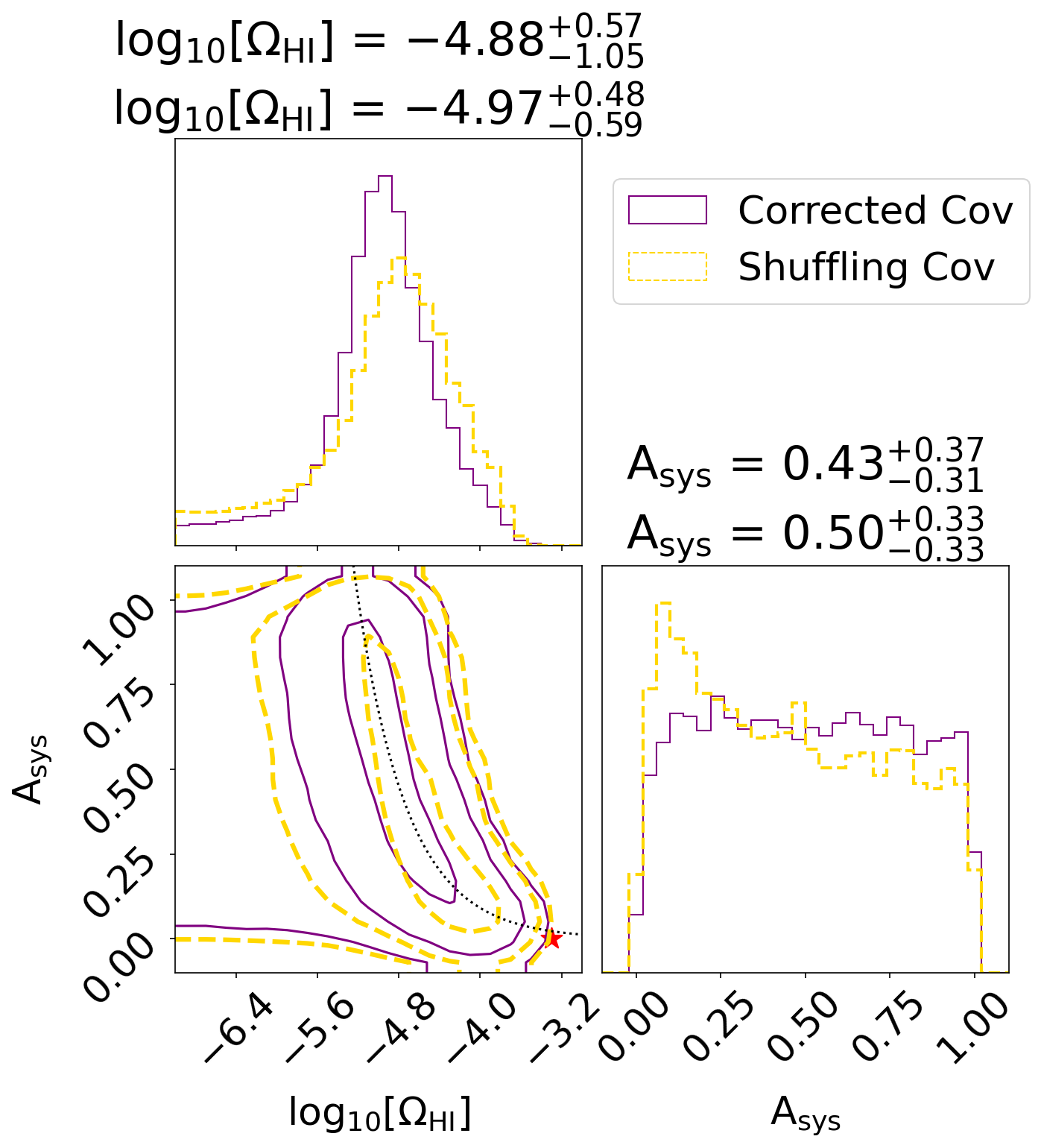}
    \caption{The 2D posterior distribution between the inferred \hi\ density $\Omega_{\hi}$ and the amplitude of the systematics, for the cases with corrected covariance and shuffling covariance of the four-parameter model. The outer, middle and inner contours denote the $3\sigma$, $2\sigma$, and $1\sigma$ regions. From top to bottom, the titles shown are for the shuffling covariance and the corrected covariance, respectively. The dotted line in the 2D posterior shows an illustrative direction of parameter degeneracy. The red star denotes $\Omega_{\rm \hi} = 0.5\times 10^{-3},\,A_{\rm sys} = 0.0$.}
    \label{fig:projection}
\end{figure}

We further explore the degeneracy between $\bar{M}_{\rm \hi}$ and $A_{\rm sys}$.
Since we are interested in the underestimation of $\Omega_{\rm \hi}$, we convert the posterior of $\bar{M}_{\rm \hi}$ to $\Omega_{\rm \hi}$ and show the results in \autoref{fig:projection}.
We include the contour of the $3\sigma$ region to fully visualise the posterior.
As shown, the degeneracy direction between the two parameters has a turn, illustrated by the black dotted line in  \autoref{fig:projection}.
When the systematics amplitude is large with $A_{\rm sys}\gtrsim 0.2$, an increase in $\Omega_{\hi}$ leads to a sharp decrease in $A_{\rm sys}$.
As $A_{\rm sys}$ further decreases, the \hi\ density increases significantly.
In the ideal case of a stacking experiment, we expect that no systematics are present so that $\Omega_{\rm \hi} \sim 0.5\times 10^{-3}$, $A_{\rm sys}=0$ which we denote as the red star.
The $\Omega_{\rm \hi} \sim 5\times 10^{-4}$ value lies slightly outside the $3\sigma$ contour of the posterior, suggesting that there is likely a modelling imperfection that leads to the underestimation of the \hi\ density.

Furthermore, the degeneracy between the two parameters leads to strong posterior projection effects (e.g. \citealt{2022PhRvD.106f3506G}) in the model inference.
``Posterior projection effect;; refers to the issue that the marginalised 1D distribution of the posterior may be skewed due to the complicated degeneracy between the model parameters.
We first demonstrate that the 1D posterior distribution is indeed skewed.
In the brackets of the reported values of \autoref{tab:results}, we show the MAP estimation\footnote{Since we adopt flat priors for all model parameters, in our case, the MAP estimation is simply the maximum of the 1D posterior distribution.} of the parameters.
If the 1D posterior of the parameters follow ideal Gaussian distributions, it is expected that the MAP estimation and the median of the posterior should agree well with each other.
In all cases, except the two-parameter model, we find the systematic shift of \hi\ mass to higher values for the MAP estimation when comparing the median of the posterior.
In particular, there is a $>1\sigma$ discrepancy between the MAP and the median of $\bar{M}_{\rm \hi}$ for the three-parameter case, as well as in the shuffling covariance case.
It suggests that the skewed distribution of the marginalised posterior contributes to the underestimation of the \hi\ density. 
The skewing is consistent with the degeneracy direction between $A_{\rm sys}$ and $\bar{M}_{\rm \hi}$, and, therefore, is due to projection effects caused by parameter degeneracy.

\begin{figure}
    \centering
    \includegraphics[width=1.0\linewidth]{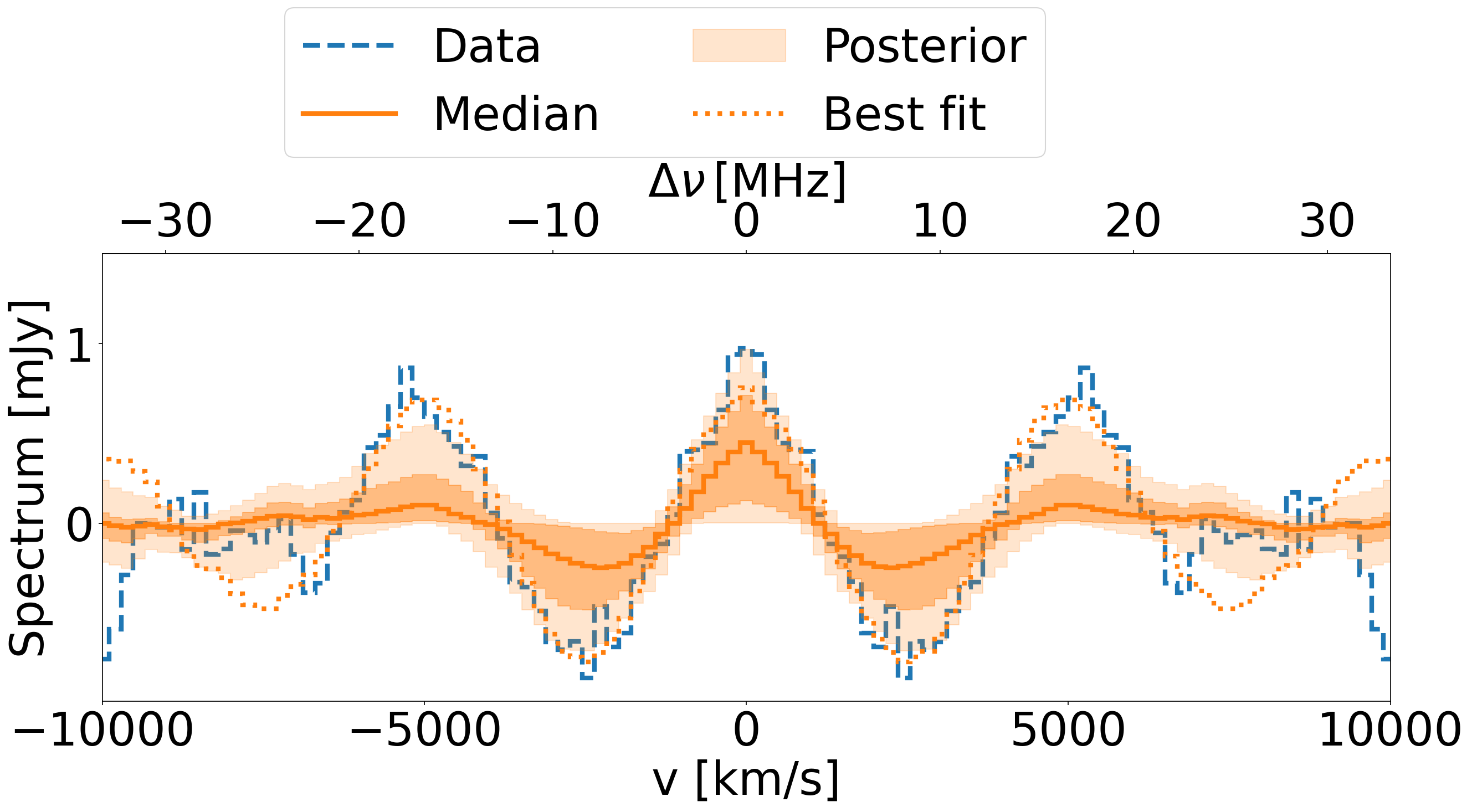}
    \caption{The posterior distribution of the fitted spectrum. The orange solid line denotes the median values of the model spectrum (``Median''). The dark shaded region shows the 68\,\% interval of the fitted spectrum, and the light shaded region shows the 95\,\% percentile. The orange dotted line shows the model spectrum of the highest log-likelihood (``Best fit''). The stacked spectrum is shown as the blue solid line for reference (``Data'').}
    \label{fig:spectrumpost}
\end{figure}

The projection effects can be better understood in the posterior of the fitted spectrum, which we show in \autoref{fig:spectrumpost}.
Due to the relatively low signal-to-noise ratio, the posterior of the model spectrum favours a small overall amplitude.
The median of the posterior has a large deviation from the best-fit model.
Around the secondary peaks of systematic oscillations, the best-fit model can be outside the 95\,\% confidence interval.
Recall that in \autoref{fig:projection}, the expected level of $\Omega_{\rm \hi}$ is also at the boundary of the $3\sigma$ confidence region.
Large \hi\ signal will produce a better fit in the central area $v\sim0$, which is weighted less compared to the secondary peaks of the systematics due to the signal covariance correction.
This leads to a stronger degeneracy between $\Omega_{\rm \hi}$ and $A_{\rm sys}$, since a lower \hi\ amplitude can always be compensated by a higher level of systematics to fit peaks and troughs of the oscillating systematics.
Therefore, when shuffling covariance is considered, higher values of $\bar{M}_{\rm \hi}$ are preferred, as the central region is weighted more by the covariance.
Since the secondary peaks are the convolution of systematics and the \hi\ signal, in the ideal case, we should expect the amplitude of the variance will also follow the oscillations.
However, as we do not have any prior information on the underlying \hi\ signal, we resort to using the current covariance.
Note that due to the signal-to-noise ratio of the measurement, we do not expect the posterior of $A_{\rm sys}$ to be well constrained.
Therefore, while a better covariance estimation gives a more accurate correlation matrix of the measured stacked spectrum, we do not expect the parameter space of $A_{\rm sys} - \Omega_{\rm \hi}$ to be tightened in the posterior.

Regardless of the covariance estimation, the constraints on the oscillation frequency $\nu_{\rm sys}$ remain robust and provide strong evidence that the systematics originate from the beam ripple of the instrument.

\section{Discussion}
\label{sec:discussion}
In this paper, we showed that the emission-line stacking of the \hi\ intensity maps can be used to examine the systematics in the data and infer the \hi\ density of the survey volume.
Using forward modelling of the stacked spectrum, Bayesian inference can be used to constrain the systematics as well as the \hi\ signal.
The findings of this paper in the context of stacking have implications on the cosmological analysis of the \hi\ clustering signal, which we discuss below.

In the power spectrum analysis, it is common to fully utilise the 3D information on the clustering in $\bm{k}$-space to maximise the information content, for example, by using the multipole clustering wedges (e.g. \citealt{2017MNRAS.467.2085G}).
For the stacked signal cube, 
the measured signal is the \hi\ density at a given separation to the galaxy position, which is similar to the two-point cross-correlation function.
It is natural to conclude that the optimal summary statistics is to model a cylindrical signal by averaging the stacked cube into $\Delta\theta-\Delta\nu$ space \citep{2025arXiv250321743D}.
However, as we explore in \secref{sec:covest}, the covariance of such a signal will be difficult to model, due to the complicated correlation between the angular pixels.
If the primary purpose of stacking is to constrain the systematics and the overall \hi\ amplitude, the stacked spectrum is sufficient and is easier to model.

In \secref{sec:results}, we explore in detail how the estimation of the \hi\ density is biased by the systematics due to the degeneracy between the two.
From this conclusion, we can envision that if there is no additional mitigation of systematics, a lower bound of the amplitude of \hi\ clustering will not be robust.
The parameter space where the \hi\ signal is extremely low and the signal is mainly driven by the systematics is within the 95\,\% interval of the posterior.
We note that the lack of constraining power on the lower limit can be resolved by simply having deeper observations.
As shown in \autoref{fig:spectrumpost}, the median of the posterior from the full four-parameter fitting scenario is much lower than the measured stacked spectrum due to the relatively low signal-to-noise ratio, and a higher detection significance would naturally exclude the parameter space of extremely low \hi\ density. 
{It is expected that current and forthcoming observations by MeerKLASS will achieve this noise requirement.} 
However, the biasing from the posterior projection effects will persist, and improvements in the data analysis to lower the systematics are needed to resolve this issue.

\begin{figure*}
    \centering
    \includegraphics[width=1.0\linewidth]{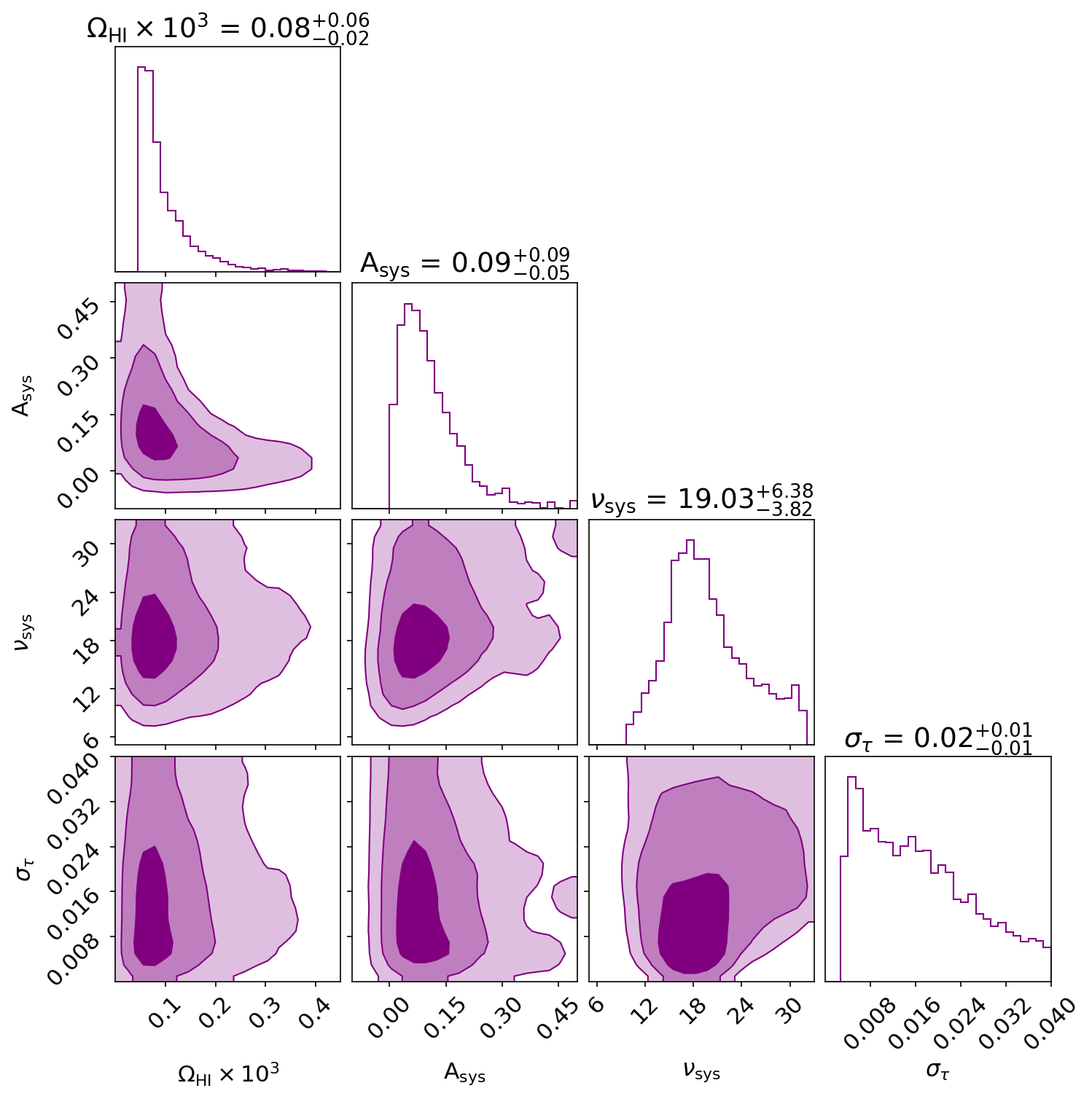}
    \caption{The same as the four-parameter case in \autoref{fig:posterior}, but with an additional prior of $\Omega_{\rm \hi} \in [5\times10^{-4},0.05]$. The posterior of $\bar{M}_{\rm HI}$ has been converted to $\Omega_{\rm \hi}$.  The outer, middle and inner contours denote the 3$\sigma$, 2$\sigma$, and 1$\sigma$ regions, respectively.}
    \label{fig:posthiprior}
\end{figure*}

{Due to the strong degeneracy, we can put mildly tighter priors on} $\Omega_{\rm \hi}$ {and obtain a constraint on $A_{\rm sys}$. We present the model fitting results for assuming a prior of} $\Omega_{\rm \hi} \in [5\times10^{-4},0.05]$ {in \autoref{fig:posthiprior}. In this case, a constraint on the systematics amplitude is achieved at $A_{\rm sys}=0.09^{+0.09}_{-0.05}$. However, as we have discussed in \secref{subsec:beamosci}, the constraint cannot be directly converted to a measurement of the beam ripple. Future work should be focused on understanding the propagation of the beam ripple to the map-level systematics, and using the measurements of the beam ripple to put informative priors on $A_{\rm sys}$, which will enable accurate measurements of} $\Omega_{\rm \hi}$.

We note that, throughout this paper, the cosmological model is fixed.
The \hi\ signal and the covariance estimation are cosmology-dependent, and varying the \hi\ galaxy clustering statistics is not discussed, as we do not have observation-driven priors on the \hi\ power spectrum.
For future data analysis of MeerKLASS survey and SKA-Mid, it is expected that the detection of \hi\ auto-power will allow a more thorough analysis of the impact of cosmology on the \hi\ stacking.
Moreover, if the modelling of the signal is good enough to estimate a covariance among the different summary statistics, we can jointly model the \hi\ stacking, the cross-power spectrum, and the \hi\ power spectrum to consistently marginalise over the clustering model parameters.
It can also maximise the science output, as additional information on the \hi\ density helps break the degeneracy between the amplitude of the matter power and the average brightness temperature.

We have shown that the covariance estimation in this work can be further improved.
If a better constraint of the amplitude of the oscillating systematics can be achieved in the future, we can simulate mock observations with the systematics to build the covariance.
Furthermore, the systematics can be expressed as an operator in the data vector, and the data covariance can be modelled analytically in the quadratic estimator formalism together with the PCA cleaning matrix (see, e.g. \citealt{2021MNRAS.501.1463K,2023MNRAS.518.2971C}).
The quadratic estimator can in turn be used to reconstruct the systematics and clean the data, as suggested in \citet{2022PhRvD.106d3534W}.

{For the mock simulation, we have ignored the redshift evolution of the stacking subsample, by assuming a constant redshift kernel and randomly assigning} \hi\ {mass from a fixed HIMF. While it is sufficient for our work given the narrow sub-band and the signal-to-noise ratio, future work should consider the redshift distribution of the sources, as well as the impact of the magnitude limit on the} \hi\ {mass distribution of the selected targets.}

In this work, we have observed that the systematics modulate the covariance of the data for both the noise and the \hi\ signal.
Therefore, in power spectrum analysis, the impact of systematics needs to be considered for modelling the covariance as well.
In \citetalias{2025MNRAS.537.3632M}, transfer function realisations are used to calculate the covariance of the power spectrum.
Since the mock realisations are injected into the data and then PCA cleaned, we expect that the systematics are included in the covariance estimation.
However, similar to the shuffling covariance in this work, the systematic effects may be distorted in covariance estimation.
The impact of covariance estimation with the existence of systematics needs to be carefully studied in the future for the power spectrum analysis. 

The properties of the systematics obtained from this work have implications for the power spectrum analysis. 
For instance, we find that the secondary peaks sourced by systematics extend up to $\sim 15-20$ MHz, corresponding to $75-100$ Mpc; hence, it is expected that the measured power spectrum at the corresponding Fourier modes $k_\parallel \sim0.08\,{\rm Mpc^{-1}}$ may be affected by these systematics.
In the ongoing data analysis (MeerKLASS collaboration, in preparation), we adopt several techniques that can mitigate the impact of systematics on the power spectrum estimation.
First, we adopt a multiscale ``mPCA'' technique where the large and small scales of the intensity maps are cleaned separately, as outlined in \citet{2024arXiv241206750C}.
Second, we perform internal cross-correlation, i.e. cross-correlating the \hi\ intensity maps obtained from different observation blocks and/or different dishes.
In \secref{subsec:beamosci}, we discussed how the beam ripple may cause calibration errors that lead to the systematics in the data.
The calibration solutions are time- and dish-dependent, and the chromatic errors may be reduced through cross-correlating datasets taken from different blocks and dishes.
Finally, in the averaging of the 1D power spectrum, we exclude the small $k_\parallel$ regions that contain the peak.

{This work sets the path to the measurement of one-point statistics from line-intensity maps. 
one-point statistics are estimators of the probability distribution function of the measured brightness temperature~\citep{2017MNRAS.467.2996B, 2024PhRvD.109d3517B} and are very sensitive to the line luminosity function. Correctly combined with the power spectrum~\citep{2019ApJ...871...75I, 2022PhRvD.106j3534S}, they can break the degeneracies between astrophysical and cosmological parameters, boosting the constraining power of line-intensity mapping~\citep{2022arXiv220901223B, 2024PhRvD.110b3507S}. 
However, these measurements are very prone to being contaminated by observational systematics: compared with stacking, they do not involve the average around resolved sources and actually account for the full temperature distribution, rather than the mean at two points. 
Nonetheless, they have in common that they use directly the line-intensity maps. 
Although there are proposals to deal with the foreground contamination in one-point statistics using one-point cross correlations~\citep{2023MNRAS.525.1824B, 2023MNRAS.520.5305C} or conditional statistics~\citep{2019PhRvL.123w1105B}, the characterisation of the systematic errors at map level that can be obtained from analysis like the one presented in this work will be key for a correct measurement of one-point statistics from the observations.}

\section{Conclusion}
\label{sec:conclusion}
In this paper, we present a comprehensive data analysis and validation pipeline for performing emission-line stacking of MeerKLASS \textit{L}-band deep-field survey intensity maps onto the positions of GAMA spectroscopic galaxies.

The MeerKLASS \textit{L}-band intensity maps at $z\sim0.4$ were observed in single-dish mode using the MeerKAT telescope, with a primary beam of $\sim1\,$deg.
The comoving transverse scale of $\sim30\,$Mpc of the beam is larger than the typical scales of dark matter halos, complicating the stacked signal of \hi\ emission.
Meanwhile, the overlapping GAMA G23 field has a number density of galaxies that is much lower than the expected number density of \hi\ galaxies, making it highly incomplete  for the stacking purposes.
We build a mock pipeline to generate the population of \hi\ galaxies and simulate the stacking of the intensity maps onto a GAMA-like galaxy sample, and find the following:
\begin{itemize}
    \item 
    Due to the large physical scale of the primary beam, the stacked cubelet of each source contains contribution from emissions from other sources, as the \hi\ emission is extended along the line of sight by the velocity dispersion and extended along the angular plane by the primary beam.
    As a result, the stacked cube contains extra \hi\ emission for all voxels within, leading to severe double counting.
    At the same time, the central region of the stacked cube has an excess \hi\ signal against the background, which is the desired \hi\ emission from the target sources.
    \item
    The extra emission from double counting is removed by the PCA cleaning procedure used to clean the foregrounds.
    When collapsed along the frequency direction into an angular image, the stacked signal after PCA cleaning shows no extra emission away from the centre pixel.
    The excess \hi\ signal in the central area follows the shape of the primary beam.
    \item 
    The stacked signal can also be averaged along the angular plane into a stacked spectrum.
    Comparing the input \hi\ emission-line profile with the output stacked spectrum, we find that the amplitude of the stacked signal is massively amplified by the extra emission from double counting and the clustering of \hi\ sources.
    The amplified signal then suffers signal loss from PCA cleaning.
    The differences in the final stacked spectrum compared to the input require forward modelling.
    \item 
    Given the depth of the MeerKLASS \textit{L}-band deep-field survey, we find that the stacked measurement is feasible, and the \hi\ signal can be detected with high statistical significance.
\end{itemize}

The viability of the stacking detection and the requirement for forward modelling call for detailed study into the modelling of the signal and its covariance.
We run 100 independent realisations of the mock observation and find that:
\begin{itemize}
    \item 
    The \hi\ signal in the stacked spectrum can be forward modelled under a simplified assumption, by assuming that the GAMA galaxy sample contains all the \hi\ mass in the survey volume and the \hi\ mass is evenly distributed among the galaxies.
    Using this assumption to rerun the stacking simulation while keeping the overall \hi\ density identical, we find that the differences between the stacked spectra of the original simulation and the simplified forward modelling are within the variance among different realisations.
    \item 
    The mock covariance can be calculated using the realisations.
    We find that, for the stacked image, the pixels in the central region are highly correlated due to the signal covariance, as the \hi\ signal is smoothed by the primary beam.
    The covariance of the stacked spectrum also shows correlation between different velocity channels, due to the double counting as the same intensity map pixel will be sampled multiple times into different channels.
    \item 
    Due to the complicated correlation of the stacked signal and the systematics in the data, a method of covariance estimation based on the data is needed.
    We test covariance estimation using random shuffling of galaxy positions.
    Each random shuffle produces a reference stacked cube that is consistent with null detection on average, and the covariance of the reference signal can be used to estimate the mock covariance.
    We find that the shuffling estimate correctly reproduces the variance away from the central region of the stacked cube where the noise variance dominates, but fails to take into account the \hi\ signal variance near the centre.
    \item 
    We find that the resulting correlation matrix in the stacked image does not match the true correlation in the mock, as the shuffling does not contain an excess signal convolved with the primary beam.
    For the stacked spectrum, the correlation matrices match relatively well, as long as the stacked spectrum is symmetrised so that only positive values of $\Delta \nu$ are considered.
    We can construct a corrected covariance estimate by using the correlation matrix from the shuffling and rescaling the variance based on the mock.
\end{itemize}

We then use the MeerKLASS intensity mapping data to perform the stacking analysis and find that:
\begin{itemize}
    \item 
    A stacking signal is detected in both the angular stacked image and the stacked spectrum.
    In the angular stacked image, we find that there is an excess signal in the central region, consistent with the primary beam.
    Using the corrected covariance estimate, we find the detection significance to be $8.66\sigma$.
    \item 
    The stacked spectrum shows a clear excess signal peak around $\Delta\nu\sim0$, while also having a clear systematics component with oscillating features with a period of $\sim 20\,$MHz.
    The detection significance is found to be $7.45\sigma$ for the unsymmetrised stacked spectrum and $5.29\sigma$ for the symmetrised spectrum.
\end{itemize}

The stacking measurement reveals a clear feature of oscillation in the spectral direction.
We investigate in detail the origin of the systematics and conclude the following:
\begin{itemize}
    \item
    The systematics most likely originates from the chromaticity of the beam.
    The diffractive interference between the primary and secondary reflector of the MeerKAT telescope modulates the beam, causing a ripple of the beam size across frequencies.
    \item 
    In the stacked spectrum, the systematics are of the same order of magnitude as the \hi\ signal.
    \item 
    The systematics is not an additive component of the data vector.
    Using the random shuffling as a null test, we do not find any feature of the systematics in the reference stacked signal.
    \item 
    The systematics not only modulate the data vector of the stacked spectrum, they also change the data covariance.
    The covariance estimation shows that in the reference image and the reference spectrum, the data vector is correlated in a way that is not seen in the mock simulation.
    Therefore, the systematics must have been convolved with the map data, which then contributes to higher-order statistics in the covariance.
    \item 
    By comparing the structure of the beam ripple and the stacked spectrum in Fourier space, we find that the peaks of the two overlap with each other, which shows that the systematics and the beam ripple share the same oscillating frequency.
    \item 
    The beam ripple induces systematics into the map data in a convoluted way that requires further investigation in the calibration and map-making pipeline.
    If the chromaticity is simply due to the beam convolving with the sky signal, we should not expect the noise covariance to be affected by the systematics.
    However, the shuffling covariance using the noise-dominated reference signal shows a clear imprint of systematics.
    We conclude that the systematic effects happen at stages of data processing prior to the foreground cleaning.
    The structure of the systematics is then introduced to the eigenmodes of the frequency-frequency covariance, which then affects the PCA cleaning.
    The PCA cleaning matrix is then operated on the entire data vector including the noise.
    \item 
    The systematics can be modelled effectively by convolving an error function with the map data along the frequency direction.
    The oscillating feature of the systematics can be parameterised in Fourier space, with varying amplitude, oscillating frequency, and the shape of the oscillation.
\end{itemize}

Including the systematics in the forward modelling, we perform Bayesian inference on the stacked spectrum using the importance nested sampling technique.
We impose wide flat priors and use the posterior of the parameter fitting to conclude the following:
\begin{itemize}
    \item 
    The model fitting routine gives constraints on the average \hi\ mass of the GAMA galaxies under the simplified assumption, which can be converted to an effective constraint on the \hi\ density in the survey volume.
    The amplitude of the systematics is not well constrained, with the posterior occupying the wide prior volume.
    \item 
    By varying the number of free parameters in the systematics modelling, we find that the constraints on the \hi\ density are consistent under different modelling complexity.
    The Bayesian evidence for using the shape of the beam ripple and fitting just the amplitude of the systematics is consistent with the case of varying the oscillation frequency.
    It further supports the fact that the beam ripple describes the systematics well.
    \item
    The fitting gives a constraint on the oscillating frequency of the systematics that is consistent with the beam ripple reported in previous literature.
    If the full shape of the systematics is varied instead of just the frequency, the constraining power degrades while the posterior of the oscillation frequency is consistent, suggesting the robustness of the constraints.
    The conservative case gives an estimation of the frequency to be $\nu_{\rm sys}={17.90}_{-4.27}^{+6.53}\,$MHz.
    \item 
    The estimation of the \hi\ density is found to be lower than expected based on the measurements at similar redshifts.
    We find that the covariance estimation impacts the estimation of the systematics amplitude.
    Due to the strong degeneracy between the systematics amplitude and the \hi\ density, the deviation of the estimated covariance and the true covariance may contribute to the underestimation.
    \item 
    In the parameter space of the systematics amplitude and the \hi\ density, the orientation of the degeneracy changes with the systematics amplitude, leading to strong posterior projection effects.
    In the posterior of the fitted stacked spectrum, we find that the best-fit spectrum is near the boundary of the 95\,\% confidence interval of the posterior at the secondary peaks of the systematics.
    Due to the limited signal-to-noise ratio, there is a lack of constraining power in the systematics amplitude, and therefore, the constraints on the \hi\ density are not robust.
\end{itemize}

Our findings provide strong incentive to include the stacking analysis in the cosmological analysis of the \hi\ intensity mapping data in cross-correlation with optical galaxies.
The stacking measurement is a powerful tool to validate the detection of the \hi\ signal and examine the quality of the data for residual systematics.
The stacked spectrum can be modelled to infer the \hi\ density, providing a unique window for measuring the evolution of cosmic \hi\ across different redshifts using future \hi\ intensity mapping data.
As the data quality and depth improve for future MeerKLASS survey and SKAO, we expect that the stacking analysis can help disentangle the \hi\ density and the matter clustering amplitude, while also allowing for the modelling of residual systematics as nuisance parameters in the cosmological analysis.
It will serve as a robust summary statistic to maximise the constraining power of the data, and our work provides the first analysis of its kind as a starting point to build robust inference methods toward future SKAO.

\section{Acknowledgments}
ZC and AP are funded by a UKRI Future Leaders Fellowship [grant
MR/X005399/1; PI: Alkistis Pourtsidou].
%
SCu acknowledges support from the UKRI Stephen Hawking Fellowship (grant reference EP/U536751/1) and was also supported by a UKRI Future Leaders Fellowship grant [MR/V026437/1].
JLB acknowledges funding from the Ram\'on y Cajal Grant RYC2021-033191-I, financed by MCIN/AEI/10.13039/501100011033 and by the European Union ``NextGenerationEU''/PRTR, as well as the project UC-LIME (PID2022-140670NA-I00), financed by MCIN/AEI/10.13039/501100011033/FEDER, UE.
SCa acknowledges support from the Italian Ministry of University and Research (\textsc{mur}), PRIN 2022 `EXSKALIBUR – Euclid-Cross-SKA: Likelihood Inference Building for Universe's Research', Grant No.\ 20222BBYB9, CUP D53D2300252 0006, from the Italian Ministry of Foreign Affairs and International
Cooperation (\textsc{maeci}), Grant No.\ ZA23GR03, and from the European Union -- Next Generation EU. 
IPC is supported by the European Union within the Next Generation EU programme [PNRR-4-2-1.2 project No. SOE\textunderscore0000136, RadioGaGa].
JF acknowledges support of Funda\c{c}\~{a}o para a Ci\^{e}ncia e a Tecnologia through the Investigador FCT Contract No. 2020.02633.CEECIND/CP1631/CT0002, and the research grants UIDB/04434/2020 and UIDP/04434/2020.
MGS acknowledges support from the South African Radio Astronomy Observatory and National Research Foundation (Grant No. 84156).
The MeerKAT telescope is operated by the South African Radio Astronomy Observatory, which is a facility of the National Research Foundation, an agency of the Department of Science and Innovation. We acknowledge the use of the Ilifu cloud computing facility, through the Inter-University Institute for Data Intensive Astronomy (IDIA).

We thank Ludwig Schwardt, Mattieu de Villiers and Dirk de Villiers for discussions on the primary beam of the MeerKAT telescope.

%

\vspace{5mm}
\facility{MeerKAT}


\software{
\textsc{numpy} \citep{2020Natur.585..357H},
\textsc{scipy} \citep{2020NatMe..17..261V},
\textsc{astropy} \citep{2022ApJ...935..167A},
\textsc{matplotlib} \citep{Hunter:2007}
}



\appendix

\section{Impact of the covariance estimation on the systematics}
\label{apdx:covanalytic}
In this appendix, we briefly discuss the limits of the covariance estimation routine presented in \secref{sec:covest}.
Specifically, we aim to examine the impact of the mock-corrected covariance in \autoref{eq:covcorrected}.

The stacked signal can be expressed as a data vector $\vec{d}$.
In the mock, the true stacked signal can be written as a combination of the \hi\ signal and noise,
\begin{equation}
    \vec{d}_{\rm mock} = \vec{d}_{\rm mock}^{\rm \: \hi} + \vec{d}_{\rm mock}^{\rm \: n},
\end{equation}
assuming that the foreground has been sufficiently removed.
The true mock covariance can then be expressed as a combination of the \hi\ covariance and the noise covariance,
\begin{equation}
    \langle \vec{d}_{\rm mock} \vec{d}_{\rm mock}^{\rm \: T} \rangle = \mathbf{C}^{\rm mock}_{\rm \hi} + \mathbf{C}^{\rm mock}_{\rm n},
\end{equation}
where $\langle \rangle$ denotes the assemble average.

On the other hand, the shuffled data vector $\vec{d}_{\rm shuffle}$, on average, contains only the noise component.
Assuming that the shuffling, on average, reflects the sampling of the pixels of the real galaxy catalogue, then the covariance of the shuffled data vector is simply
\begin{equation}
    \mathbf{C}^{\rm mock,shuffle} = \langle \vec{d}_{\rm mock, shuffle} \vec{d}_{\rm mock,shuffle}^{\rm \: T} \rangle = \mathbf{C}^{\rm mock}_{\rm n}.
\end{equation}

The discrepancy between the two covariances is in the amplitude as well as the correlation, as we have discussed in \secref{subsec:shuffle}.
In \autoref{eq:covcorrected}, we defined a correction, which can be written as
\begin{gather}
    \mathbf{R} = {\rm diag}[\vec{r}],\\
    \vec{r}_i = \sqrt{\mathbf{C}^{\rm mock}_{ii}/\mathbf{C}^{\rm mock,shuffle}_{ii}},\\
    \hat{\mathbf{C}}^{\rm mock} =  \langle \mathbf{R}\, \vec{d}_{\rm mock,shuffle} \vec{d}_{\rm mock,shuffle}^{\rm \: T} \, \mathbf{R}^{\rm T} \rangle = \mathbf{R} \mathbf{C}^{\rm mock}_{\rm n} \mathbf{R}^{\rm T},
\end{gather}
where diag[$\vec{r}$] denotes a diagonal matrix with $\vec{r}$ as its diagonal elements.
It is straightforward to see that the diagonal elements of $\hat{\mathbf{C}}^{\rm mock}$ are equal to those of the true covariance ${\mathbf{C}}^{\rm mock}$, whereas the correlation follows $\mathbf{C}^{\rm mock,shuffle}$ and therefore $\mathbf{C}^{\rm mock}_{\rm n}$.
This introduces a slight underestimation of correlation at intermediate intervals of $\Delta \nu$ as seen in \autoref{fig:corrmix}.

We then apply the correction to the data.
The data vector can be written as a multiplicative systematic operator $\mathbf{S}$ on the underlying \hi\ and noise data\footnote{Note that, in reality, the systematics operators on the \hi\ data and the noise should be different, with the noise only having systematic effects through the PCA cleaning matrix. Here, for simplicity, we write them as one matrix $\mathbf{S}$. Note that the derivation for the distortion of covariance is not affected by this simplification.}
\begin{equation}
    \vec{d}_{\rm data} = \mathbf{S} \big(\vec{d}_{\rm data}^{\rm \: \hi} + \vec{d}_{\rm data}^{\rm \: n} \big).
\end{equation}
The covariance of the data is then
\begin{equation}
    \mathbf{C}^{\rm data} = \langle \vec{d}_{\rm data} \vec{d}_{\rm data}^{\rm \: T}\rangle = \mathbf{S} \big(\mathbf{C}^{\rm data}_{\rm \hi}+\mathbf{C}^{\rm data}_{\rm n}\big)\mathbf{S}^{\rm T}.
\label{eq:cdata}
\end{equation}

The reference stacked signal $\vec{d}_{\rm data,shuffle}$, on the other hand, contains only the systematics and the noise. The estimated covariance matrix can then be written as
\begin{equation}
    \hat{\mathbf{C}}^{\rm data} = \langle \mathbf{R}\, \vec{d}_{\rm data,shuffle}\vec{d}_{\rm data,shuffle}^{\rm \: T} \, \mathbf{R}^{\rm T} \rangle 
    =  \mathbf{R}\, \mathbf{S}\,\mathbf{C}^{\rm data}_{\rm n}\, \mathbf{S}^{\rm T} \mathbf{R}^{\rm T}.
\end{equation}

We can rewrite $\hat{\mathbf{C}}^{\rm data}$ so that
\begin{equation}
   \hat{\mathbf{C}}^{\rm data} = \mathbf{R}\, \mathbf{S}\,\mathbf{R}^{-1}\,\mathbf{R}\,\mathbf{C}^{\rm data}_{\rm n}\,\mathbf{R}^{\rm T}\,(\mathbf{R}^{\rm T})^{-1}\, \mathbf{S}^{\rm T} \mathbf{R}^{\rm T} 
   = \tilde{\mathbf{S}}_{\mathbf{R}}\,\mathbf{R}\,\mathbf{C}^{\rm data}_{\rm n}\,\mathbf{R}^{\rm T}\,\tilde{\mathbf{S}}_{\mathbf{R}}^{\rm T},
\label{eq:chatdata}
\end{equation}
where we have defined a new matrix $\tilde{\mathbf{S}}_{\mathbf{R}} = \mathbf{R}\, \mathbf{S}\,\mathbf{R}^{-1} $.
Assuming that the mock correctly reflects the amplitude of the \hi\ and noise signal in the data, 
$\mathbf{R}\,\mathbf{C}^{\rm data}_{\rm n}\,\mathbf{R}^{\rm T} \approx \mathbf{C}^{\rm data}_{\rm \hi}+\mathbf{C}^{\rm data}_{\rm n}$ which is the target data covariance without the systematics.
Comparing \autoref{eq:cdata} with \autoref{eq:chatdata}, we can see that the covariance estimate indeed includes the systematics.
However, the systematics are distorted by the correction matrix $\mathbf{R}$.

Note that $\mathbf{R}$ is diagonal, so that
\begin{equation}
    \big(\tilde{\mathbf{S}}_{\mathbf{R}}\big)_{ij} = \frac{r_i}{r_j} \mathbf{S}_{ij}.
\end{equation}
The values of $\vec{r}$ are consistent with 1 at large values of $|\Delta \nu|$ and larger than 1 at small values, as we have shown in \autoref{fig:spectralmockcov}.
As a result, $\tilde{\mathbf{S}}_{\mathbf{R}}$ has a mismatch with the true $\mathbf{S}$, therefore biasing the covariance and the subsequent inference.

As we have no prior knowledge on the amplitude of the systematics, we resort to using the shuffling for covariance estimation as a way of including the systematics blindly in the data.
Alternatively, if no correction is made, the underlying data covariance is distorted, which biases the covariance and the inference differently.
By quantifying the differences between the two covariances, we can examine the effect of the imperfect covariance estimation, which we discuss in \secref{subsec:diffcov}.

\bibliography{sample631}{}
\bibliographystyle{aasjournal}


\end{CJK*}
\end{document}